\documentclass[IOP]{emulateapj}
\usepackage{hyperref}
\usepackage{natbib}
\usepackage{amsmath}
\usepackage{nicefrac}
\usepackage[usenames,dvipsnames]{xcolor}
\usepackage{graphicx}
\usepackage{footnote}
\usepackage{rotating}
\usepackage{slashbox}
\shorttitle{LGRB World Model}
\shortauthors{Shahmoradi}

\def\gtrsim{\mathrel{\hbox{\rlap{\hbox{\lower4pt\hbox{$\sim$}}}\hbox{$>$}}}}
\def\lessim{\mathrel{\hbox{\rlap{\hbox{\lower4pt\hbox{$\sim$}}}\hbox{$<$}}}}
\newcommand{\liso}{L_{iso}}
\newcommand{\eiso}{E_{iso}}
\newcommand{\epkz}{E_{p,z}}
\newcommand{\durz}{T_{90,z}}
\newcommand{\pbol}{P_{bol}}
\newcommand{\sbol}{S_{bol}}
\newcommand{\epk}{E_{p}}
\newcommand{\dur}{T_{90}}
\newcommand{\pph}{P_{50-300}}

\makeatletter
\providecommand*{\diff}{\@ifnextchar^{\DIfF}{\DIfF^{}}}\def\DIfF^#1{\mathop{\mathrm{\mathstrut d}}\nolimits^{#1}\gobblespace}\def\gobblespace{\futurelet\diffarg\opspace}\def\opspace{\let\DiffSpace\!\ifx\diffarg(\let\DiffSpace\relax\else\ifx\diffarg\[\let\DiffSpace\relax\else\ifx\diffarg\{\let\DiffSpace\relax\fi\fi\fi\DiffSpace}

\begin{document}

\setlength{\pdfpageheight}{\paperheight}
\setlength{\pdfpagewidth}{\paperwidth}

\title{A Multivariate Fit Luminosity Function and World Model for Long GRBs}
\author{Amir Shahmoradi}
\affil{Institute for Fusion Studies, The University of Texas at Austin, Texas 78712, USA}
\email{Email: amir@physics.utexas.edu}

\begin{abstract}
    It is proposed that the luminosity function, the rest-frame spectral correlations and distributions of cosmological Long-duration (Type-II) Gamma-Ray Bursts (LGRBs) may be very well described as multivariate log-normal distribution. This result is based on careful selection, analysis and modeling of LGRBs' temporal and spectral variables in the largest catalog of Gamma-Ray Bursts available to date: 2130 BATSE GRBs, while taking into account the detection threshold and possible selection effects. Constraints on the joint rest-frame distribution of the isotropic peak luminosity ($\liso$), total isotropic emission ($\eiso$), the time-integrated spectral peak energy ($\epkz$) and duration ($\durz$) of LGRBs are derived. The presented analysis provides evidence for a relatively large fraction of LGRBs that have been missed by BATSE detector with $\eiso$ extending down to $\sim10^{49}~ [erg]$ and observed spectral peak energies ($\epk$) as low as $\sim5~[keV]$. LGRBs with rest-frame duration $\durz\lesssim1[s]$ or observer-frame duration $\dur\lesssim2[s]$ appear to be rare events ($\lesssim0.1\%$ chance of occurrence). The model predicts a fairly strong but highly significant correlation ($\rho=0.58\pm0.04$) between $\eiso$ \& $\epkz$ of LGRBs. Also predicted are strong correlations of $\liso$ \& $\eiso$ with $\durz$ and moderate correlation between $\liso$ \& $\epkz$. The strength and significance of the correlations found, encourage the search for underlying mechanisms, though undermine their capabilities as probes of Dark Energy's equation of state at high redshifts. The presented analysis favors -- {\it but does not necessitate} -- a cosmic rate for BATSE LGRBs tracing metallicity evolution consistent with a cutoff ${Z/Z_{\Sun}}\sim0.2-0.5$, assuming no luminosity-redshift evolution.
\end{abstract}


\maketitle

\section{Introduction}
    \label{sec:intro}

    Ever since the discovery of the first Gamma-Ray Burst (GRB) by the Vela Satellites in 1967 \citep{klebesadel_Observations_1973}, there has been tremendous effort and attempts to constrain the energetics, luminosity function and the underlying mechanism responsible for these events. Early observations of Konus \citep{mazets_recent_1981} and Ginga \citep{fenimore_interpretations_1988,nishimura_what_1988} gamma/X-ray instruments suggested a possible link between GRBs and Neutron Stars with output energy ranges of the order of $\sim10^{40}~[erg]$. With the launch of Compton Gamma-Ray Observatory (CGRO), Burst And Transient Source Experiment (BATSE) onboard CGRO dramatically changed the understanding of GRBs. While previous catalogs \citep[e.g.,][]{atteia_second_1987} indicated an isotropic distribution of GRB sources, the BATSE observations extended this isotropy down to the weakest bursts. The non-homogenous \citep[e.g.,][]{fenimore_intrinsic_1993} and isotropic spacial event distribution \citep[e.g.,][]{meegan_spatial_1992,briggs_dipole_1993,fishman_first_1994} provided, for the first time, strong support for a cosmological versus galactic origin of GRBs, undermining neutron stars in the local universe as the potential candidates for some -- if not all -- classes of gamma-ray events. Furthermore, the joint duration-hardness distribution of GRBs provided a direct evidence for at least two separate classes of Gamma-Ray Bursts: Long-soft vs. short-hard \citep[e.g.,][also Figure \ref{fig:classification} here]{kouveliotou_identification_1993}.

    The possibility of a cosmological origin for GRBs, indicated an enormous output energy on the order of $\sim10^{51}~[erg]$ \citep[e.g.,][]{dermer_statistics_1992}. Nevertheless, an accurate description of the GRB Luminosity Function (LF) also required a knowledge of GRB cosmic rate, an information that could not be extracted from BATSE observations alone. This became possible only with the launch of the Italian-Dutch X-ray satellite BeppoSax \citep{boella_bepposax_1997} and the identification of the first GRB with firmly measured cosmological redshift \citep{metzger_spectral_1997} that marked the beginning of the afterglow era in the field of Gamma-Ray Bursts. The launch of Swift satellite \citep{gehrels_swift_2004} was another milestone that revolutionized the study of GRBs by facilitating the X-ray afterglow observations \citep{burrows_swift_2005} and further ground-based follow-ups for redshift measurement.

    Alongside the observational triumphs over a few decades, several theoretical models have stood up against the rivals based on the available evidence and GRB data. Most prominently, the Collapsar model \citep[e.g.,][]{woosley_gamma-ray_1993} has been relatively successful in linking the Long-duration class of Gamma-Ray Bursts (LGRBs) to the final stages in the lives of massive stars while the Short-duration class of bursts (SGRBs) is generally attributed to the coalescence of compact binary systems \citep[e.g.,][and references therein]{paczynski_gamma-ray_1986, nakar_short-hard_2007}. The two classes of SGRBs \& LGRBs in this work correspond to Type I \& Type II GRBs respectively, according to the physical classification scheme of \citet{zhang_making_2007} \& \citet{bloom_gamma-ray_2008}. Further refinement of the potential candidates as the progenitors and the emission mechanism for both classes, requires more rigorous analysis of observational data in all possible energy frequencies. In particular, the prompt gamma-ray emission of LGRBs has been subject of intense observational and theoretical studies:

    Beginning with BATSE observations, numerous authors have examined the prompt emission of LGRBs searching for potential underlying correlations among the spectral parameters \citep[e.g.,][]{nemiroff_gross_1994, fenimore_gamma-ray_1995, mallozzi_nu_1995, Petrosian_fluence_1996, Brainerd_cosmological_1997, Dezalay_hardness-intensity_1997, Petrosian_cosmological_1999, lloyd_cosmological_2000,norris_long-lag_2005}. The lack of known redshifts for BATSE events and poor knowledge of LGRB cosmic rates, however, strongly limited the prediction power of such analyses. Instead, the first direct evidence for potential correlations and constraints on the distributions of the prompt emission spectral parameters came with a few LGRBs detected by BATSE, BeppoSax, IPN or HETE-II satellites with measured redshifts \citep[e.g.,][]{reichart_construction_2001, amati_intrinsic_2002, ghirlanda_collimation-corrected_2004, yonetoku_gamma-ray_2004} and was further developed by the inclusion of Swift LGRBs \citep[e.g.,][]{schaefer_hubble_2007, gehrels_gamma-ray_2009}. Such findings, however, have been criticized for relying primarily on a handful of events with high Signal-to-Noise Ratio (SNR) required for spectral analysis with afterglows sufficiently bright for redshift measurement, arguing that the proposed joint distributions of the spectral properties, do not represent the entire underlying population of LGRBs \citep[e.g.,][]{band_testing_2005, nakar_outliers_2005, li_redshift_2007, butler_complete_2007, butler_generalized_2009, shahmoradi_how_2009, shahmoradi_hardness_2010, shahmoradi_possible_2011}. Responding to criticisms, attempts were made to model the effects of different gamma-ray instruments' detection thresholds and the limiting effects of spectral analysis \citep[e.g.,][c.f. \citet{shahmoradi_possible_2011} for a review of relevant literature]{ghirlanda_e_2008, nava_peak_2008}.

    Despite significant progress, difficulties in modeling the complex effects of detector threshold on the multivariate distribution of the prompt-emission spectral properties and the lack of a sufficiently large sample of uniformly detected LGRBs has led the community to focus on individual spectral variables, most importantly the luminosity function \citep[e.g.,][]{petrosian_interpretation_1993, schmidt_luminosities_1999, kommers_intensity_2000, kumar_energetics_2000, band_energy_2001, porciani_association_2001, sethi_luminosity_2001, schmidt_luminosity_2001, stern_evidence_2002, guetta_luminosity_2005, salvaterra_gamma-ray_2007, salvaterra_evidence_2009, schmidt_gamma-ray_2009, campisi_redshift_2010, wanderman_luminosity_2010, salvaterra_complete_2012}.  Recently, \citet[][hereafter B10]{butler_cosmic_2010} presented an elaborate multivariate analysis of Swift LGRBs, including the potential correlations among three temporal and spectral variables: total isotropic energy emission ($\eiso~[erg]$), a definition of duration ($T_{r45}~[s]$), and the spectral peak energy ($\epkz~[keV]$) of the bursts. Focusing their analysis on the $\eiso-\epkz-T_{r45}$ interrelation, B10 find a strong and significant correlation -- but with a {\it broad} scatter -- between the isotropic emission and the peak energy of the Swift LGRBs. Also realized by B10, is the possibility of a positive -- and perhaps strong -- correlation between the duration and the total energy output of the bursts. Moreover, to accommodate the potential existence of a large population of sub-luminous events, a broken power-law luminosity function (LF) is used by B10.

    It is known that the energy budget of Gamma-Ray Bursts must be limited and a turnover in the LF at low-luminosity tail of the population is expected. However, the choice of the broken power-law as the candidate LF is justified by the fact that current observational data cannot constrain this turnover point -- expected to be far below the detection threshold of current gamma-ray instruments. The existence of a turnover point in the LF can have important clues for the underlying physics of LGRBs. The low-luminosity tail of the LF in such case would be the result of the convolution of the Stellar Mass distribution with the LGRB rate as a function of the properties of massive dying stars, imposed by the mechanism.
    \\
    Motivated by the search for the potential shape of the LGRB LF at the low-luminosity and the flurry of recent reports on the possible existence of correlations among LGRBs' temporal and spectral variables, the author of this manuscript has considered a wide variety of different statistical models that could incorporate and explain all observed correlations and spectral distributions, while preserving the prediction power of the model for the spectral properties of potentially large fraction of LGRBs that could go undetected by the current gamma-ray detectors. Here it will be shown that it is indeed possible to construct an LGRB world model capable of describing most (if not all) prompt-emission spectral and temporal properties observed in the current data catalogs. Towards this, the presented analysis is mainly focused on the largest catalog of GRBs available to date: BATSE catalog of $2130$ GRBs \citep[][]{paciesas_fourth_1999}. The author shows that there is still an enormous amount of information buried in BATSE observations that need to be dug out by GRB researchers.\\

    In the following sections, an example of such data mining on BATSE catalog will be presented: Section \ref{sec:samsel} presents an elaborate method for classifying Gamma-Ray Bursts into the two known subclasses of Short and Long duration (SGRBs \& LGRBs). In Section \ref{sec:MC} an LGRB world model capable of describing BATSE observations is presented, followed by a discussion of the procedure for fitting the model to data in Section \ref{sec:MF}. Univariate and multivariate goodness-of-fit tests are performed to ensure the model predicts BATSE data accurately (Section \ref{sec:GOF}). The results from model fitting and implications on the multivariate distribution of the prompt emission spectral properties are discussed in Section \ref{sec:RD}. A summary of the analysis and important outcomes of the proposed LGRB world model will be presented in Section \ref{sec:SCR}.

\section{GRB World Model}
    \label{sec:GWM}

    \subsection{Sample Selection}
    \label{sec:samsel}

\begin{figure}
            \includegraphics[scale=0.31]{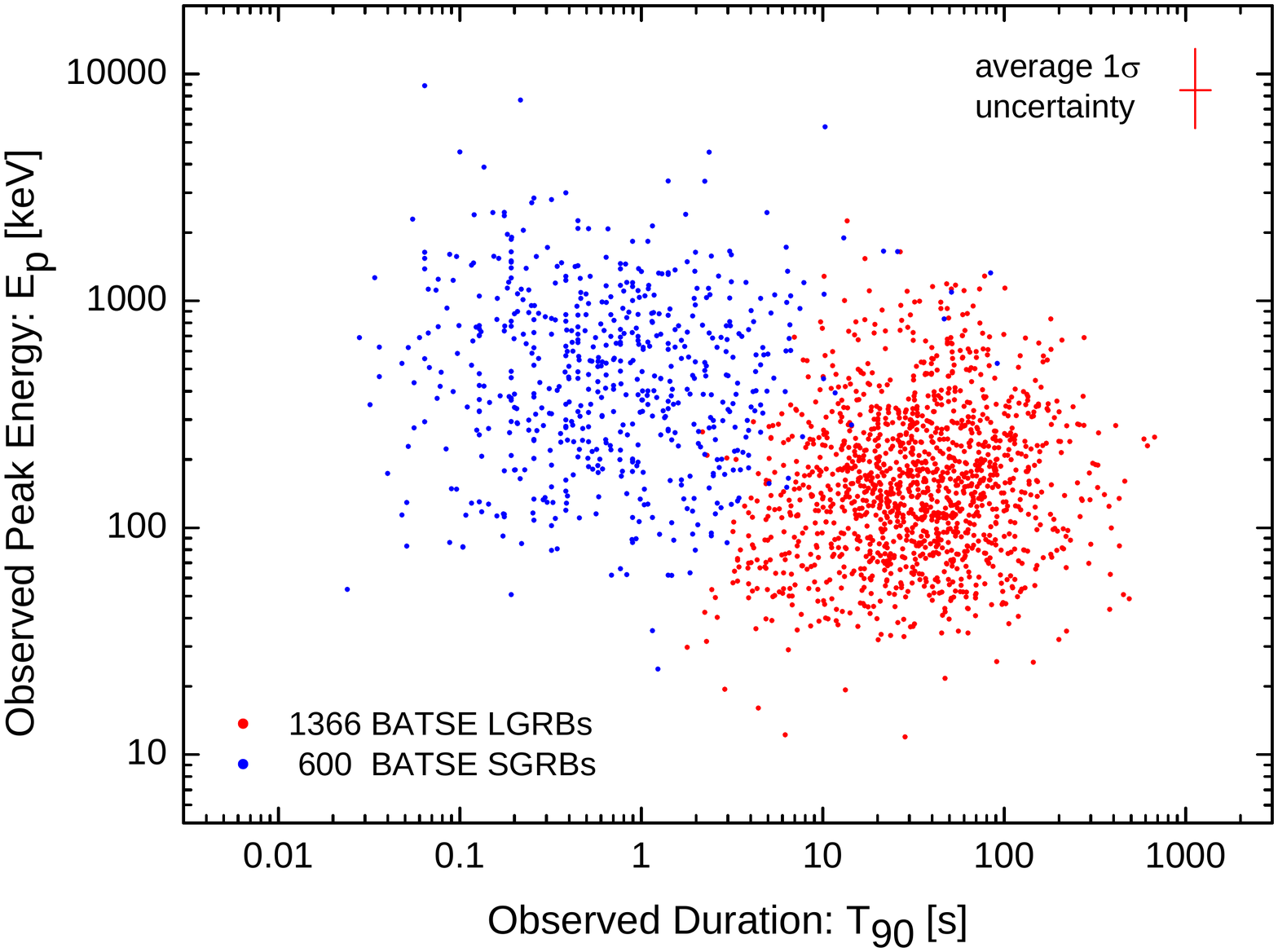}
            \includegraphics[scale=0.31]{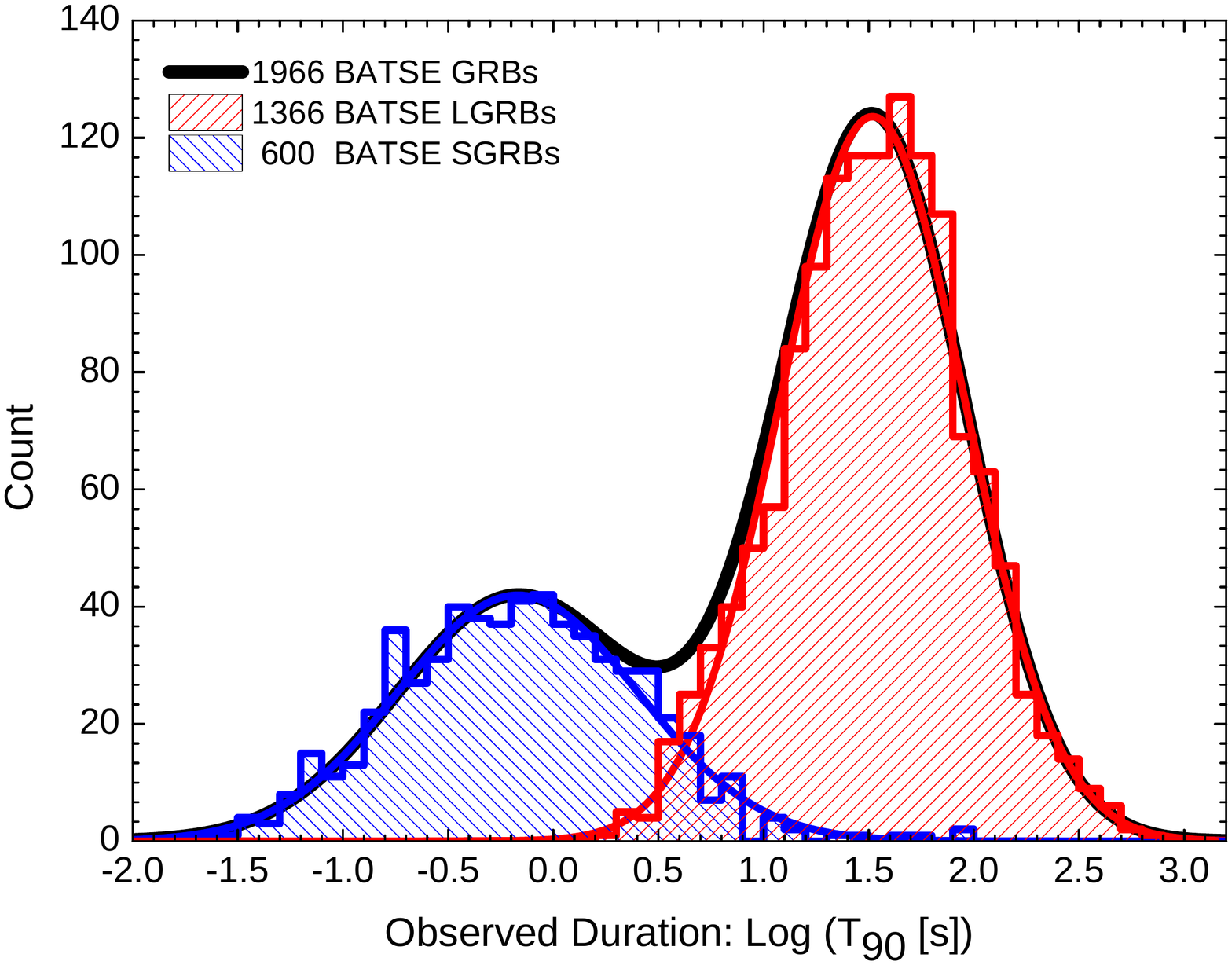}
            \includegraphics[scale=0.31]{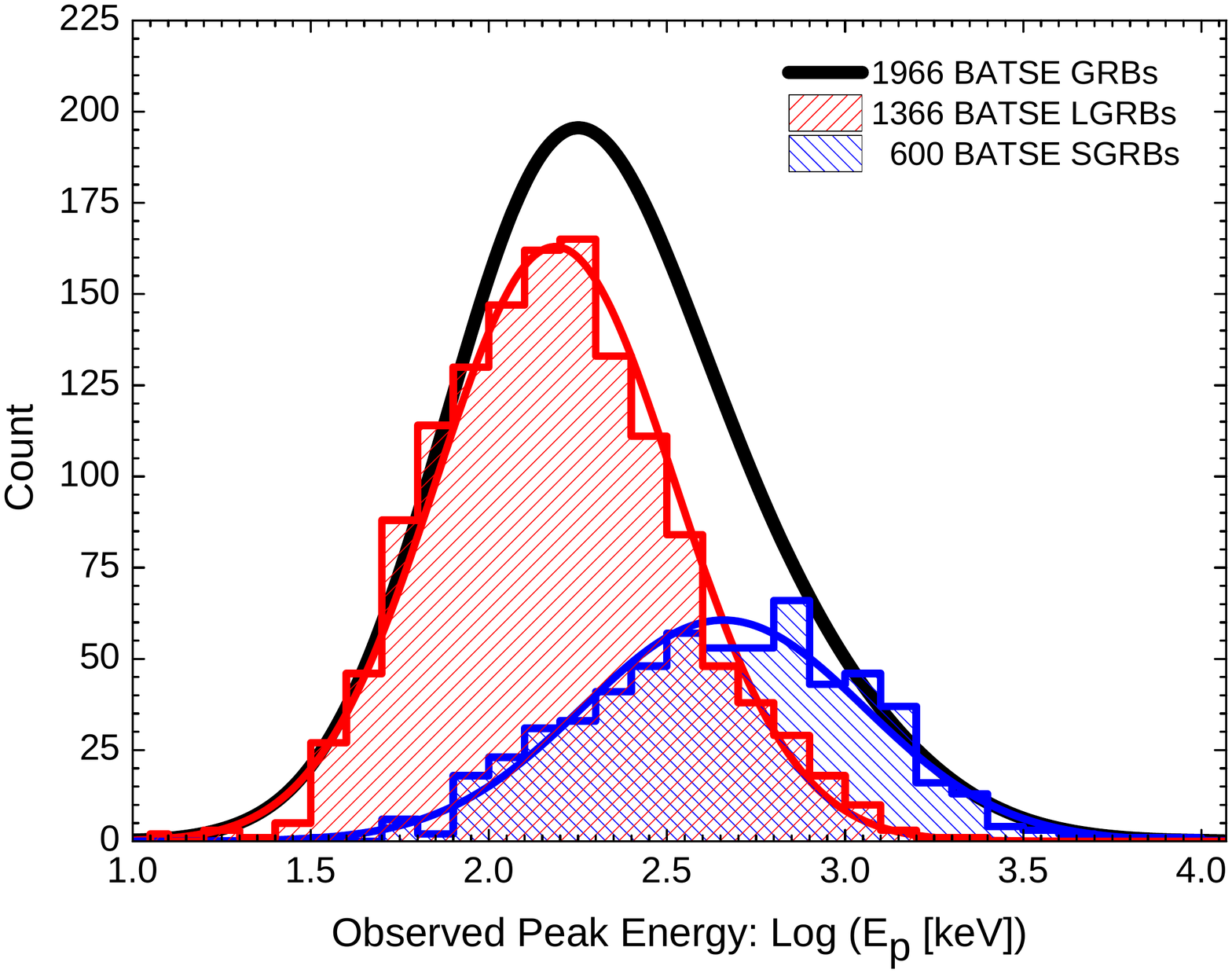}
            \caption{Classification of $1966$ BATSE LGRBs according to the most suitable clustering algorithm and set of GRB variables (Section \ref{sec:samsel}): Fuzzy C-means classification on $\epk$ \& $\dur$.  \textcolor{red}{Red} \& \textcolor{blue}{blue} colors represent \textcolor{red}{LGRB} \& \textcolor{blue}{SGRB} classes respectively, in all three plots. {\it Top}: The joint $\dur-\epk$ distribution. The uncertainties in LGRBs are derived from the Empirical Bayes model discussed in Appendix \ref{sec:appC}. {\it Center}: $\dur$ distribution. {\it Bottom}: $\epk$ distribution. $\epk$ estimates are taken from \citet{shahmoradi_hardness_2010}. Compare this plot to the plot of Figure \href{http://adsabs.harvard.edu/abs/2010MNRAS.407.2075S}{(13)} of \citet{shahmoradi_hardness_2010}, where the entire univariate $\epk$ distribution of BATSE GRBs were fit by a 2-component Gaussian mixture. \label{fig:classification}}
        \end{figure}

        Depending on their detection criteria, some gamma-ray detectors might facilitate detection of one class of bursts over the others. For example, the specific detector sensitivity of the Burst Alert Telescope (BAT) onboard Swift satellite \citep{gehrels_swift_2004} results in better detections of LGRBs over SGRBs \citep[e.g.,][]{band_comparison_2003, band_postlaunch_2006}. Therefore, a simple GRB classification method, such as a cutoff in the observed duration distributions of GRBs ($\dur\sim3[s]$) generally results in LGRB samples that are minimally contaminated by SGRBs \citep[e.g.,][]{butler_cosmic_2010}. Compared to BAT, however, BATSE Large Area Detectors (LAD) had an increased {\it relative} sensitivity to SGRBs \citep[e.g.,][]{band_comparison_2003}. Knowing that many temporal and spectral properties of Long \& Short -duration GRBs, most importantly $\dur$, overlap, identification of LGRBs in the BATSE catalog solely based on their $\dur$ will likely result in significant number of misclassifications (e.g., Figure \ref{fig:classification}, {\it center panel}).

        Thus, to ensure a correct analysis of long-duration class of GRBs, it is first necessary to collect the least biased sample of events, all belonging to LGRB category. The word `bias' here refers to the bias that might be introduced when using the traditional definition of GRB classes, based on a sharp cutoff on the duration variable $\dur$ \citep[][]{kouveliotou_identification_1993}, as it is generally assumed by many GRB researchers \citep[e.g.,][]{guetta_luminosity_2005, butler_cosmic_2010, campisi_redshift_2010, wanderman_luminosity_2010} or not assumed or explicitly discussed \citep[e.g.,][]{salvaterra_gamma-ray_2007, salvaterra_evidence_2009, salvaterra_complete_2012}.  This goal is achieved by testing extensive varieties of classification and clustering methods, most importantly, the fuzzy clustering algorithms of \citet{rousseeuw_fuzzy_1996} and the method of fuzzy C-means \citep{dunn_fuzzy_1973, bezdek_pattern_1981}. Each BATSE-catalog GRB is assigned a probability (i.e., class coefficient) of belonging to LGRB (versus SGRB) population according to the choice of clustering algorithm and the set of GRB variables used. This can be any combinations of the peak flux ($\pph~[photon/s/cm^2]$) in BATSE detection energy range, in three different time scales: $64$, $256$ \& $1024~[ms]$; bolometric fluence ($\sbol~[erg/cm^2]$); the spectral peak energy ($\epk~[keV]$) estimates by \citet{shahmoradi_hardness_2010}; duration ($\dur$~[s]) and the Fluence-to-Peak-Flux Ratio ($FPR~[s]$). Then GRBs with LGRB class coefficient $>0.5$ are flagged as Long-duration class bursts. Overall, the fuzzy C-means classification method with the two GRB variables $\epk$ \& $\dur$ is preferred over other clustering methods and sets of GRB variables (c.f. Appendix \ref{sec:appA}). This leads to the selection of $1376$ events as LGRBs out of 1966 BATSE GRBs having measured temporal and spectral parameters mentioned above. \footnote{Data for 1966 BATSE GRBs with firmly measured peak flux, fluence and duration are taken from The BATSE Gamma Ray Burst Catalogs: \url{http://www.batse.msfc.nasa.gov/batse/grb/catalog/}. The spectral peak energy ($\epk$) estimates of these events are taken from \citet{shahmoradi_hardness_2010}, also available for download at: \url{https://sites.google.com/site/amshportal/research/aca/in-the-news/lgrb-world-model}.}

        As a further safety check to ensure minimal contamination of the sample by SGRBs, the lightcurves of $291$ bursts among $1966$ BATSE GRBs with LGRB class coefficients in the range of $0.3-0.7$ are visually inspected in the four main energy channels of BATSE Large Area Detectors (LAD). This leads to reclassification of $17$ events (originally flagged as LGRB by the clustering algorithm) to potentially SGRB or Soft Gamma Repeater (SGR) events, and reclassification of $7$ events (originally flagged as SGRB by the clustering algorithm) to LGRBs. The result is a reduction in the size of the original LGRB sample from $1376$ to $1366$ (Table \ref{tab:data}). It is notable that the inclusion of the uncertainties on the two GRB variables $\dur$ \& $\epk$ turns out to not have significant effects on the derived samples of the two GRB classes discussed above. Also, a classification based on $T_{50}$ instead of $\dur$ results in about the same samples for the two GRB classes with only negligible difference of $\sim0.7\%$.

    \subsection{Model Construction}
    \label{sec:MC}

        The goal of the presented analysis is to derive a multivariate model that is capable of reproducing the observational data of $1366$ BATSE LGRBs. Examples of multivariate treatment of LGRB data are rare in Gamma-Ray Burst literature, with the most recent (and perhaps the only) example of such work presented by B10. Conversely, many authors have focused primarily on the univariate distribution of the spectral parameters, most importantly on the luminosity function (LF). A variety of univariate models have been proposed as the LGRB LF and fit to data by approximating the complex detector threshold as a step function \citep{schmidt_luminosities_1999} or an efficiency grid (e.g., the four-interval efficiency modeling of \citet{guetta_luminosity_2005}) or by other approximation methods. A more accurate modeling of the LF, however, requires at least two LGRB observable incorporated in the model: the bolometric peak flux ($\pbol$) and the observed peak energy ($\epk$). The parameter $\epk$ is required, since most gamma-ray detectors are photon counters, a quantity that depends on not only $\pbol$ but also $\epk$ of the burst. This leads to the requirement of using a bivariate distribution as the minimum acceptable model to begin with, for the purpose of constraining the LF. The choice of model can be almost anything \citep[e.g.,][] {kommers_intensity_2000, porciani_association_2001, sethi_luminosity_2001, schmidt_gamma-ray_2009, campisi_redshift_2010, wang_evolving_2011}, since current theories of LGRBs prompt emission do not set strong limits on the shape and range of the luminosity function or any other LGRB spectral or temporal variables.

        Here, the multivariate log-normal distribution is proposed as the simplest natural candidate model capable of describing data. The motivation behind this choice of model comes from the available observational data that closely resembles a joint multivariate log-normal distribution for four most widely studied temporal and spectral parameters of LGRBs in the observer-frame: $\pbol$, $\sbol$ (bolometric fluence), $\epk$, $\dur$: Since most LGRBs originate from moderate redshifts $z\sim1-3$, a fact known thanks to Swift satellite \citep[e.g.,][]{butler_cosmic_2010,racusin_fermi_2011}, the convolution of these observer-frame parameters with the redshift distribution results in negligible variation in the shape of the rest-frame joint distribution of the same LGRB parameters. Therefore, the redshift-convoluted 4-Dimensional (4D) rest-frame distribution can be well approximated as a linear translation from the observer-frame parameter space to the rest-frame parameter space, keeping the shape of the distribution almost intact. This implies that the joint distribution of the intrinsic LGRB variables: the isotropic peak luminosity ($\liso$), the total isotropic emission ($\eiso$), the rest-frame time-integrated spectral peak energy ($\epkz$) and the rest-frame duration ($\durz$) might be indeed well described by a multivariate log-normal distribution.

        In general, models with higher nonzero moments than the log-normal model can also be considered for fitting, such as a multivariate Skew-lognormal \citep[e.g.,][]{azzalini_class_1985} or variants of multivariate stable distributions \citep[e.g.,][]{press_multivariate_1972}. This, however, requires fitting for higher number of free parameters which is practically impossible given BATSE data with no available redshift information.

        Following the discussion above, the process of LGRB observation can be therefore considered as a non-homogeneous Poisson process whose  mean rate parameter -- the cosmic LGRB differential rate, $\mathcal{R}_{cosmic}$ -- is the product of the differential comoving LGRB rate density $\dot\zeta(z)$ with a $p=4$ dimensional log-normal Probability Density Function (pdf), $\mathcal{LN}$, of four LGRB variables: $\liso$, $\eiso$, $\epkz$ and $\durz$, with location vector $\vec \mu$ and the scale (i.e., covariance) matrix $\Sigma$,

        \begin{eqnarray}
            \label{eq:cosmicrate}
            \mathcal{R}_{cosmic} &=& \frac{\diff N}{\diff\liso~\diff\eiso~\diff\epkz~\diff\durz~\diff z} \\
                           &\propto& \mathcal{LN}\bigg(\liso,\eiso,\epkz,\durz\big|\vec \mu,\Sigma\bigg) \nonumber  \\
                           &\times & \frac{\dot\zeta(z)\nicefrac{\diff V}{\diff z}}{(1+z)}, \nonumber
        \end{eqnarray}

        where the factor $(1+z)$ in the denominator accounts for cosmological time dilation and the comoving volume element per unit redshift, $\nicefrac{\diff V}{\diff z}$, is,

        \begin{equation}
            \label{eq:dvdz}
            \frac{\diff V}{\diff z} = \frac{C}{H_0}\frac{4\pi {D_L}^2(z)}{(1+z)^2\bigg[\Omega_M(1+z)^3+\Omega_\Lambda\bigg]^{1/2}},
        \end{equation}

        with $D_L$ being the luminosity distance,

        \begin{equation}
            \label{eq:lumdis}
            D_L(z)=\frac{C}{H_0}(1+z)\int^{z}_{0}dz'\bigg[(1+z')^{3}\Omega_{M}+\Omega_{\Lambda}\bigg]^{-1/2},
        \end{equation}

        assuming a flat $\Lambda CDM$ cosmology, with parameters set to $h=0.70$, $\Omega_M=0.27$ and $\Omega_\Lambda=0.73$ (Jarosik et al. 2011). Here, $C$ \& $H_{0}=100h~[Km/s/MPc]$ stand for the speed of light and the Hubble constant respectively.

        The 4-dimensional log-normal distribution of Equation (\ref{eq:cosmicrate}), $\mathcal{LN}$, has an intimate connection to multivariate Gaussian distribution in the logarithmic space of LGRB observable parameters (c.f. Appendix \ref{sec:appD}).

        One could generalize the LGRB rate of Equation (\ref{eq:cosmicrate}) to incorporate a redshift evolution of the LGRB variables in the form of $\mu_i(z)=\mu_{0,i}+\alpha_i\log(1+z),~(i=1,..,4)$, where $\vec\alpha$ has to be constrained by observational data. This is, however, impractical for BATSE data due to unknown redshifts, as the fitting results in degenerate values for $\vec\alpha$. Nevertheless, the multivariate analysis of Swift LGRBs presented by B10 strongly rejects the possibility of redshift evolution of the LF, a fact that further legitimizes the absence of redshift-luminosity evolution in Equation (\ref{eq:cosmicrate}).

        As for the comoving rate density $\dot\zeta(z)$, it is assumed that LGRBs trace the Star Formation Rate (SFR) in the form of a piecewise power-law function of \citet{hopkins_normalization_2006} (hereafter HB06),
        \begin{equation}
            \dot\zeta(z) = \frac{\diff N}{\diff z} \propto \begin{cases}
                                                    (1+z)^{\gamma_0} & z<z_0 \\
                                                    (1+z)^{\gamma_1} & z_0<z<z_1 \\
                                                    (1+z)^{\gamma_2} & z>z_1, \\
                                             \end{cases}
            \label{eq:zeta}
        \end{equation}

        with parameters $(z_0,z_1,\gamma_0,\gamma_1,\gamma_2)$ set to best-fit values $(0.97,4.5,3.4,-0.3,-7.8)$ of HB06, also to the best values $(0.993,3.8,3.3,0.055,-4.46)$ of an updated SFR fit by \citet{li_star_2008}.
        Alternatively, the bias-corrected redshift distribution of LGRBs derived from Swift data (B10) with best-fit parameter values $(0.97,4.00,3.14,1.36,-2.92)$ can be employed as $\dot\zeta(z)$. This parameter set is consistent with an LGRB rate scenario tracing metallicity-corrected SFR with a cutoff ${Z/Z_{\Sun}}\sim0.2-0.5$ (Figure (10) \& Equation (8) in B10; \citet{li_star_2008}). The hypothesis of LGRB rate evolving with cosmic metallicity is both predicted by the Collapsar model of long-duration Gamma-Ray Bursts \citep[e.g.,][]{woosley_progenitor_2006} and supported by observations of LGRBs host galaxies \citep[e.g.,][]{stanek_protecting_2006, levesque_host_2010}, although the metallicity-rate connection and the presence of a sharp metallicity cutoff has been challenged by few recent host galaxy observations \citep[e.g.,][]{levesque_high-metallicity_2010-1, levesque_high-metallicity_2010} and possible unknown observational biases \citep[e.g.,][]{levesque_host_2012}.

        The cosmic LGRB rate, $\mathcal{R}_{cosmic}$, in Equation (\ref{eq:cosmicrate}), although quantified correctly, does not represent the observed rate ($\mathcal{R}_{obs}$) of LGRBs detected by BATSE LADs, unless convolved with an accurate model of BATSE trigger efficiency, $\eta$, as a function of the burst redshift and rest-frame parameters, discussed in Appendix \ref{sec:appB},
        \begin{equation}
            \label{eq:obsrate}
            \mathcal{R}_{obs}=\eta(\liso,\epkz,\durz,z)\times\mathcal{R}_{cosmic}
        \end{equation}

    \subsection{Model Fitting}
    \label{sec:MF}

        Now, with a statistical model for the observed rate of LGRBs at hand (i.e., Equation \ref{eq:obsrate}), the best-fit parameters can be obtained by the method of Maximum Likelihood. This is done by maximizing the Likelihood function of the model, given the observational data, using a variant of Metropolis-Hastings Markov Chain Monte Carlo (MCMC) algorithm discussed in detail in Appendix \ref{sec:appC}. As mentioned before, the fitting is performed for three redshift distribution scenarios of HB06, B10 \& \citet{li_star_2008}.
        \begin{table}[htbp]
\begin{center}
\caption{Mean best-fit parameters of LGRB World Model, \\
 for three redshift distribution scenarios \label{tab:BFP}}
\begin{tabular}{c c c c}
\hline
\hline
Parameter & HB06 & Li (2008) & B10 \\
\hline
\multicolumn{4}{c}{Redshift Parameters (Equation \ref{eq:zeta})} \\
\hline
$z_0$      & $0.97$ & $0.993$   & $0.97$  \\
$z_1$      & $4.5$  & $3.8$     & $4.00$  \\
$\gamma_0$ & $3.4$  & $3.3$     & $3.14$  \\
$\gamma_1$ & $-0.3$ & $0.0549$  & $1.36$  \\
$\gamma_2$ & $-7.8$ & $-4.46$   & $-2.92$ \\
\hline
\multicolumn{4}{c}{Location Parameters} \\
\hline
$\log(\liso)$ & $51.35\pm0.20$ & $51.50\pm0.19$ & $51.73\pm0.19$ \\
$\log(\eiso)$ & $51.82\pm0.20$ & $51.94\pm0.20$ & $52.03\pm0.21$ \\
$\log(\epkz)$ & $2.43\pm0.05$  & $2.47\pm0.05$  & $2.54\pm0.05$  \\
$\log(\durz)$ & $0.99\pm0.03$  & $0.96\pm0.03$  & $0.80\pm0.03$  \\
\hline
\multicolumn{4}{c}{Scale Parameters} \\
\hline
$\log(\sigma_{\liso})$ & $-0.22\pm0.06$ & $-0.23\pm0.06$ & $-0.20\pm0.05$ \\
$\log(\sigma_{\eiso})$ & $-0.07\pm0.03$ & $-0.07\pm0.03$ & $-0.04\pm0.03$ \\
$\log(\sigma_{\epkz})$ & $-0.44\pm0.02$ & $-0.44\pm0.02$ & $-0.44\pm0.02$ \\
$\log(\sigma_{\durz})$ & $-0.38\pm0.01$ & $-0.39\pm0.01$ & $-0.40\pm0.01$ \\
\hline
\multicolumn{4}{c}{Correlation Coefficients} \\
\hline
$\rho_{\liso-\eiso}$ & $0.93\pm0.01$ & $0.94\pm0.01$ & $0.96\pm0.01$ \\
$\rho_{\liso-\epkz}$ & $0.47\pm0.07$ & $0.45\pm0.07$ & $0.44\pm0.08$ \\
$\rho_{\liso-\durz}$ & $0.52\pm0.08$ & $0.59\pm0.09$ & $0.75\pm0.07$ \\
$\rho_{\eiso-\epkz}$ & $0.58\pm0.04$ & $0.58\pm0.04$ & $0.59\pm0.04$ \\
$\rho_{\eiso-\durz}$ & $0.63\pm0.05$ & $0.66\pm0.05$ & $0.74\pm0.04$ \\
$\rho_{\epkz-\durz}$ & $0.34\pm0.04$ & $0.37\pm0.04$ & $0.50\pm0.04$ \\
\hline
\multicolumn{4}{c}{BATSE LGRB Detection Efficiency (Eqn. A5)} \\
\hline
$\mu_{thresh}$          & $-0.44\pm0.02$ & $-0.45\pm0.02$ & $-0.44\pm0.02$ \\
$\log(\sigma_{thresh})$ & $-0.88\pm0.05$ & $-0.90\pm0.05$ & $-0.88\pm0.05$ \\
\hline
\hline
\end{tabular}
\end{center}
{Note.--- The full Markov Chain sampling of the above parameters from the 16-dimensional parameter space of the likelihood function (Appendix \ref{sec:appC}) are available for download at \url{https://sites.google.com/site/amshportal/research/aca/in-the-news/lgrb-world-model} for each of the three redshift distributions.}
\end{table}

        It is also known that $\dur$ of LGRBs are potentially subject to estimation biases. To ensure that the reported $\dur$ of BATSE LGRBs do not bias the fitting results for the rest of the parameters, model fitting was also performed by considering only three spectral variables of BATSE LGRBs: the bolometric 1-sec peak flux ($\pbol$), bolometric fluence ($\sbol$), observed peak energy ($\epk$), excluding duration ($\durz$) variable from the model, thus reducing the dimension of the model by one. Only after the fitting was performed, it became clear that the inclusion of $\dur$ of BATSE LGRBs in fitting does not significantly affect the resulting best-fit parameters of the model. Therefore, here only results from the full model fitting are presented, as in Table (\ref{tab:BFP}).

        Due to lack of redshift information for BATSE GRBs, the resulting parameters of the model exhibit strong covariations with each other. This is illustrated in the example correlation matrix of LGRB world model in Table (\ref{tab:correlations}). All location parameters appear to correlate strongly positively with each other, and so do the scale parameters. The location parameters however negatively correlate with the scale parameters, meaning that an increase in the average values of the rest-frame parameters reduces the half-width of the corresponding distributions of the variables. In general, it is also observed that the correlations among the four variables weaken with increasing the location parameters. An exception to this, is the correlation of $\durz$ with $\liso$ \& $\eiso$ which tends to increase with location parameters. Since an excess in the cosmic rates of LGRBs at high redshifts, generally results in an increase in the values of location parameters, it can be said that {\it `given BATSE LGRBs data, a higher rate for LGRBs at distant universe, generally implies weaker $\eiso-\epkz$ \& $\liso-\epkz$ correlations and stronger covariation of $\durz$ with the three other parameters'}.

    \subsection{Goodness-of-Fit Tests}
    \label{sec:GOF}
        In any statistical fitting problem, perhaps more important than the model construction is to provide tests showing how good is the model fit to input data. For many univariate studies of the GRB LF, this is done by employing well-established statistical tests such as the Kolmogorov-Smirnov \citep[e.g.,][]{kolmogoroff_confidence_1941, smirnov_table_1948} or Pearson's $\chi^2$ \citep[e.g.,][]{fisher_conditions_1924} tests. In the case of multivariate studies (e.g., B10), a combination of visual inspection of the fitting results, KS test on the marginal and bivariate distributions and variants of $\chi^2$ (e.g., likelihood ratio) tests have been used.

        In general, univariate tests on the marginal distributions of multivariate fits provide only necessary -- but not sufficient -- evidence for a good {\it multivariate} fit. Alternatively one could assess the similarity by the use of nonparametric multivariate Goodness-of-Fit (GoF) tests. Such tests, although exist, have been rarely discussed and treated in statistics due to difficulties in the interpretation of the test statistic \citep[e.g.,][]{peacock_two-dimensional_1983, press_numerical_1992, justel_multivariate_1997}. Ideally, one can always use the Pearson's $\chi^2$ GoF test for any multivariate distribution. However, for the special case of BATSE $1366$ LGRBs, one would need an observed sample consisting of $N\gg1366$ observations to avoid serious instabilities that occur in $\chi^2$ tests due to small sample sizes \citep[e.g.,][]{cochran_methods_1954}.

        To ensure a good fit to the observational data in all -- and not only univariate -- levels of the multivariate structure of data, an assessment of similarity can be obtained by scanning and comparing the model and data along their principal axes, in addition to univariate tests on the marginal distributions. Although statistically not sufficient condition for the multivariate similarity of the model prediction to data, this can provide strong evidence in favor of a good fit, at much higher confidence than tests performed only on the marginal distributions.

\begin{figure*}
            \center{\includegraphics[scale=0.31]{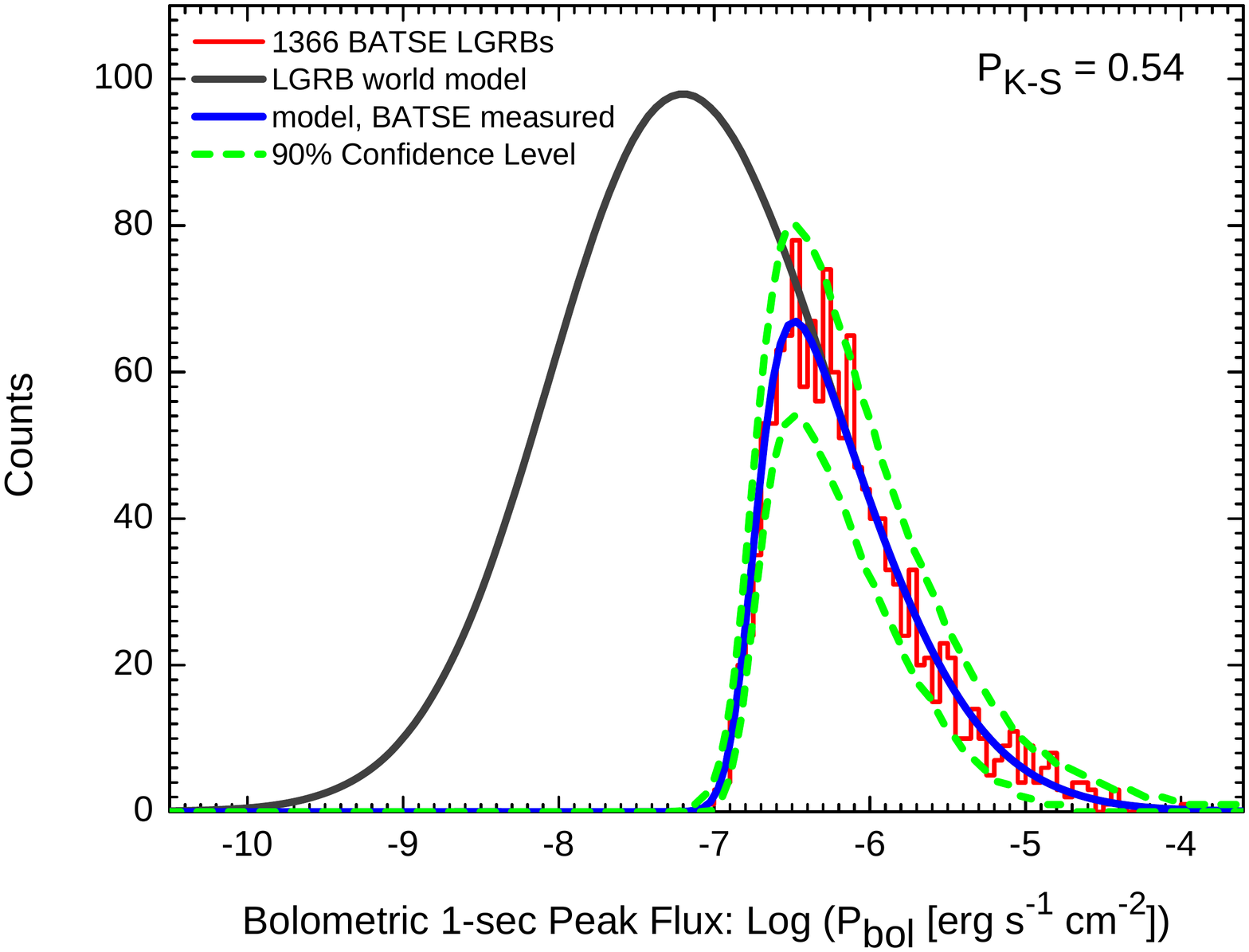}
            \includegraphics[scale=0.31]{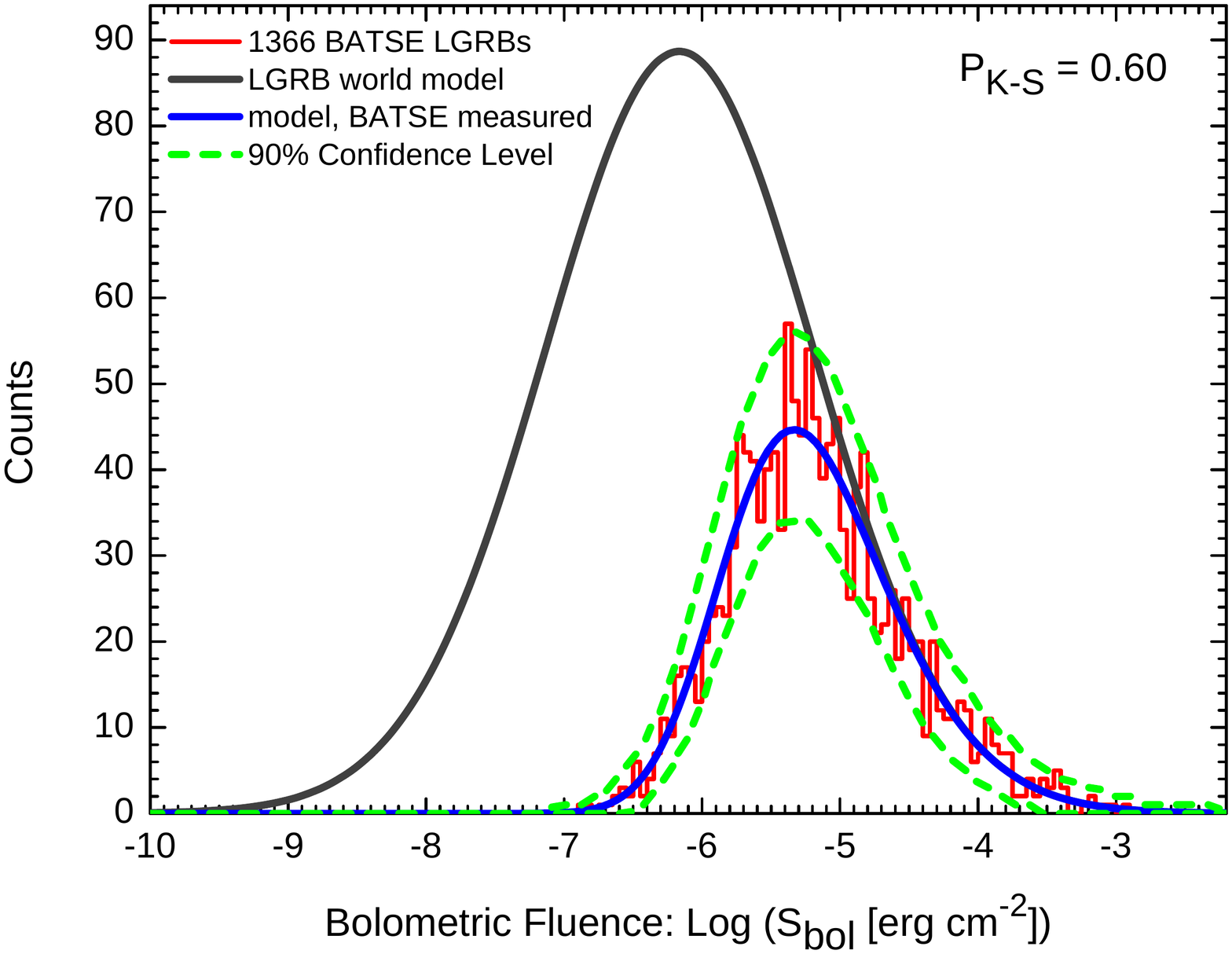}}
            \center{\includegraphics[scale=0.31]{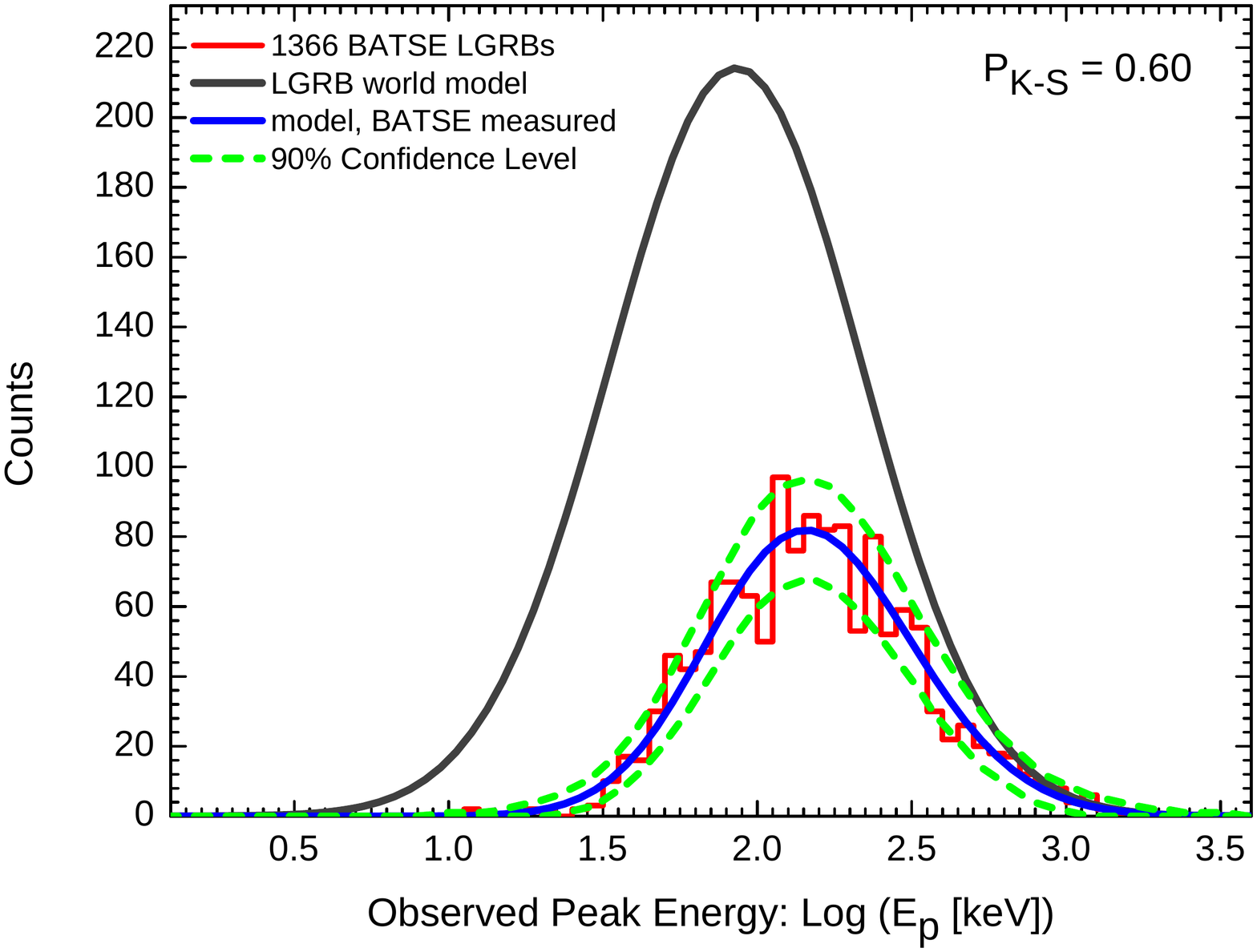}
            \includegraphics[scale=0.31]{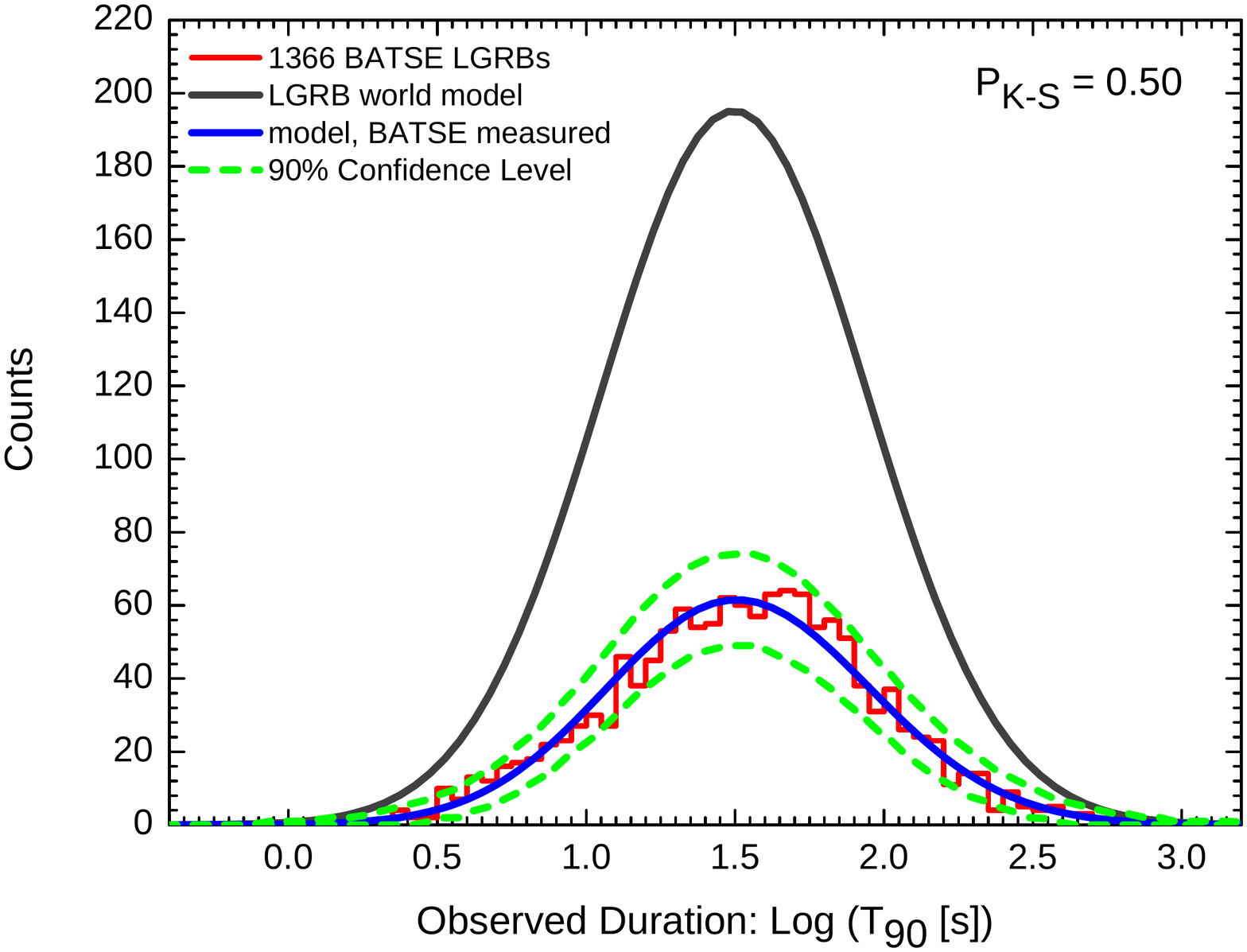}}
            \center{\includegraphics[scale=0.31]{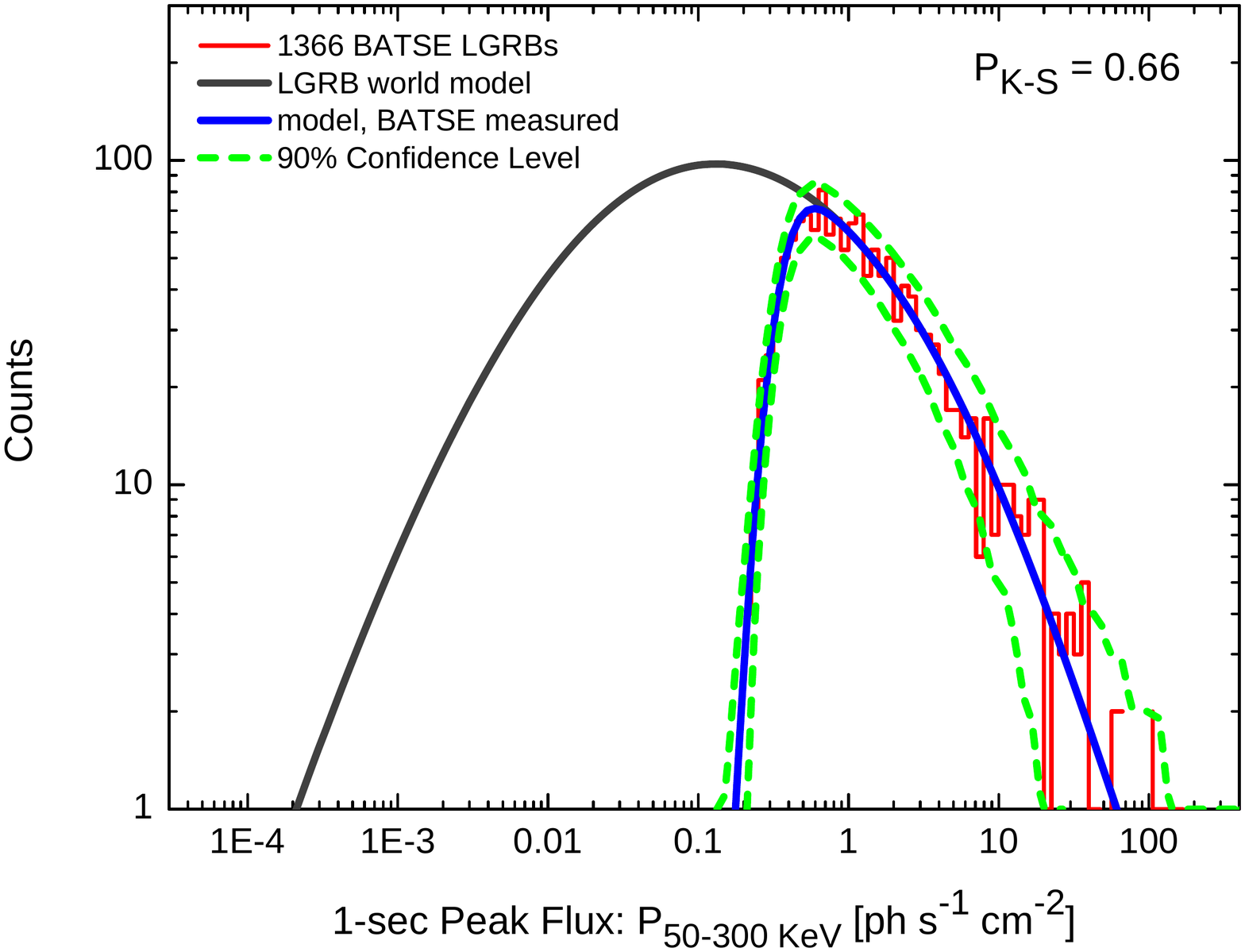}
            \includegraphics[scale=0.31]{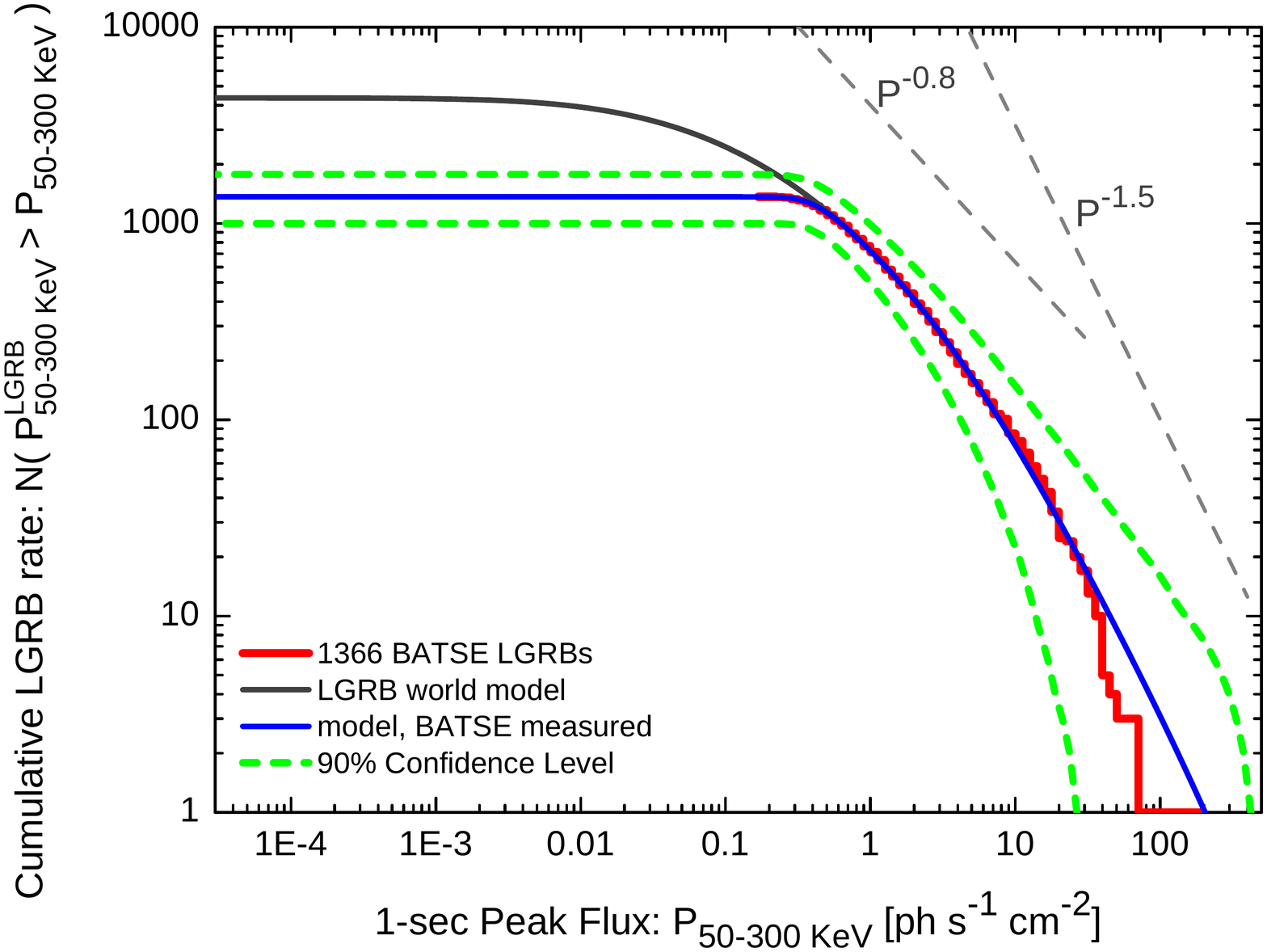}}
            \caption{The marginal distribution predictions (\textcolor{blue}{solid blue lines}) of the LGRB world model given BATSE detection efficiency, superposed on BATSE $1366$ LGRB data (\textcolor{red}{red histograms}). The \textcolor{gray}{solid grey lines} represent the model predictions for the entire LGRB population (detected and undetected), with {\it no} correction for BATSE sky exposure and the beaming factor ($f_b$). The $90\%$ confidence intervals (\textcolor{green}{dashed green lines}) represent random poisson fluctuations expected in BATSE LGRB-detection process. {\it Bottom:} The differential ({\it left panel}) \& cumulative ({\it right panel}) rate of LGRBs as a function of peak photon flux in the BATSE nominal detection energy range ($50-300~[keV]$). For a comparison with Swift sample of LGRBs and the proposed multivariate model of B10, the two \textcolor{gray}{dashed grey lines} -- taken from Figure (1) in B10 -- represent the predictions of the broken power-law luminosity function in Equation (2) of B10, in the observer frame. The Kolmogorov-Smirnov (KS) test probabilities for the goodness-of-fit of the model predictions to BATSE data are reported at the top-right corner of each plot. \label{fig:OFmarginals}}
        \end{figure*}

        Following the lines above, Figure (\ref{fig:OFmarginals}) presents the model predictions for marginal distributions of the four LGRB variables in the observer-frame. The Kolmogorov-Smirnov (KS) test probabilities for the similarity of the model predictions to the marginal distributions of BATSE LGRB variables are also reported on the top right of each plot. All three redshift-distribution parameters of HB06, Li (2008) \& B10 (Equation \ref{eq:zeta}) result in relatively similar fits to BATSE data in the observer frame. Thus, for brevity, only plots for one representative (median) redshift distribution \citep[i.e.,][]{li_star_2008} are presented.

\begin{figure*}
            \center{\includegraphics[scale=0.31]{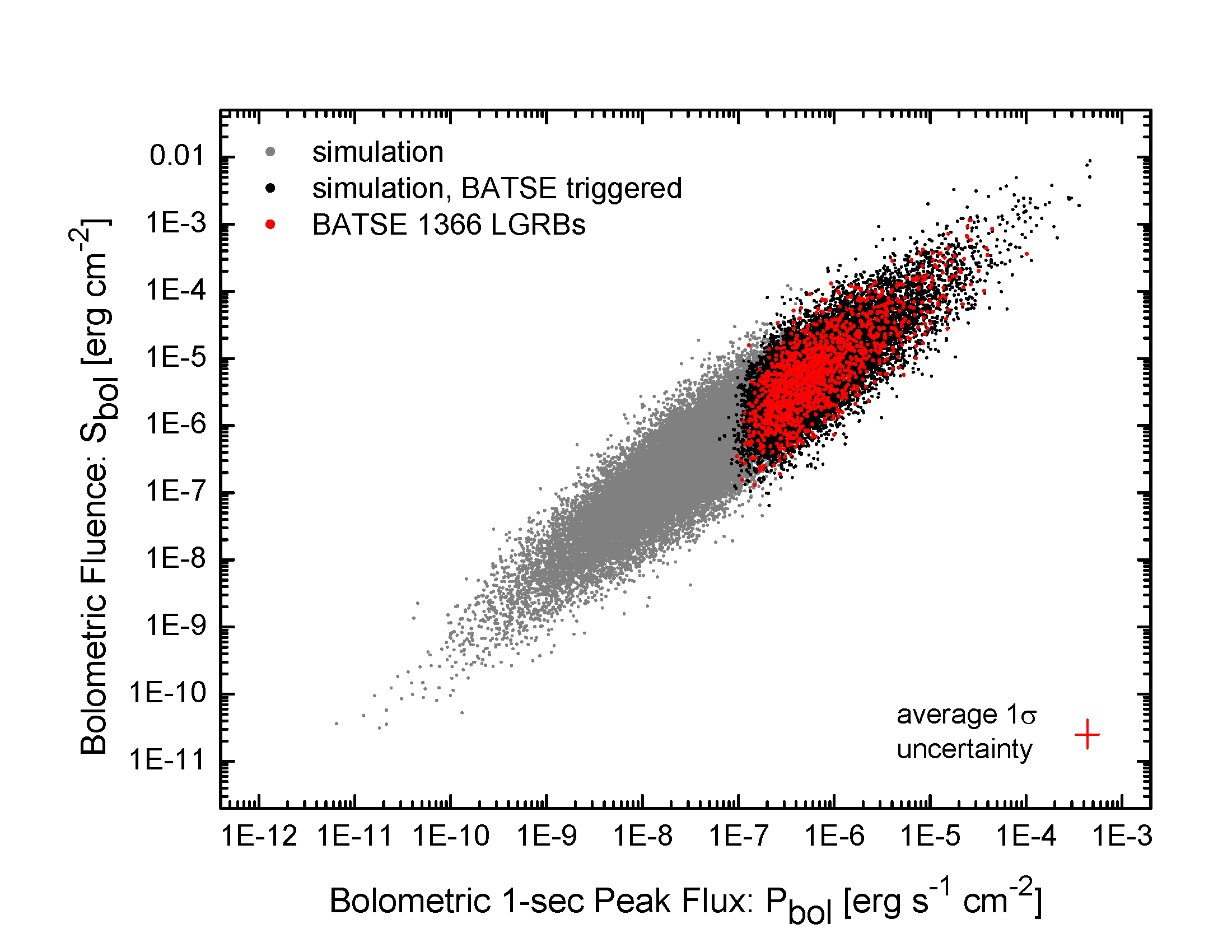}
            \includegraphics[scale=0.31]{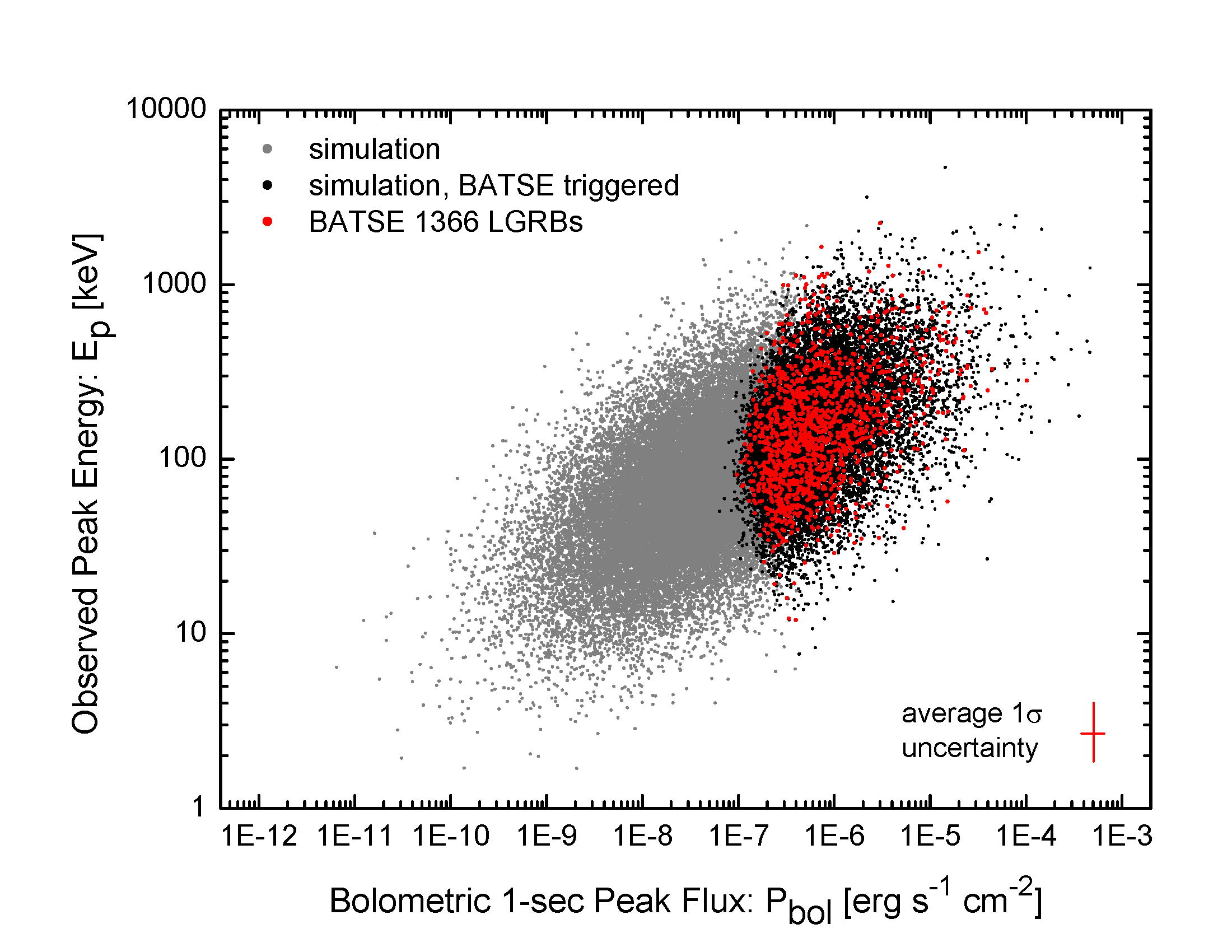}}
            \center{\includegraphics[scale=0.31]{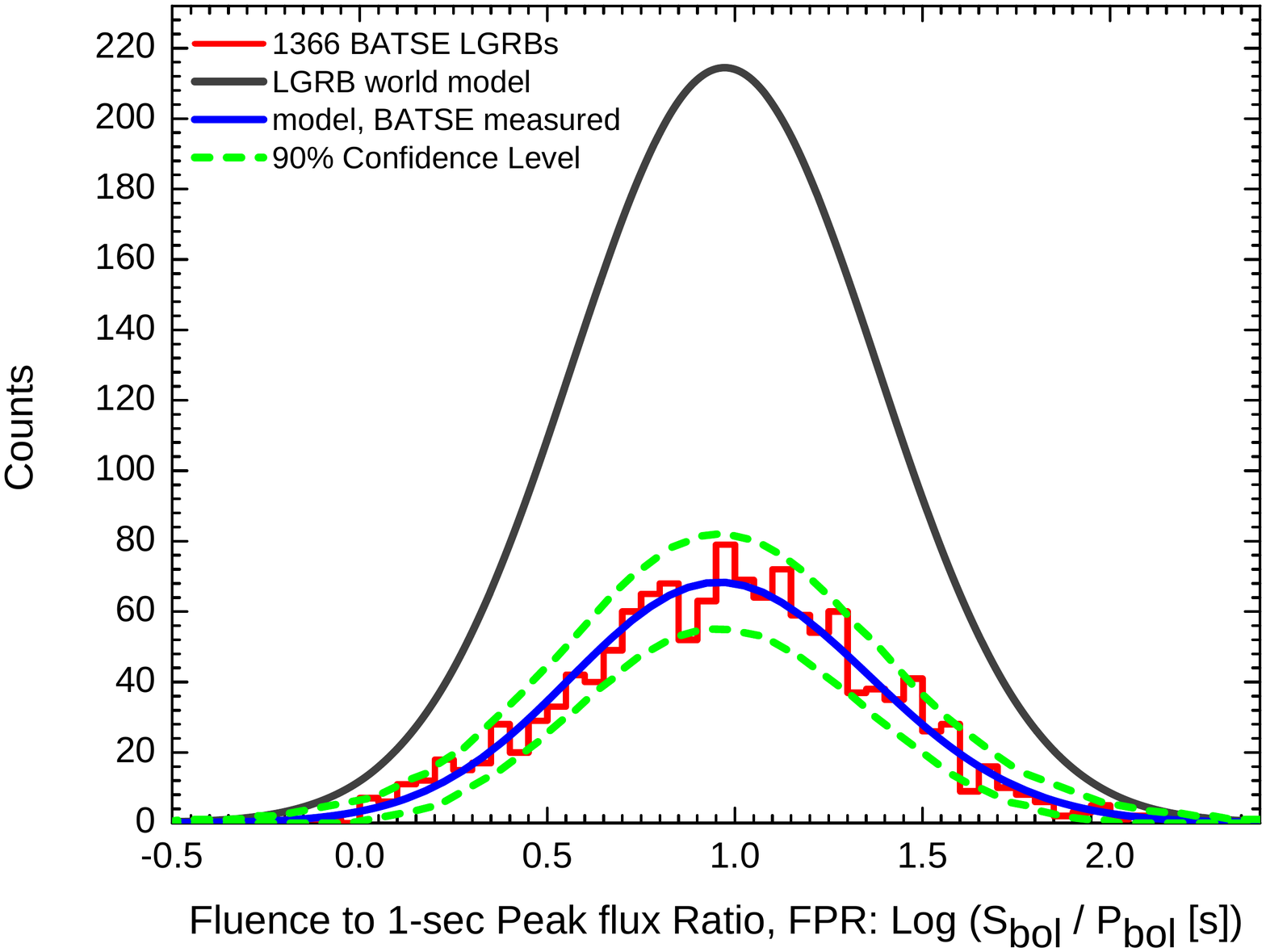}
            \includegraphics[scale=0.31]{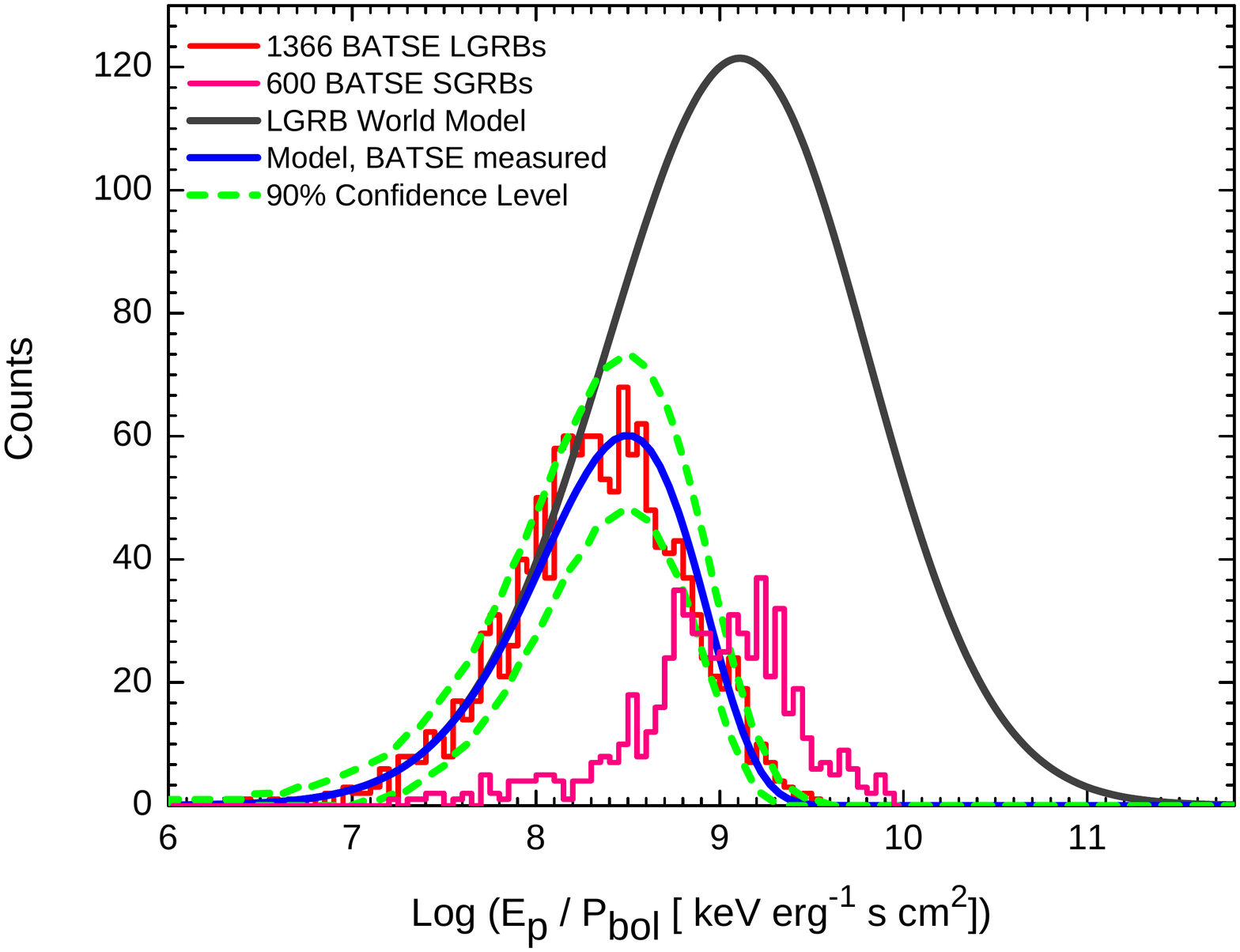}}
            \center{\includegraphics[scale=0.31]{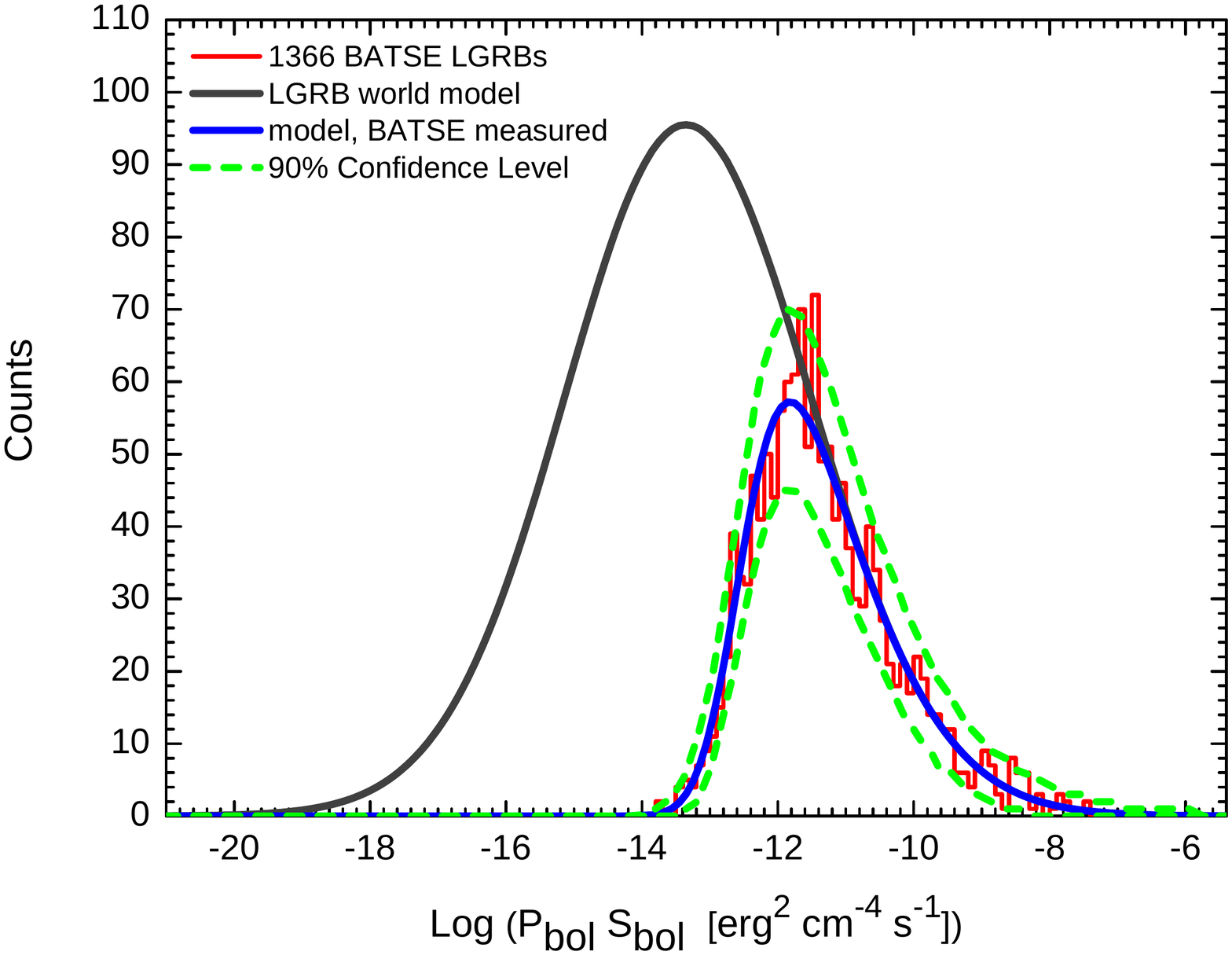}
            \includegraphics[scale=0.31]{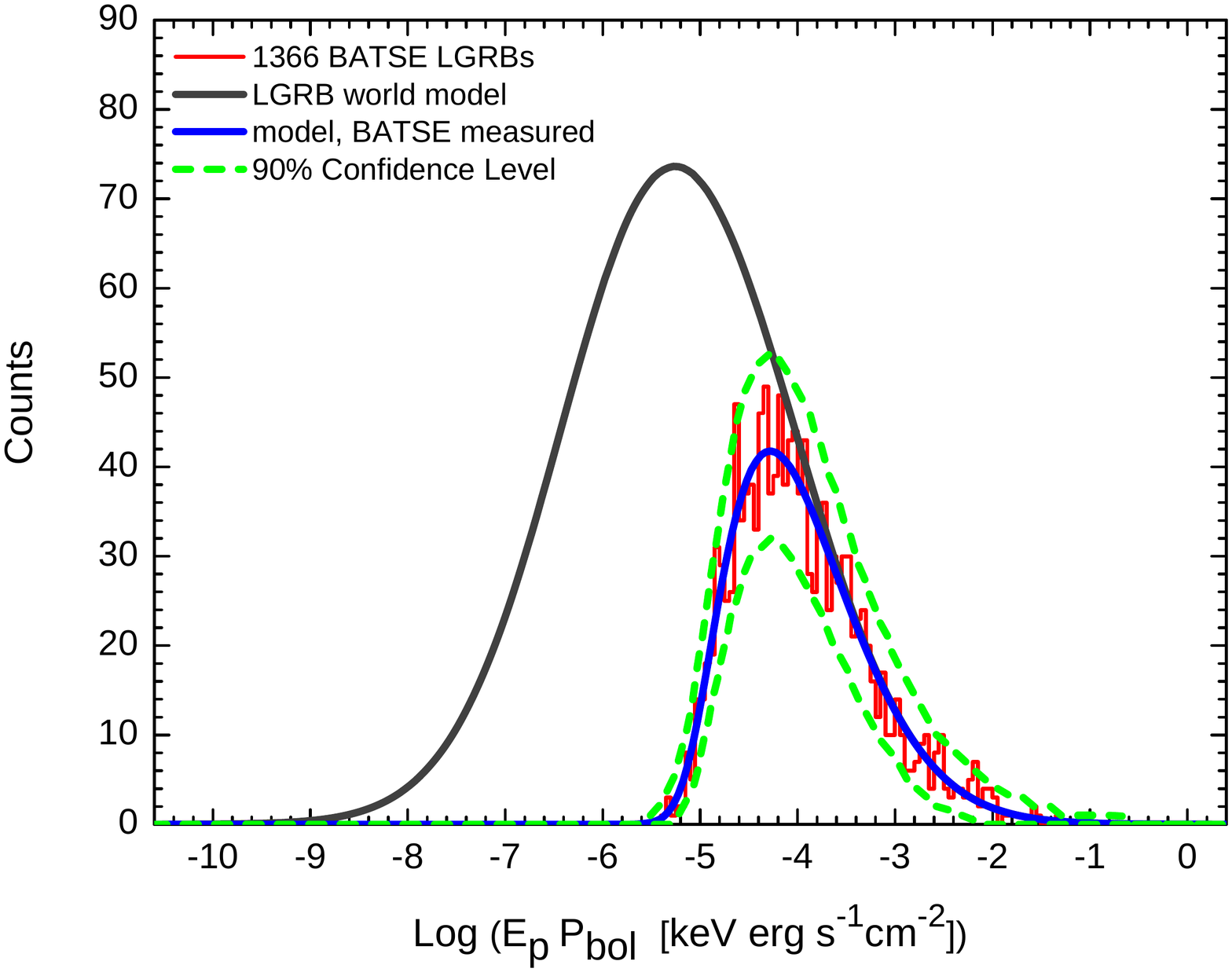}}
            \caption{{\it Top}: BATSE $1366$ LGRB data (\textcolor{red}{red dots}) superposed on the joint bivariate distribution predictions (\textcolor{black}{black dots}) of the LGRB world model for BATSE detection efficiency. The background \textcolor{gray}{grey dots} represent the model predictions for the entire LGRB population (detected and undetected). The uncertainties in BATSE LGRBs' variables are derived from the Empirical Bayes model discussed in Appendix \ref{sec:appC}. {\it Center \& Bottom}: Gauging the goodness-of-fit of the {\it bivariate} model to data by scanning and comparing the joint distributions of the model \& BATSE data along their principal axes. Colors bear the same meaning as the top panels, in addition to the $90\%$ confidence intervals (\textcolor{green}{dashed green lines}) that represent random poisson fluctuations expected in BATSE LGRB-detection process.  \label{fig:bivariates1}}
        \end{figure*}

\begin{figure*}
            \center{\includegraphics[scale=0.31]{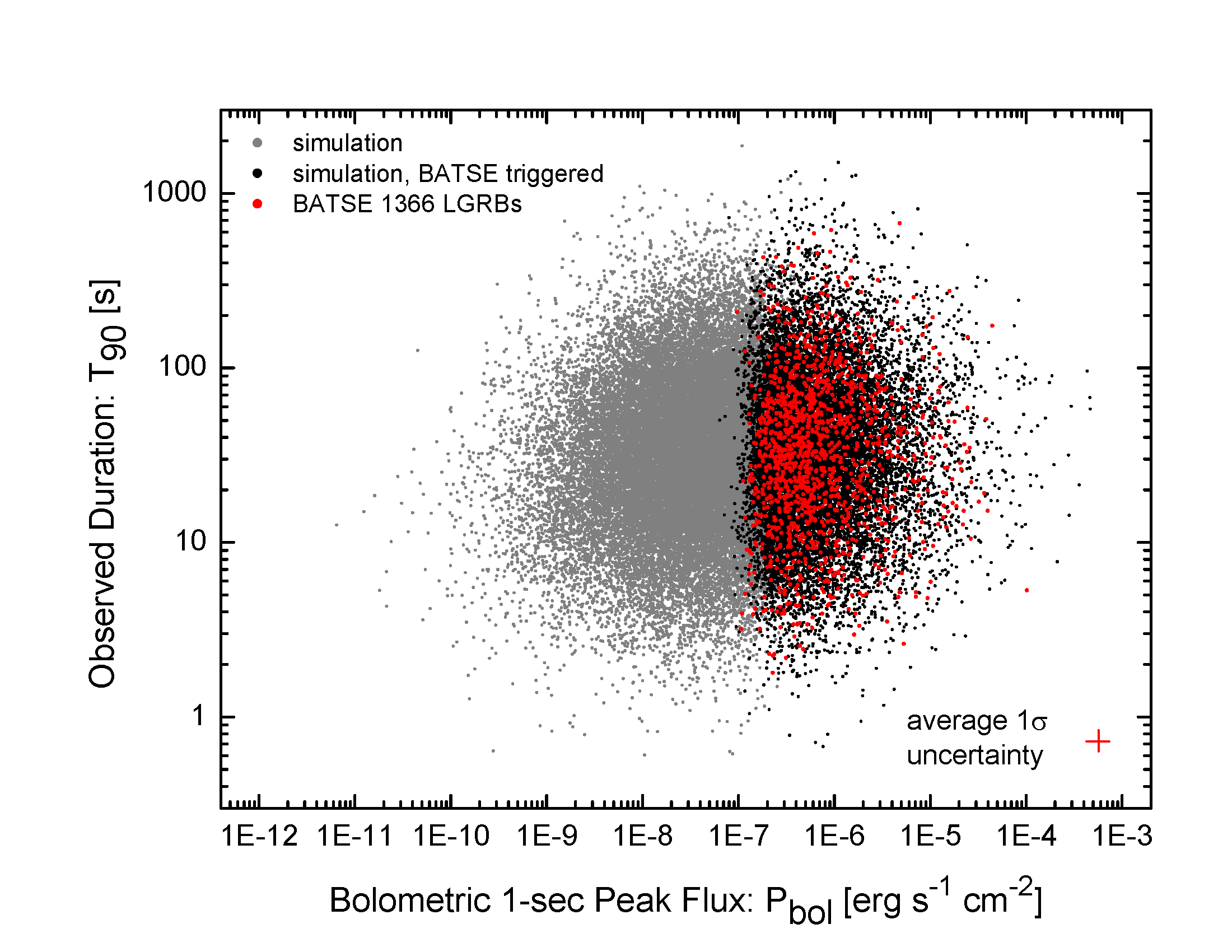}
            \includegraphics[scale=0.31]{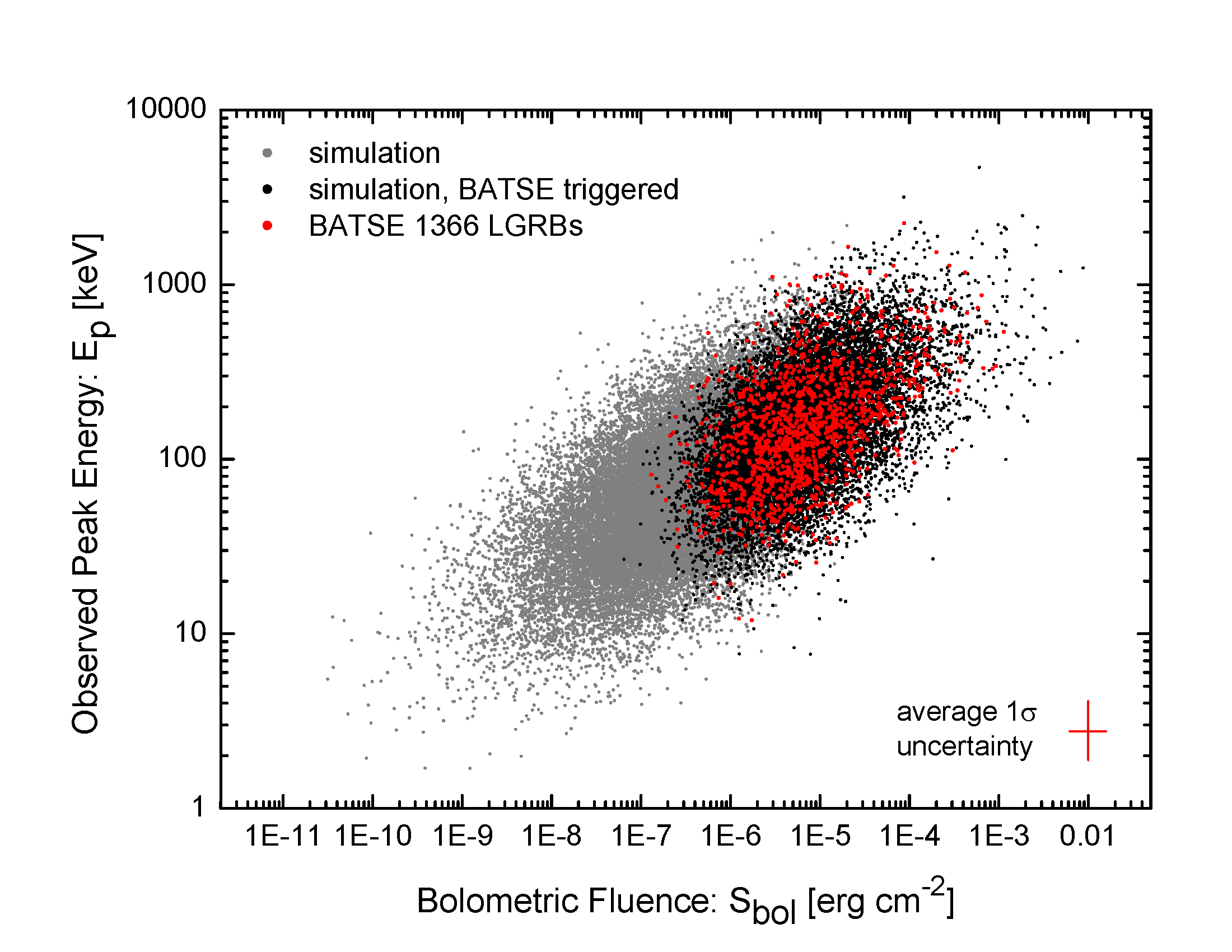}}
            \center{\includegraphics[scale=0.31]{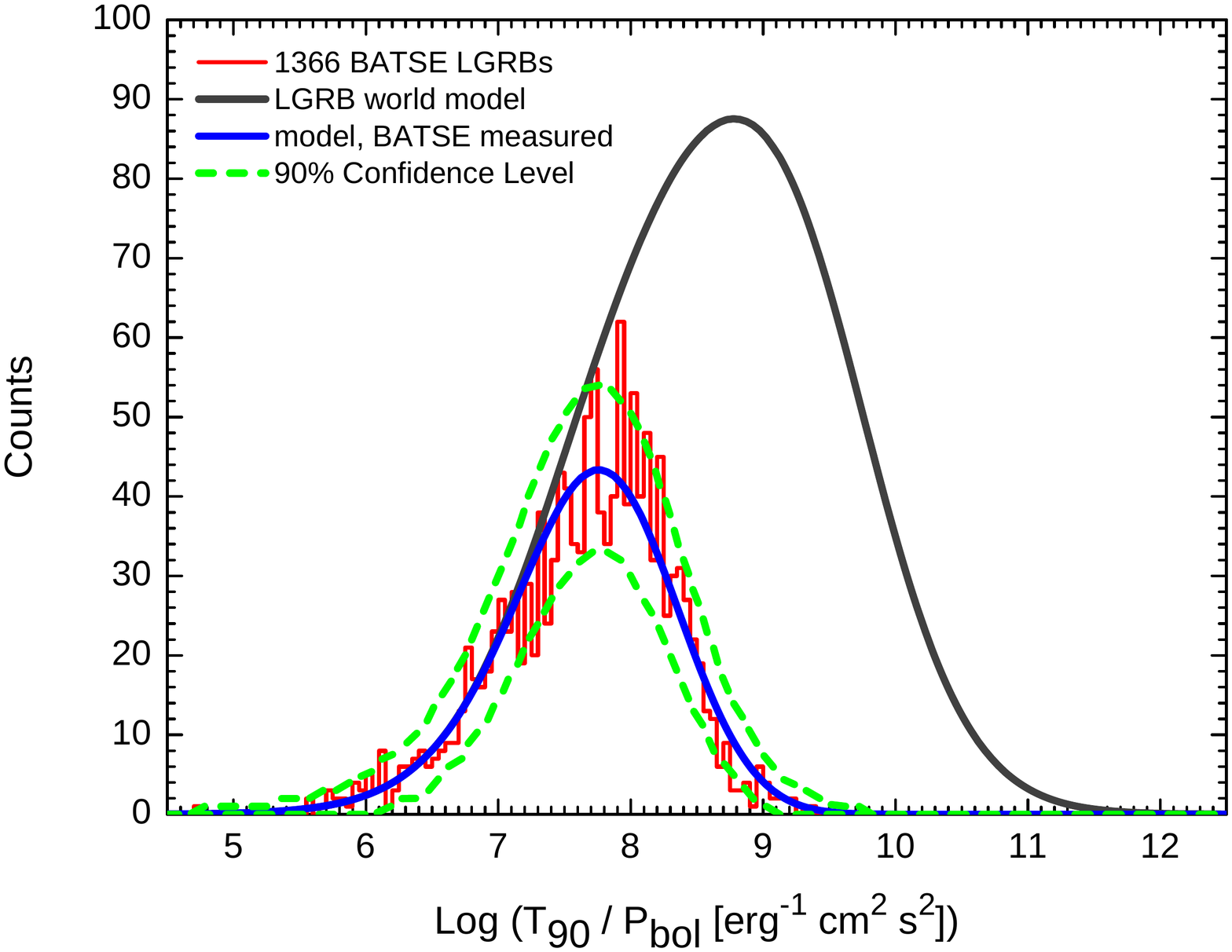}
            \includegraphics[scale=0.31]{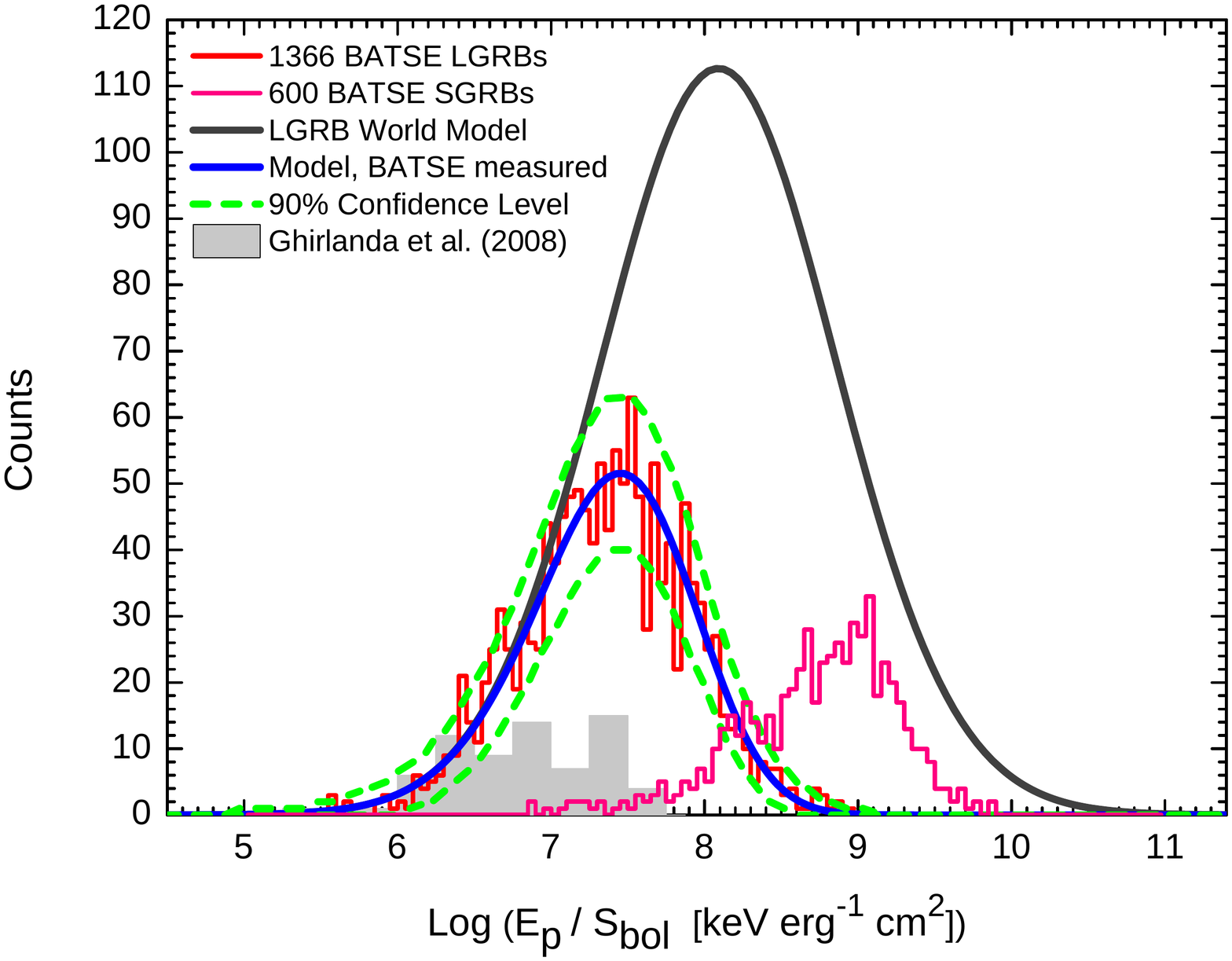}}
            \center{\includegraphics[scale=0.31]{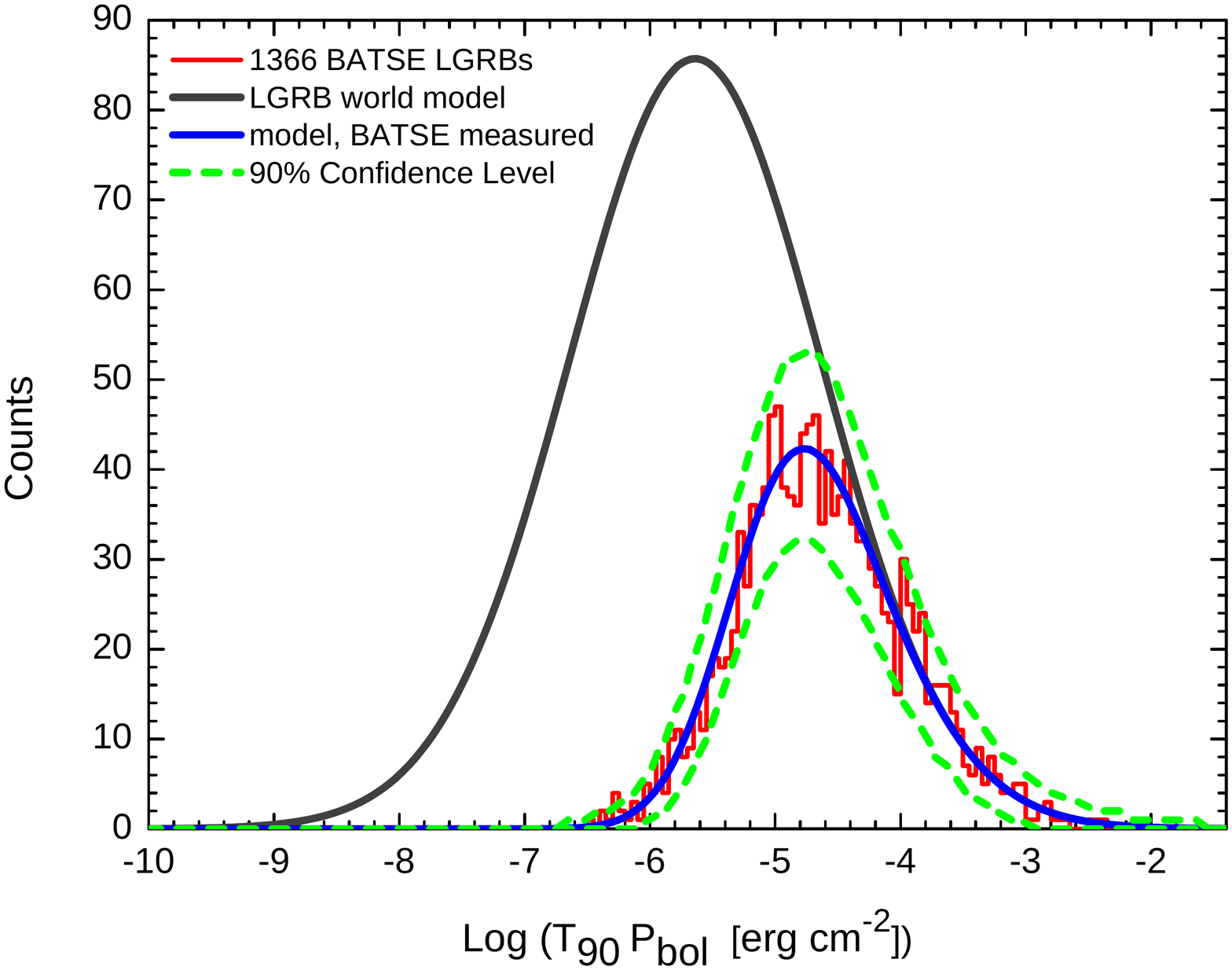}
            \includegraphics[scale=0.31]{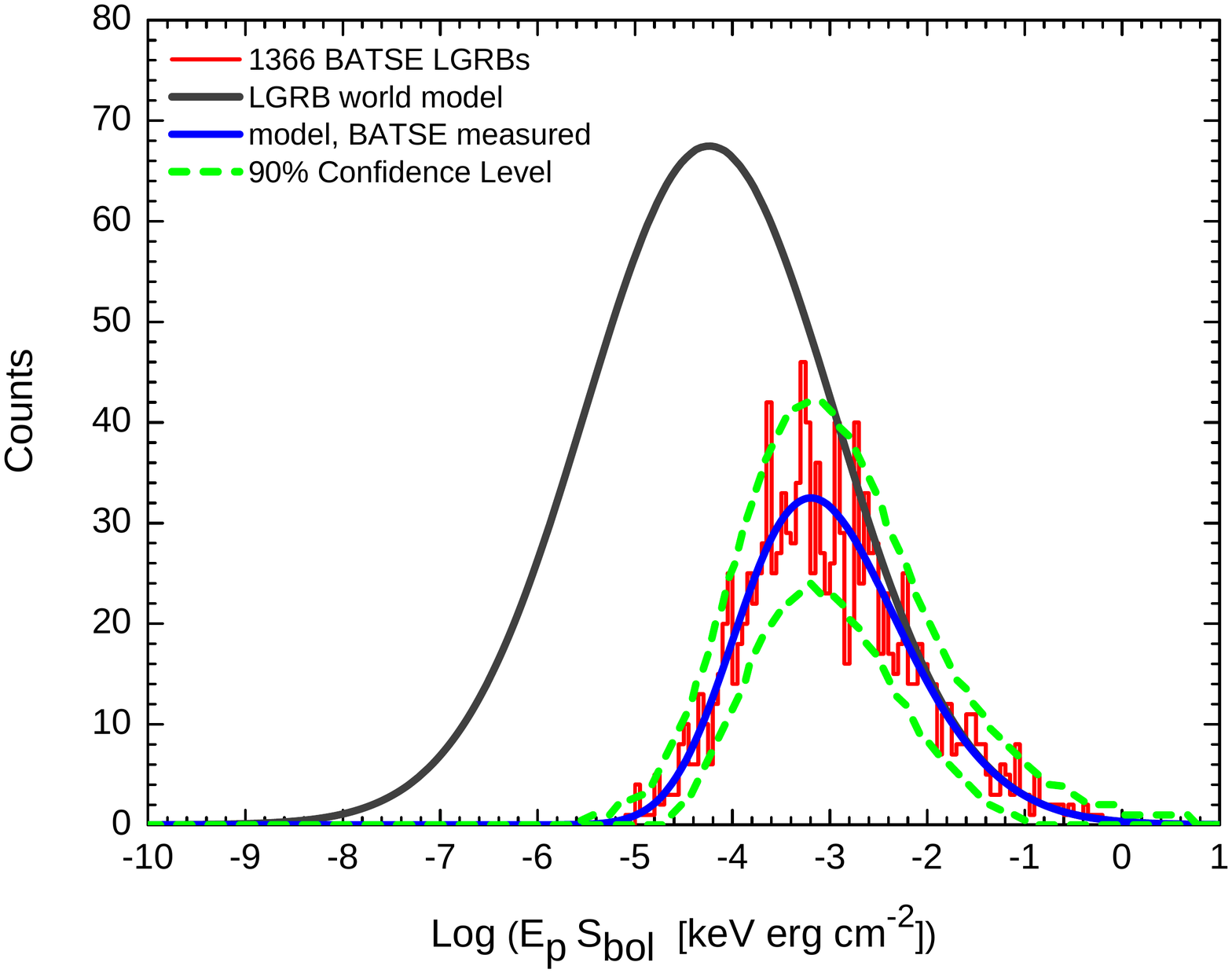}}
            \caption{{\it Top}: BATSE $1366$ LGRB data (\textcolor{red}{red dots}) superposed on the joint bivariate distribution predictions (\textcolor{black}{black dots}) of the LGRB world model for BATSE detection efficiency. The background \textcolor{gray}{grey dots} represent the model predictions for the entire LGRB population (detected and undetected). The uncertainties in BATSE LGRBs' variables are derived from the Empirical Bayes model discussed in Appendix \ref{sec:appC}. {\it Center \& Bottom}: Gauging the goodness-of-fit of the {\it bivariate} model to data by scanning and comparing the joint distributions of the model \& BATSE data along their principal axes. Colors bear the same meaning as the top panels, in addition to the $90\%$ confidence intervals (\textcolor{green}{dashed green lines}) that represent random poisson fluctuations expected in BATSE LGRB-detection process.  \label{fig:bivariates2}}
        \end{figure*}

\begin{figure*}
            \center{\includegraphics[scale=0.31]{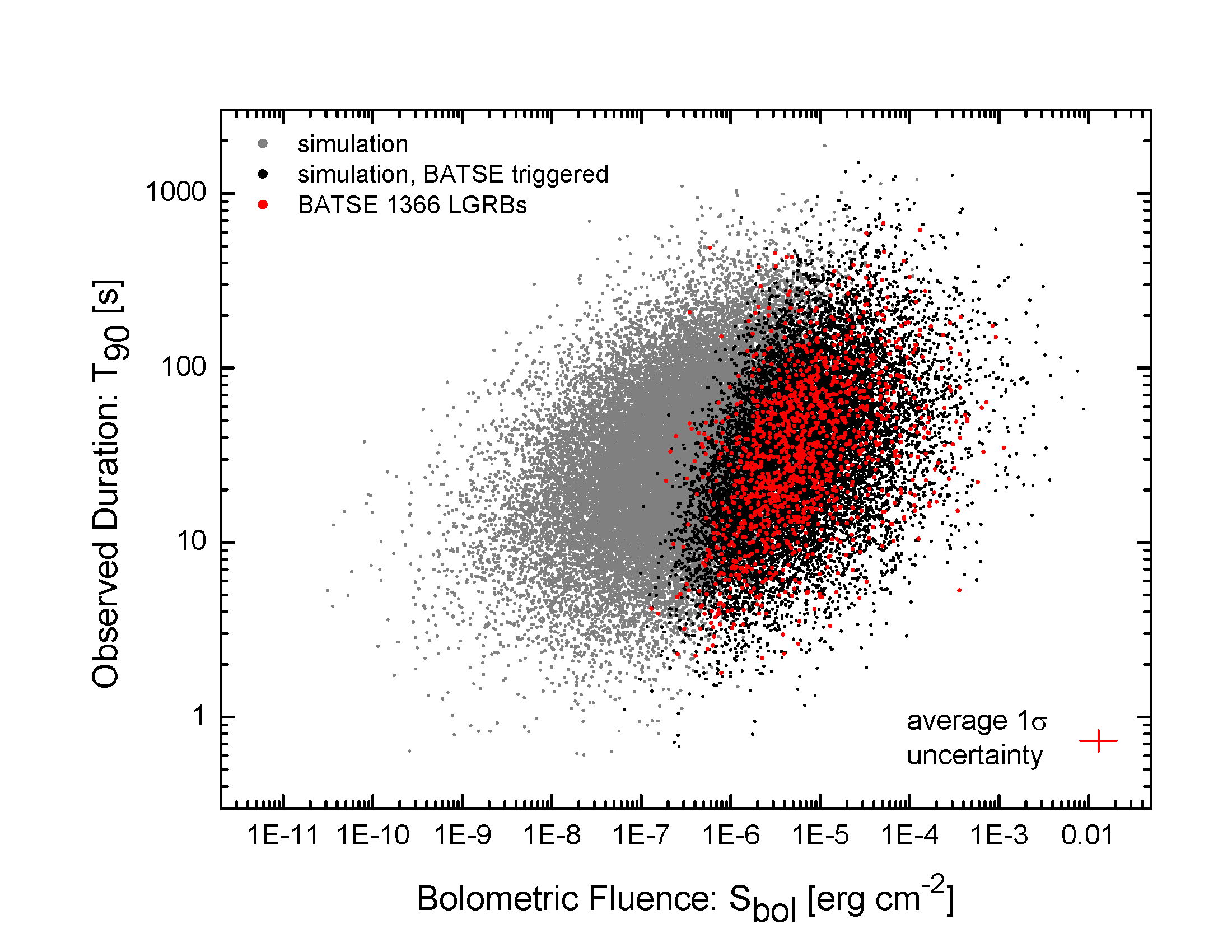}
            \includegraphics[scale=0.31]{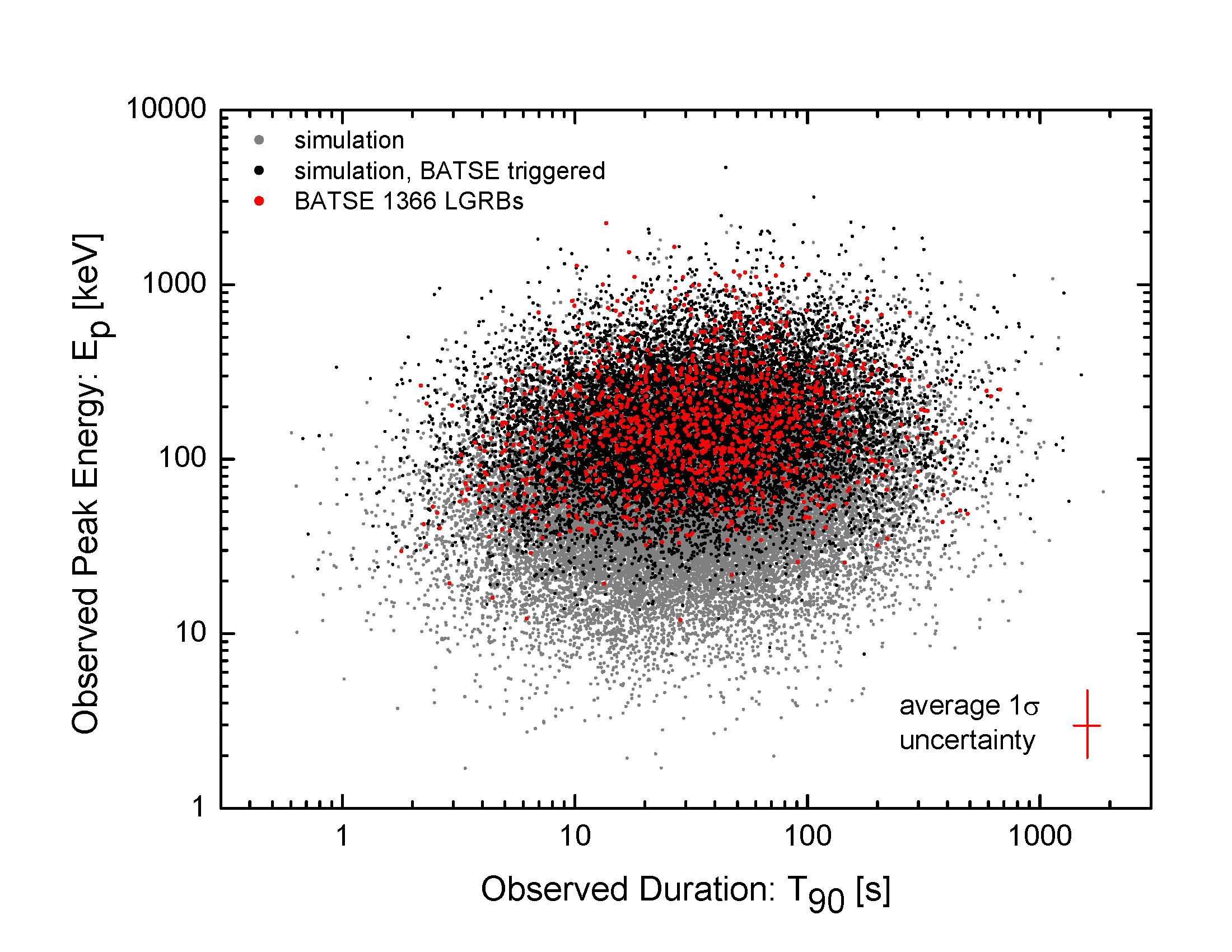}}
            \center{\includegraphics[scale=0.31]{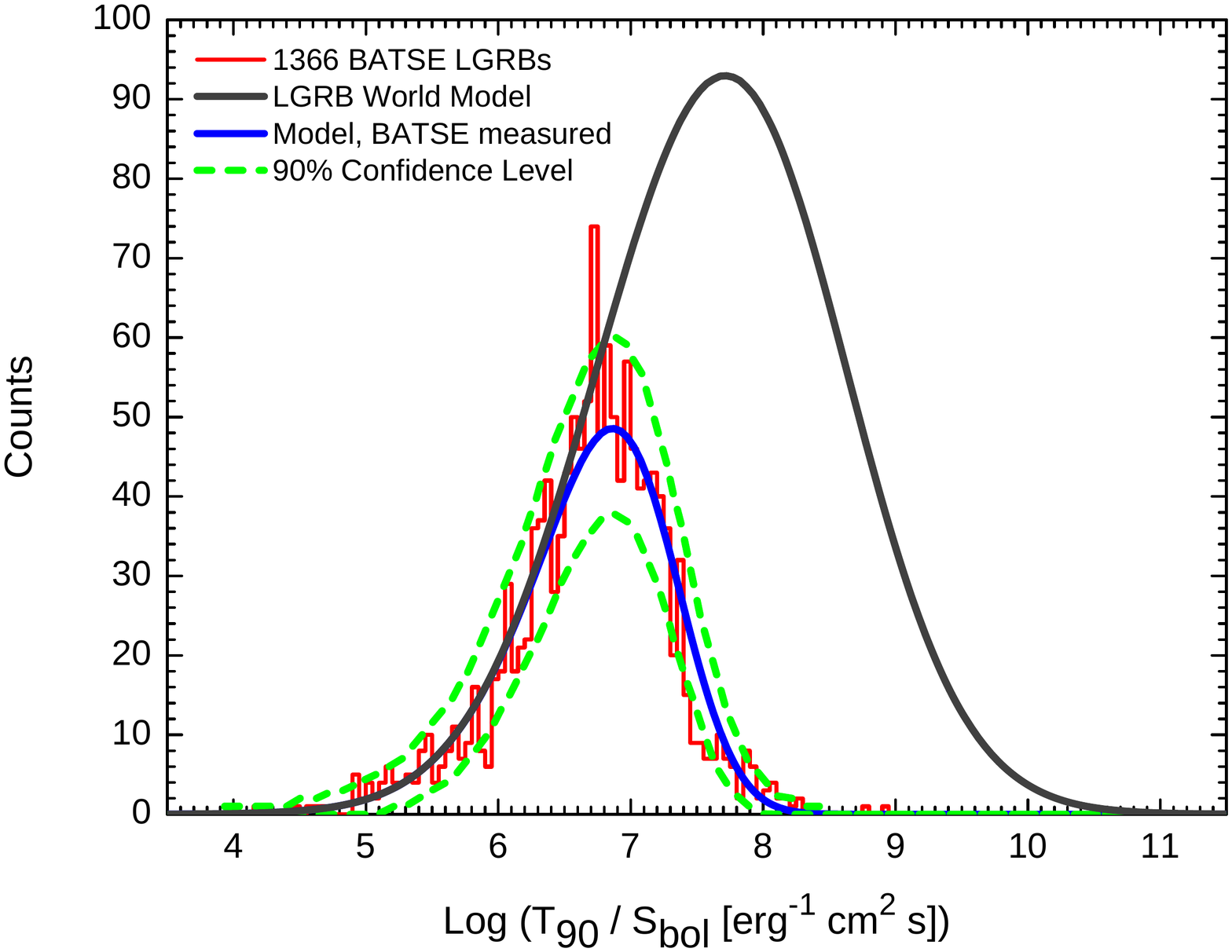}
            \includegraphics[scale=0.31]{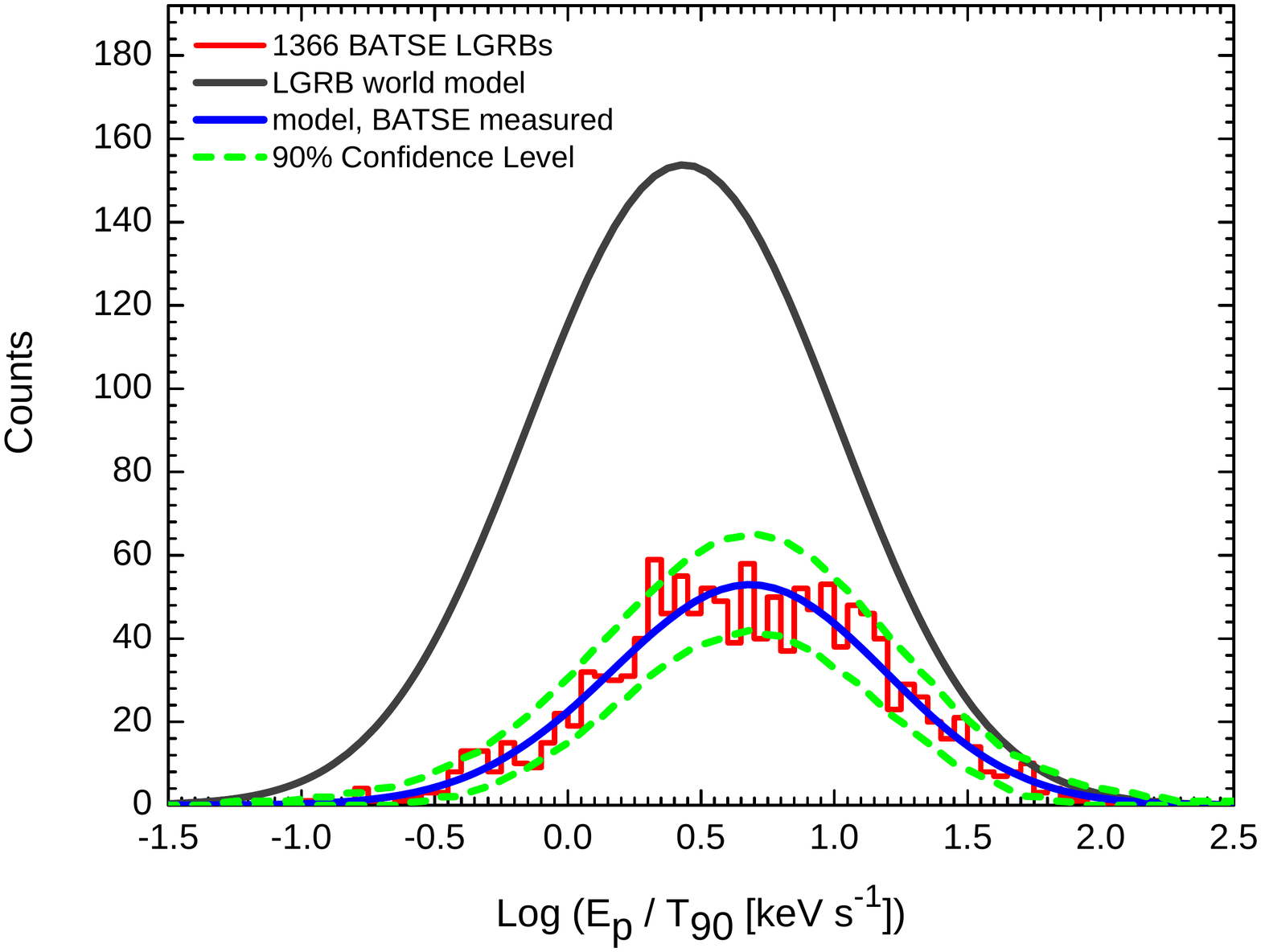}}
            \center{\includegraphics[scale=0.31]{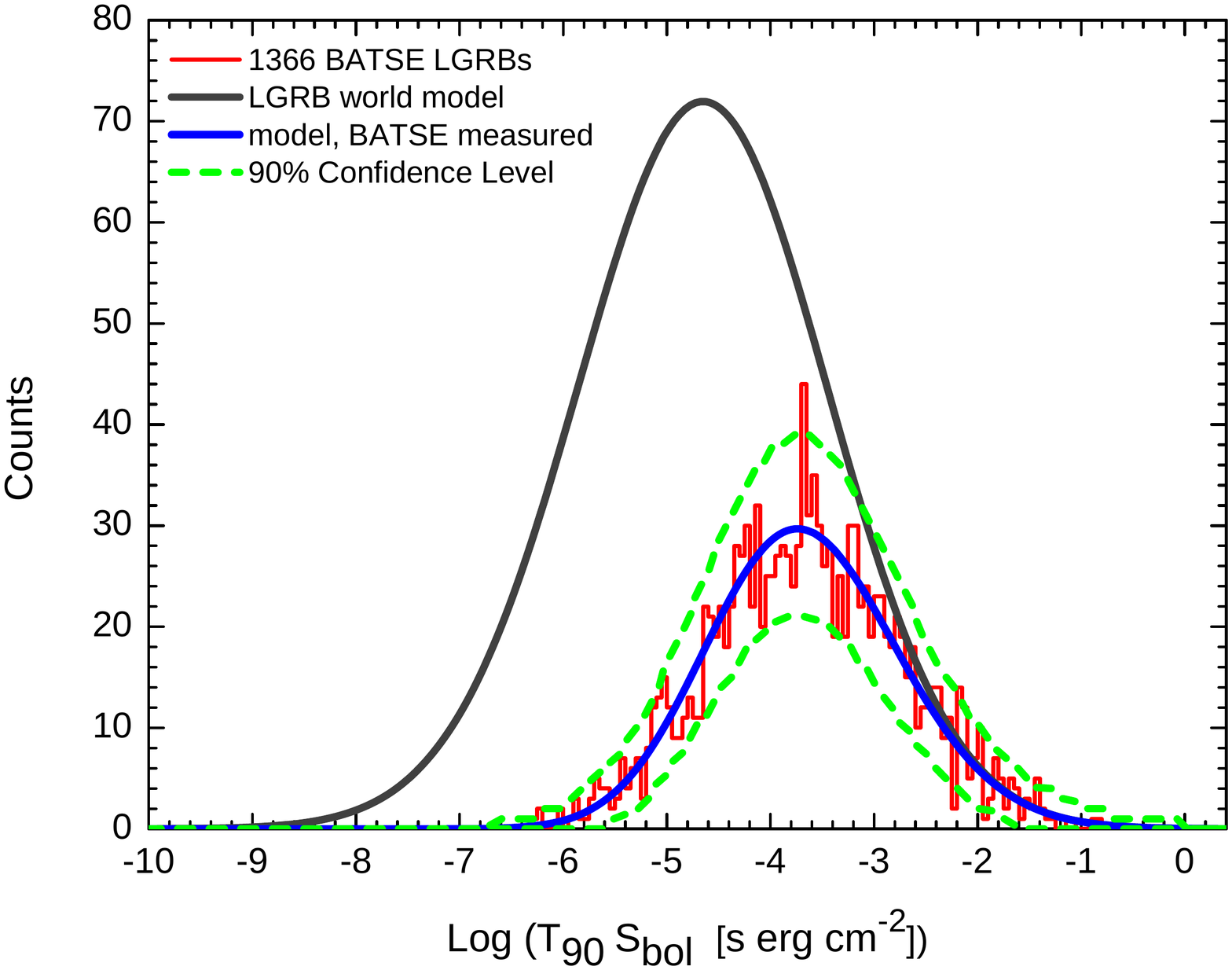}
            \includegraphics[scale=0.31]{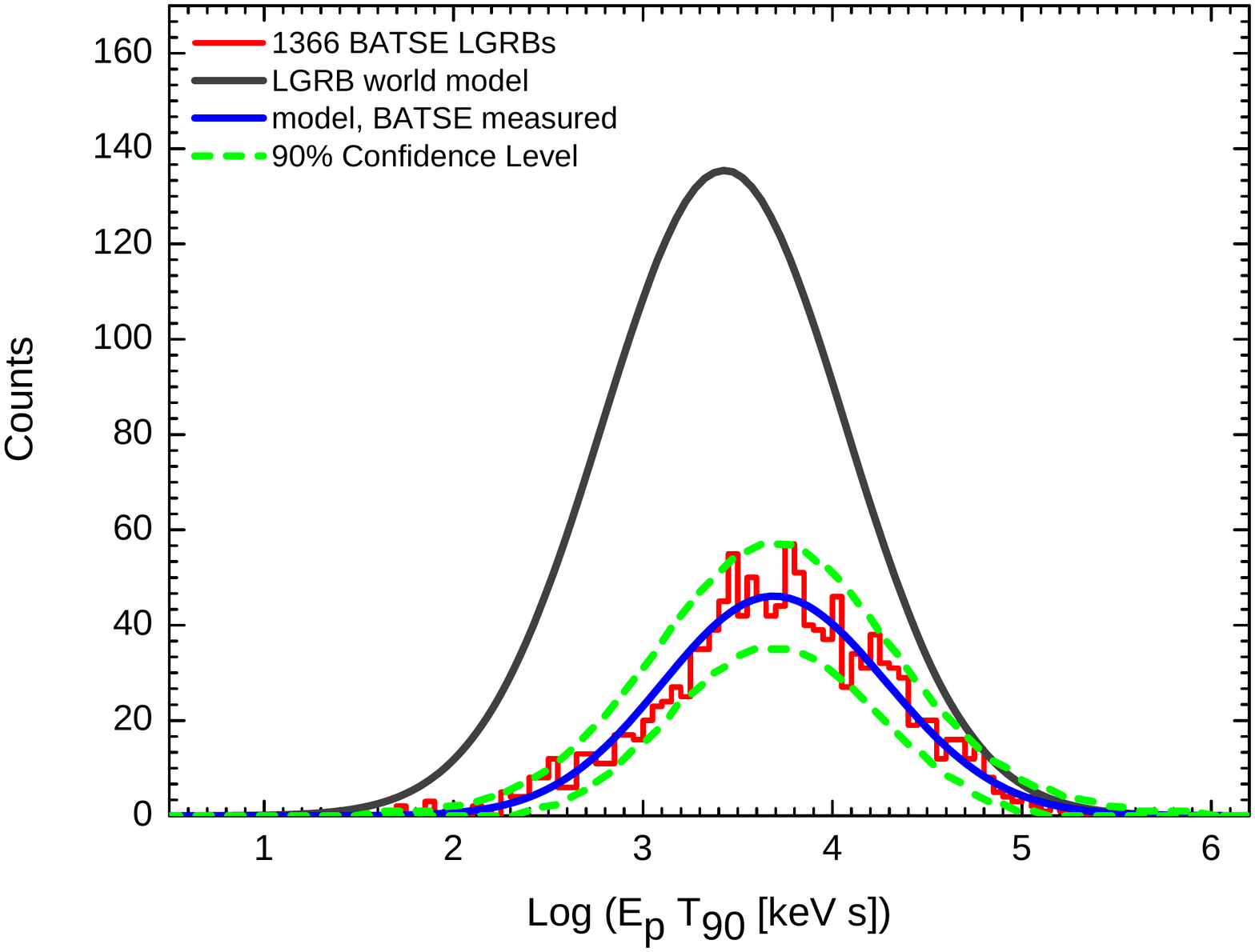}}
            \caption{{\it Top}: BATSE $1366$ LGRB data (\textcolor{red}{red dots}) superposed on the joint bivariate distribution predictions (\textcolor{black}{black dots}) of the LGRB world model for BATSE detection efficiency. The background \textcolor{gray}{grey dots} represent the model predictions for the entire LGRB population (detected and undetected). The uncertainties in BATSE LGRBs' variables are derived from the Empirical Bayes model discussed in Appendix \ref{sec:appC}. {\it Center \& Bottom}: Gauging the goodness-of-fit of the {\it bivariate} model to data by scanning and comparing the joint distributions of the model \& BATSE data along their principal axes. Colors bear the same meaning as the top panels, in addition to the $90\%$ confidence intervals (\textcolor{green}{dashed green lines}) that represent random poisson fluctuations expected in BATSE LGRB-detection process.  \label{fig:bivariates3}}
        \end{figure*}

        As the second level of GoF tests, the joint bivariate model predictions are compared to BATSE LGRB data, presented in Figures (\ref{fig:bivariates1}), (\ref{fig:bivariates2}) \& (\ref{fig:bivariates3}). This method of scanning model and data along the principal axes of the joint bivariate distributions can be generalized to trivariate and quadruvariate joint distributions. For brevity, however, only the bivariate tests are presented.

\section{Results and Discussion}
\label{sec:RD}
    It is observed in the plots of Figures (\ref{fig:OFmarginals}), (\ref{fig:bivariates1}), (\ref{fig:bivariates2}) \& (\ref{fig:bivariates3}) that the model provides excellent fit to data, within the uncertainties caused by random poisson fluctuations in the BATSE LGRBs observed rate. These random fluctuations in BATSE detections are encompassed in each graph by the dashed-green lines that represent the $90\%$ Confidence Intervals (CI) on BATSE LGRB detections (blue solid lines), derived by repeated sampling from the model.

    Unfortunately, same methods for a comparison of data and model cannot be applied in LGRBs rest-frame, due to lack of redshift information for BATSE sample of LGRBs. Nevertheless, a comparison of the model with observational data of other instruments, -- with measured redshifts -- can provide clues on the underlying joint distribution of LGRBs temporal and spectral variables in the rest-frame compared to LGRB detections of different gamma-ray instruments, as will be done in the following sections.

    \subsection{LGRB Luminosity Function \& $\log(N)-\log(P)$ diagram}
    \label{sec:LF}
        The $\log(N)-\log(P)$ diagram of Gamma-Ray Bursts has been subject of numerous studies in the BATSE era, primarily for the purpose of finding signatures of cosmological- (vs. galactic-) origins in the LGRB rate. The cosmological origin of LGRBs is now well established. Nevertheless, the $\log(N)-\log(P)$ diagram can still provide useful information for future gamma-ray experiments.

        Figure (\ref{fig:OFmarginals}, {\it bottom}) depicts the prediction of the LGRB world model for the traditional $\log(N)-\log(P)$ diagram for 1-second peak photon flux in the BATSE nominal detection energy range $50-300~[keV]$, both for the differential ({\it left panel}) \& the cumulative ({\it right panel}) LGRB rate. For all three LGRB cosmic rates considered in this work -- as in Table (\ref{tab:BFP}) -- the differential $\log(N)-\log(P)$ diagram shows a peak in the rate at $\pph\sim0.1 [ph/s/cm^2]$. Such peak in the LGRB rate results in a relative flattening at the dim end of the cumulative $\log(N)-\log(P)$ diagram, as compared to its bright end. This observation has already been reported by B10 for Swift sample of LGRBs, although an entirely different luminosity function -- a broken power-law LF -- were used by B10 in their multivariate LGRB world model (c.f. Equation (2) in B10).

\begin{figure*}
            \center{\includegraphics[scale=0.31]{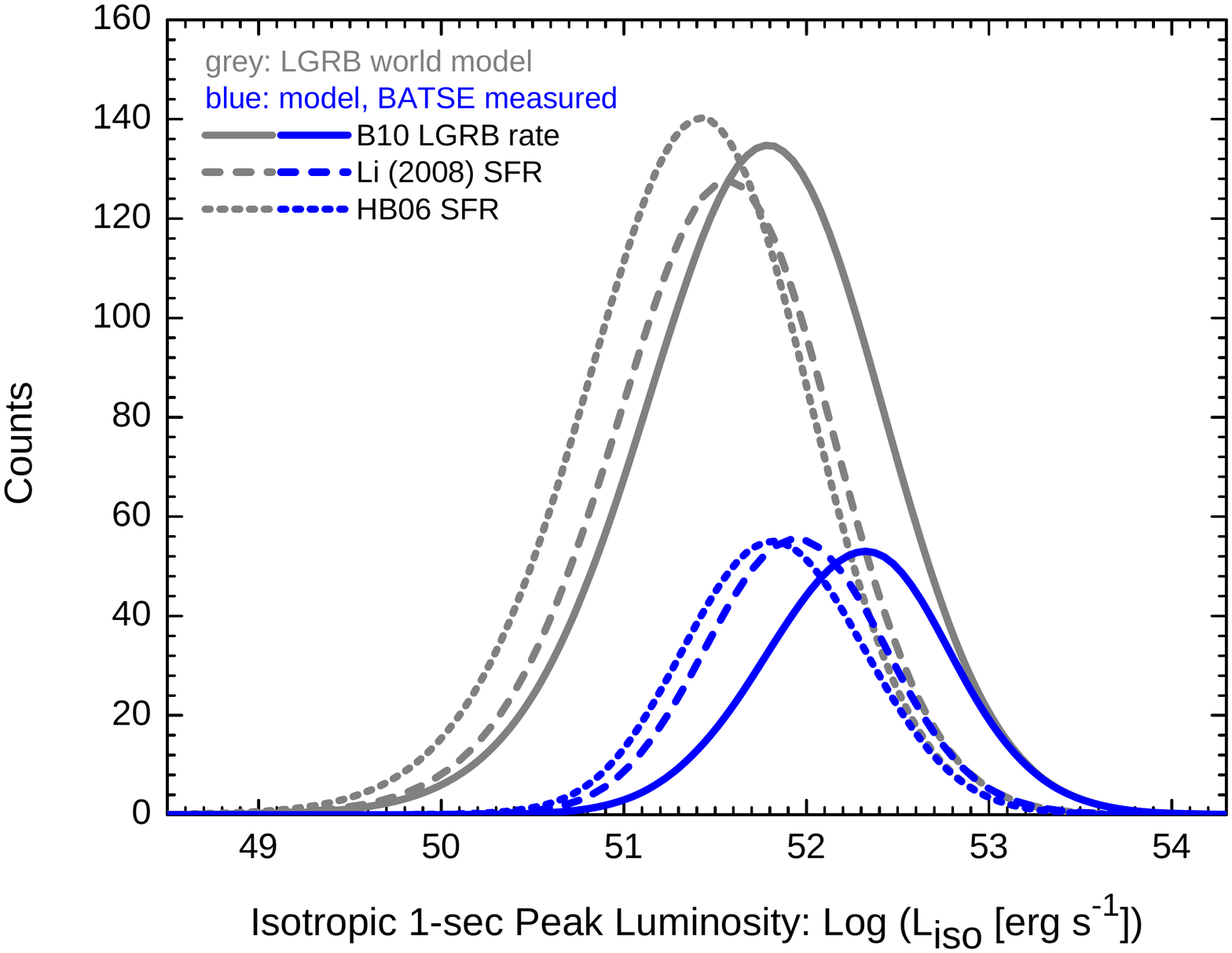}
            \includegraphics[scale=0.31]{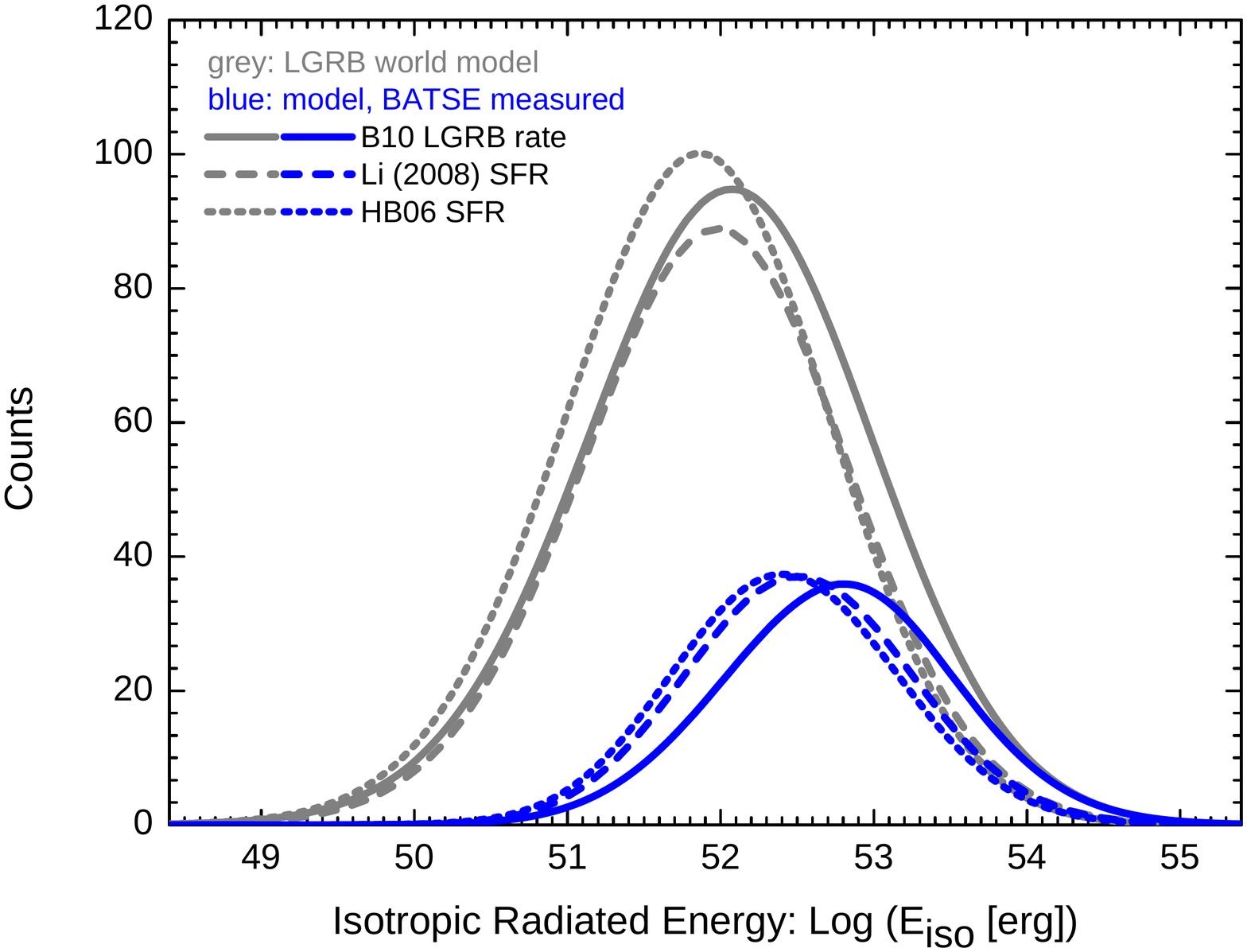}}
            \center{\includegraphics[scale=0.31]{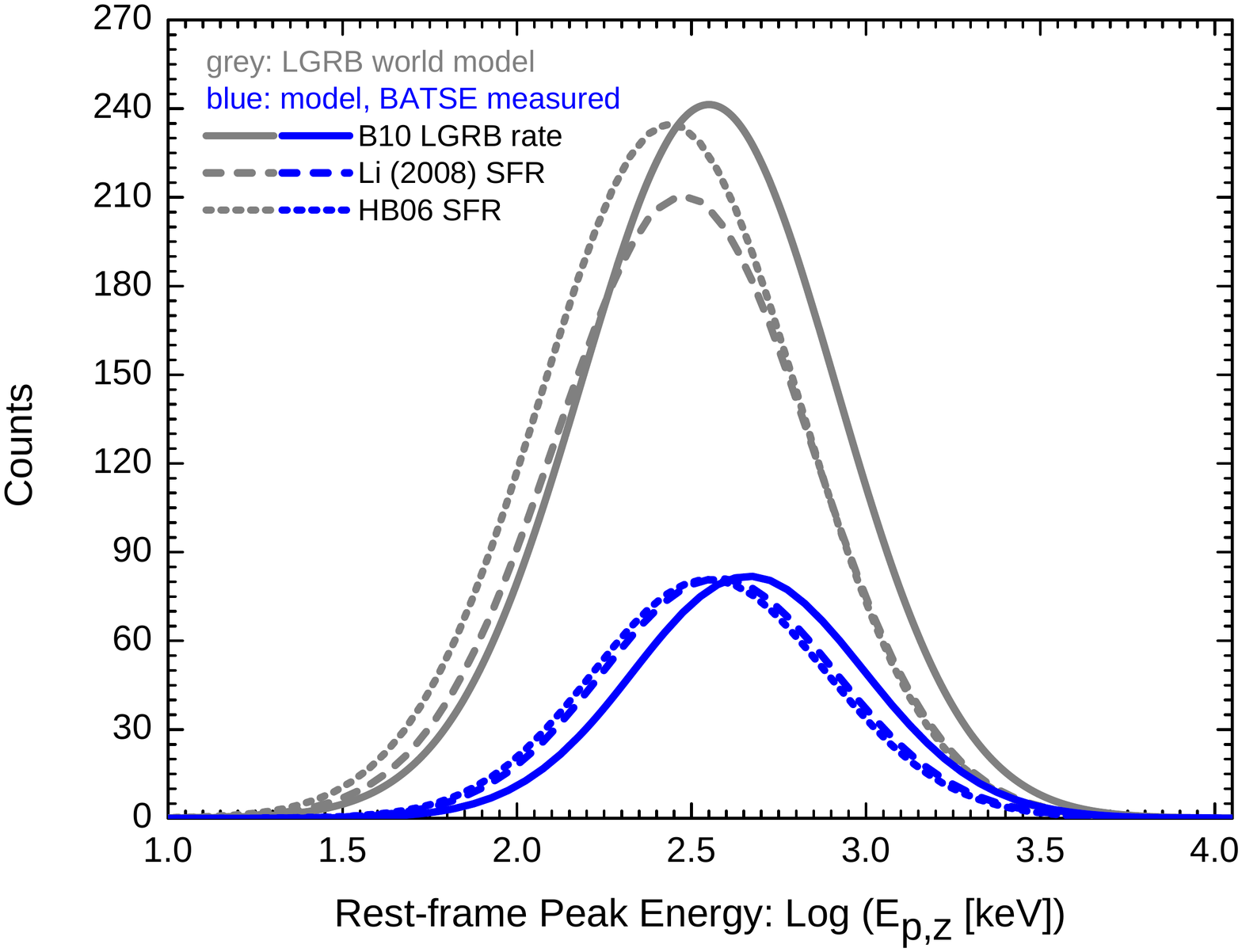}
            \includegraphics[scale=0.31]{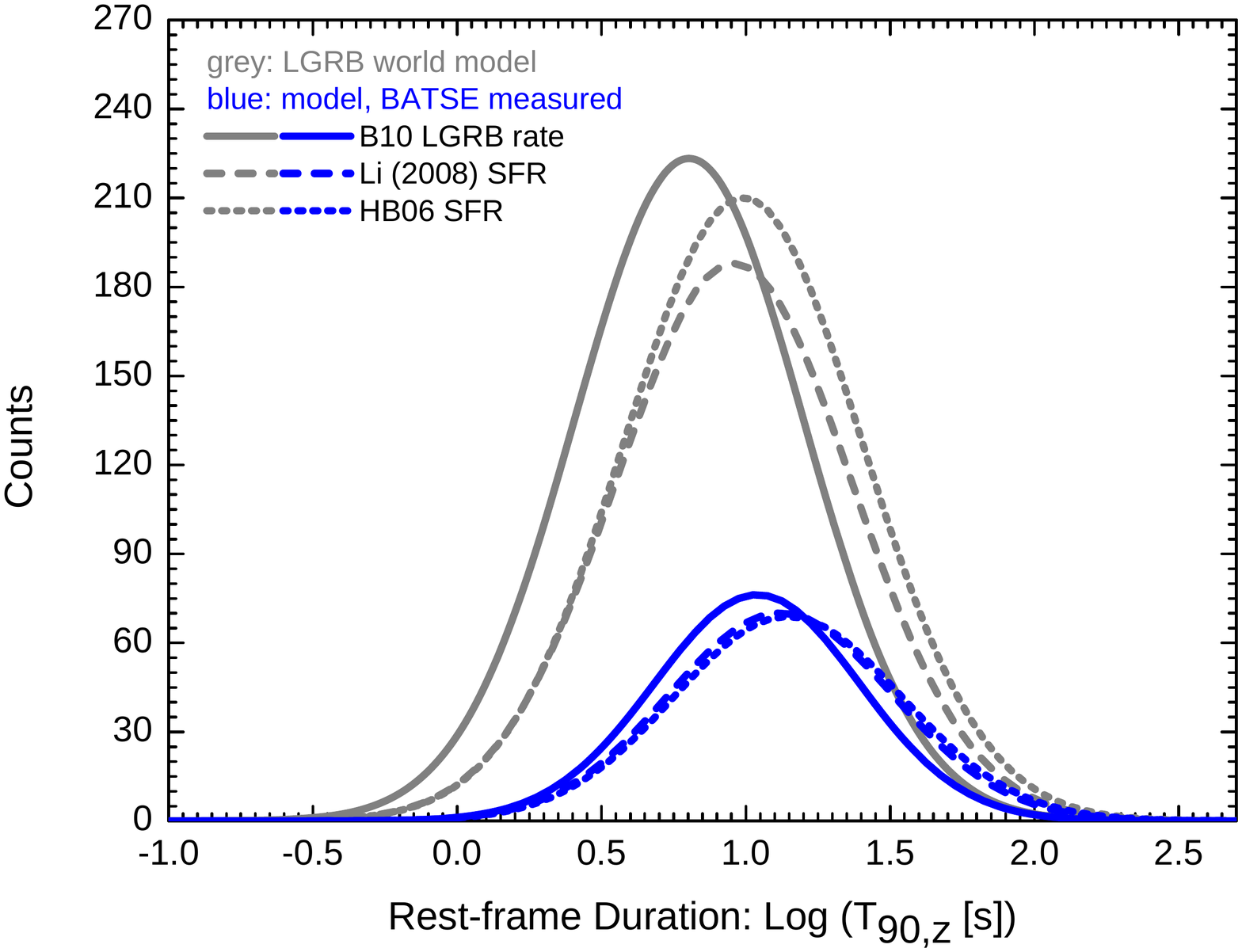}}
            \center{\includegraphics[scale=0.31]{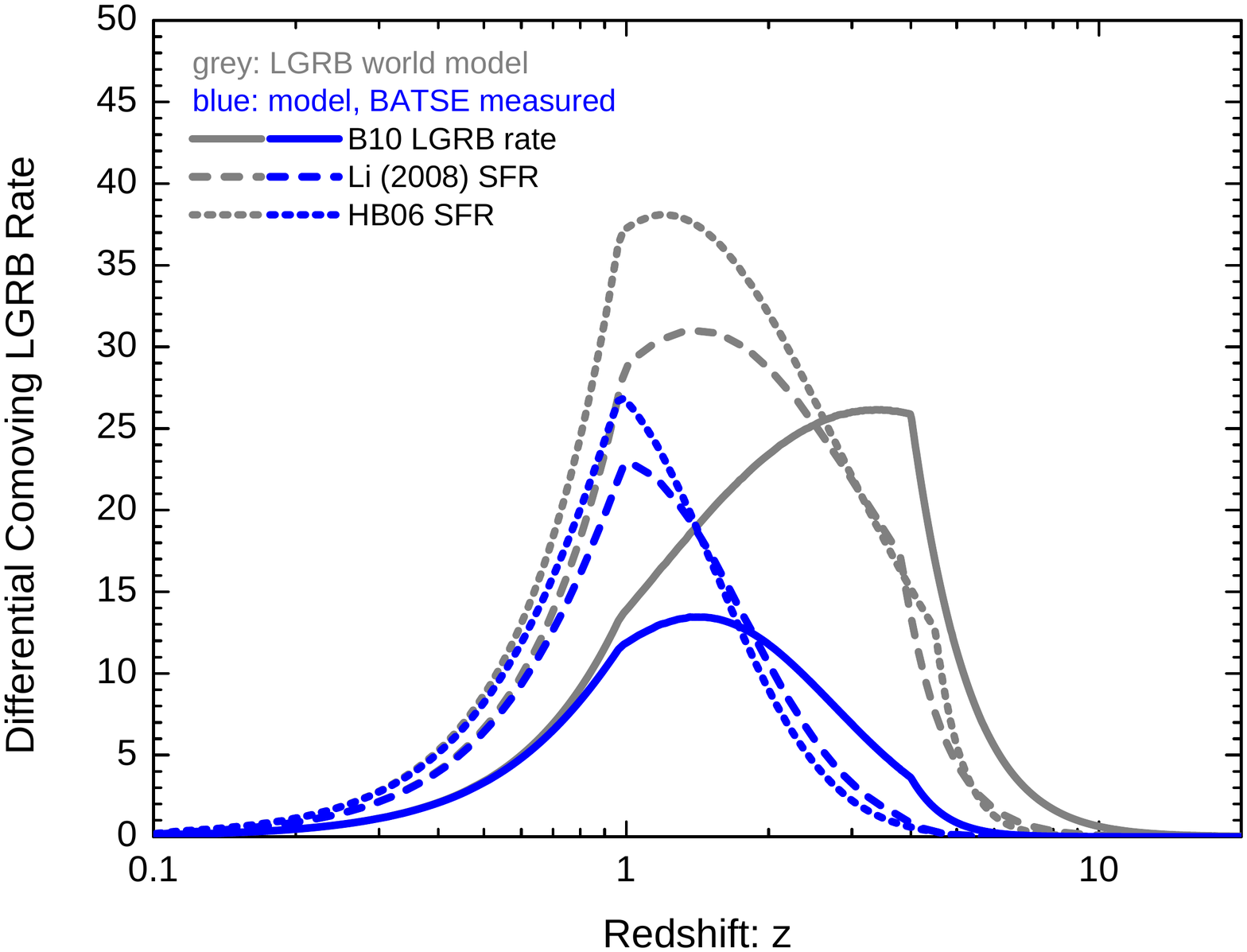}
            \includegraphics[scale=0.31]{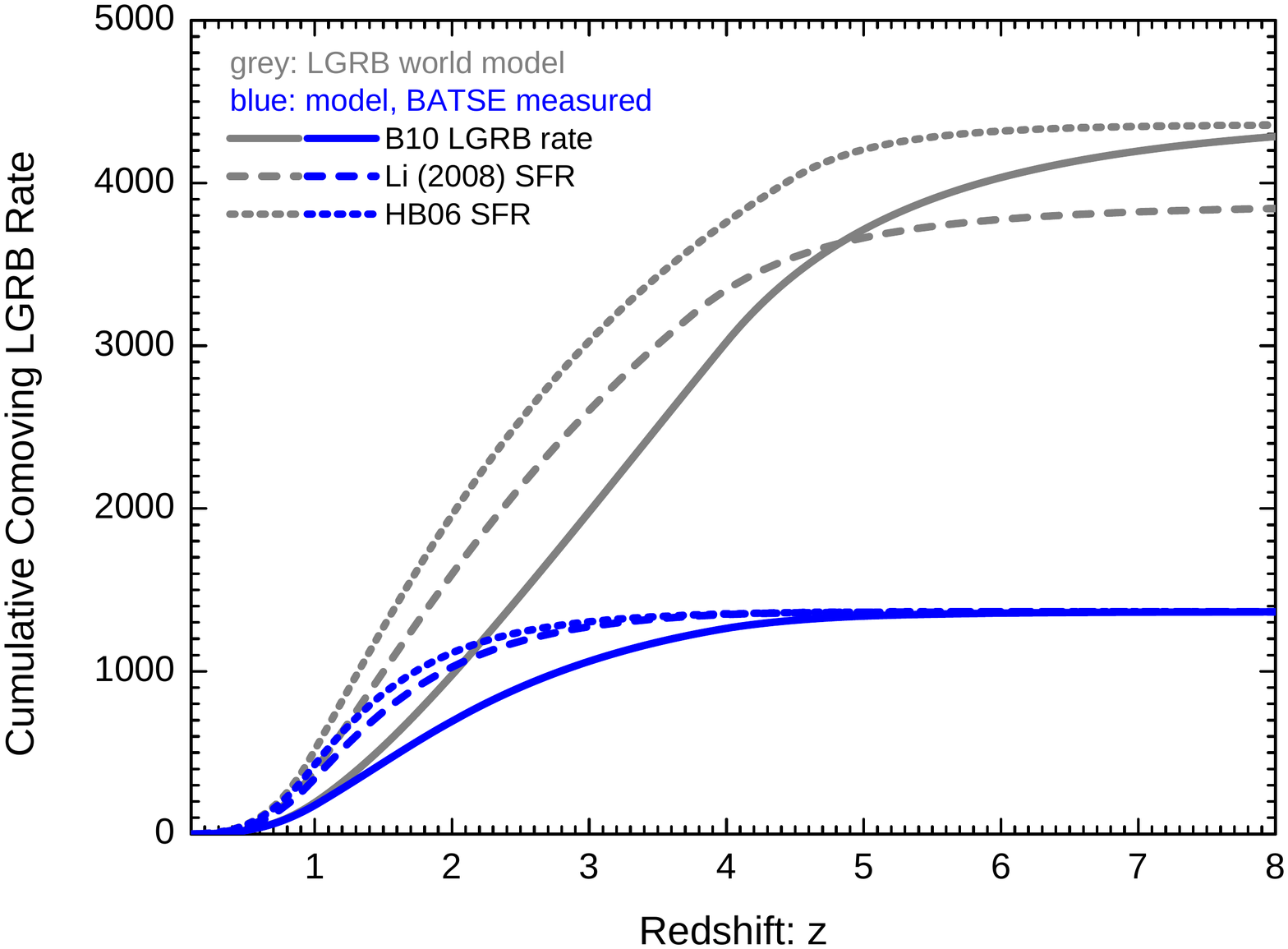}}
            \caption{{\it Top \& Center:} The marginal distribution predictions (\textcolor{blue}{blue dotted/dashed/solid lines}) of the LGRB world model for BATSE LGRBs in the burst's rest frame. The \textcolor{gray}{grey dotted/dashed/solid lines} -- corresponding to the three LGRB redshift distributions HB06/Li(2008)/B10 -- represent model predictions for the entire LGRB population (detected and undetected), with {\it no} correction for BATSE sky exposure and the beaming factor ($f_b$). For comparison with Swift sample of LGRBs, refer to Figures (2), (6) \& (7) of B10. {\it Bottom:} The three differential ({\it left panel}) \& cumulative ({\it right panel}) rates of LGRBs at a given redshift ($z$). \label{fig:CFmarginals}}
        \end{figure*}

        On the other hand, the peak in the observer-frame LGRB rate translates to three relatively different (at $\sim1\sigma$ level) peaks in the luminosity function (LF) of LGRBs. In general, it is observed that the peak of LF (i.e., the average 1-second peak luminosity of LGRBs) increases with increasing the cosmic rate of LGRBs at high redshift. This effect is well depicted in the {\it top left} panel of Figure (\ref{fig:CFmarginals}) for the three LGRB cosmic rates considered: HB06, \citet{li_star_2008}, B10.

        Compared to the predictions of B10's LGRB world model (c.f. bottom plot of Figure (6) in B10), the log-normal model suggests a lower peak for BATSE LGRBs LF ($\sim52.3$ here, vs. $\sim52.7$ in B10) for the same redshift distribution of LGRBs. Averaging over the three redshift distributions considered, the model predicts a dynamic $3\sigma$ range of observer-frame brightness $\log(\pbol [erg/s/cm^2])\in[-7.11\pm2.66]$ corresponding to $\pbol [erg/s/cm^2]\in[1.70\times10^{-10},3.58\times10^{-5}]$ for LGRBs. This translates to an average dynamic $3\sigma$ range -- in the rest-frame -- of $\log(\liso [erg/s])\in[51.53\pm1.99]$ corresponding to $\liso [erg/s]\in[3.46\times10^{49},3.38\times10^{53}]$.

    \subsection{Isotropic Emission \& Peak Energy distributions}
    \label{sec:EEPE}

        A comparison of the {\it top right} panel of Figure (\ref{fig:CFmarginals}) with Swift observations of LGRBs (e.g. B10, Figure 7 \& 8) indicates the similarity of the peak rates in BATSE \& Swift samples of total isotropic emission ($\eiso$) distributions. The distributions for both instruments show a similar peak at $\log(\eiso)\sim52.7$. The Swift $\eiso$ distribution, however, spans to relatively lower $\eiso$ as compared to BATSE. This observation is not surprising if one takes into account the strong trivariate correlations of the three LGRB variables: $\liso$, $\eiso$, $\epkz$, all of which play role in LGRB detection by Swift and BATSE. In fact, a comparison of the spectral peak energy ($\epkz$) distributions of Swift (e.g. Figure (2) in B10) with the model's prediction for BATSE LGRBs ({\it center left} panel of Figure \ref{fig:CFmarginals}) reveals the relatively high sensitivity of Swift BAT detector to dim soft LGRBs as compared to BATSE LADs. Such difference between the two detectors has already been discussed frequently by different authors \citep[e.g.][]{band_comparison_2003, band_postlaunch_2006}.

        In sum, averaging over the three redshift distributions considered, the LGRB world model predicts a dynamic $3\sigma$ range of observer-frame bolometric LGRB fluence $\log(\sbol [erg/cm^2])\in[-6.16\pm3.01]$ corresponding to $\sbol [erg/cm^2]\in[6.82\times10^{-10},7.01\times10^{-4}]$. This translates to an average $3\sigma$ range -- in the rest-frame -- of $\log(\eiso [erg])\in[51.93\pm2.71]$ corresponding to $\eiso[erg]\in[1.66\times10^{49},4.46\times10^{54}]$.

        As for the spectral peak energy ($\epk$ \& $\epkz$) distributions, the model predicts a $3\sigma$ range of observer-frame LGRB spectral peak energy $\log(\epk [keV])\in[1.93\pm1.22]$ corresponding to $\epk [keV]\in[5,1427]$. This translates to an average $3\sigma$ range -- in the rest-frame -- of $\log(\epkz [keV])\in[2.48\pm1.12]$ corresponding to $\epkz [keV]\in[23,4006]$.

    \subsection{Duration distribution}
    \label{sec:durdist}

        GRBs are traditionally flagged as long-duration class of bursts if their observed durations ($\dur$) exceed $2$ seconds \citep[e.g.,][]{kouveliotou_identification_1993}. Such classification, however, has been long known to be ambiguous close to the cutoff set at $\dur=2~[s]$. It will be therefore useful to explore how accurate such classification is for the entire LGRB population (including non-triggered LGRBs). Figure (\ref{fig:OFmarginals}, {\it center right} plot), depicts the underlying population vs. $1366$ BATSE LGRBs observed durations ($\dur$). As implied by the model, the shape of the $\dur$ distribution of LGRBs is not significantly affected by the triggering process of BATSE, since both BATSE and entire LGRB population distributions show a similar peak at $\dur\sim30~[s]$. There is however a slight difference ($<0.1~dex$) in the predicted observer-frame peak of LGRBs $\dur$ distribution, depending on the underlying LGRB redshift distribution assumed. The difference is magnified to $\sim 0.2~dex$ in the rest-frame ($\durz$) duration distribution of LGRBs, for the two extreme cases of HB06 \& B10 redshift distributions. In general, a higher LGRB rate at high redshifts (as in the case of B10 redshift distribution) results in a shift to shorter durations in the duration distribution of LGRBs, in both the observer and the rest frames (Figure \ref{fig:CFmarginals}, {\it center right} plot).

        Overall, the model predicts an average dynamic $3\sigma$ range of observer-frame $\log(\dur [s])\in[1.47\pm1.32]$ corresponding to $\dur [s]\in[1.4,620]$ for LGRBs. This translates to an average dynamic $3\sigma$ range of rest-frame $\log(\durz [s])\in[0.92\pm1.24]$ corresponding to $\durz [s]\in[0.47,145]$.

    \subsection{Temporal \& Spectral Correlations}
    \label{sec:corcoef}

        \begin{figure*}
            \center{\includegraphics[scale=0.31]{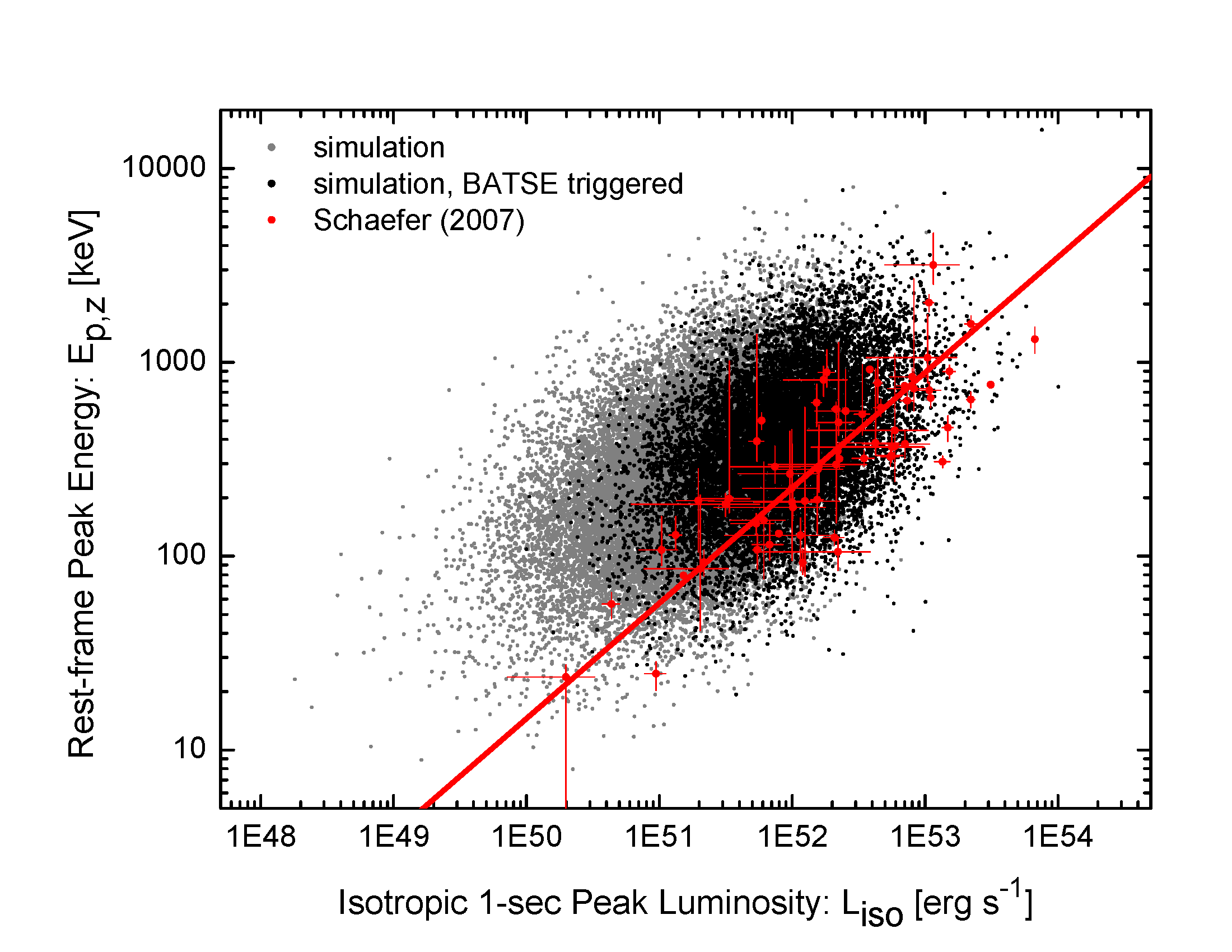}
            \includegraphics[scale=0.31]{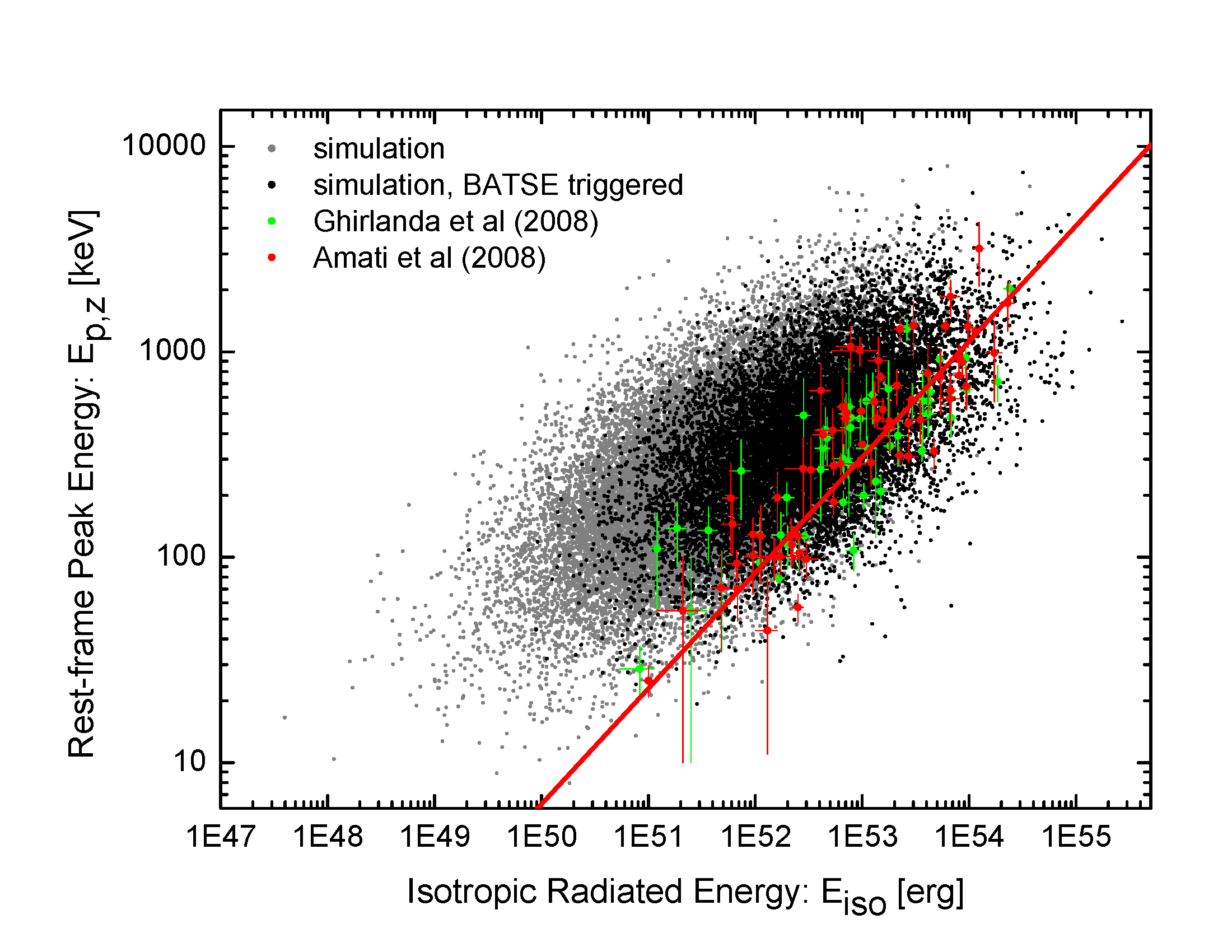}}
            \caption{The joint bivariate distribution predictions (\textcolor{black}{black dots}) of the LGRB world model for BATSE detection efficiency, assuming LGRB rate tracing the cosmic Star Formation Rate (SFR), updated by Li (2008). The background \textcolor{gray}{grey dots} represent model predictions for the entire LGRB population (detected and undetected). Superposed on the model predictions are the well-known proposed Yonetoku ($\liso-\epkz$) \& Amati ($\eiso-\epkz$) relations. The \textcolor{red}{red lines} in each plot represent the best-fit power-law relations derived by the corresponding authors. Neither the Yonetoku nor the Amati relations appear to be consistent with the predicted joint distributions for the entire LGRB population (background \textcolor{gray}{grey dots}) or 1366 BATSE LGRBs (\textcolor{black}{black dots}). This implies strong bias and selection effects in the detection process, redshift determination and spectral analysis of the LGRB samples that were used to construct the two relations. \label{fig:AYrelations}}
        \end{figure*}

        Ever since the launch of Swift Gamma-Ray detector satellite, there has been a flurry of reports on the discovery of strong and significant correlations among the spectral parameters of LGRBs, most prominently, among the rest-frame spectral peak energy and the total isotropic emission or peak luminosity of LGRBs \citep[e.g.,][]{amati_measuring_2008, ghirlanda_e_2008}. Despite the lack of measured redshifts for BATSE GRBs, signatures of such correlations had been found by earlier works in the BATSE era through careful analysis of observer-frame spectral properties of LGRBs \citep[e.g.,][]{lloyd_cosmological_2000}. Nevertheless, the strength and significance of these correlations were undermined by analyzing larger samples of BATSE catalog of GRBs \citep[e.g.,][]{nakar_outliers_2005, band_testing_2005, shahmoradi_how_2009, shahmoradi_possible_2011} or Swift sample of GRBs \citep[e.g.,][]{butler_complete_2007, butler_generalized_2009, butler_cosmic_2010}, all arguing that the sample of bursts used to construct the claimed spectral relations is representative of only bright-soft LGRBs. These arguments have been responded by others \citep[e.g.,][c.f. \citet{shahmoradi_possible_2011} for a complete history of the debate]{ghirlanda_peak_2005, ghirlanda_e_2008, nava_peak_2008}.

        A multivariate analysis of the BATSE LGRBs data that carefully eliminates potential biases at the detection process, can therefore greatly help to understand the strength and significance of the reported correlations. Figure \ref{fig:AYrelations} shows the predictions of the LGRB world model for the two widely discussed spectral correlations: $\liso-\epkz$ (the Yonetoku) and $\eiso-\epkz$ (the Amati) relations. As indicated by the model, a large fraction of BATSE LGRBs (and much larger fraction of the entire LGRB population) on the dim-hard regions of the plots appear to be under-represented by the sample of LGRBs used for the construction of these relations. Nevertheless, given the three redshift distributions considered for the model, a relatively strong (Pearson's correlation coefficient $\rho_{\eiso-\epkz}=0.58\pm0.04$) and highly significant ($>14\sigma$) correlation is predicted between $\eiso$ \& $\epkz$ of Long-duration class of GRBs. The slope of the two relations suggested by the model also differ significantly from the original reports of the relations \citep[][]{schaefer_hubble_2007, ghirlanda_e_2008}. Generally, in a regression modeling, it is assumed that there are two {\it independent} \& {\it dependent} variables. In the case of the proposed relations, none of the variables is known to depend theoretically on the other. Therefore, due to the large {\it statistically unexplained} variances of the two LGRB variables, different regression methods, such as Ordinary Least Squares: OLS($Y|X$) \& OLS($X|Y$), result in entirely different slopes for the relations. The best-fit power-law relations and the conditional variances of the regressand given the regressor can be easily obtained from the parameters of the multivariate log-normal model in Table (\ref{tab:BFP}) \citep[e.g., Section $2.11$ in][]{kutner_applied_2004}.

\begin{figure*}
            \center{\includegraphics[scale=0.31]{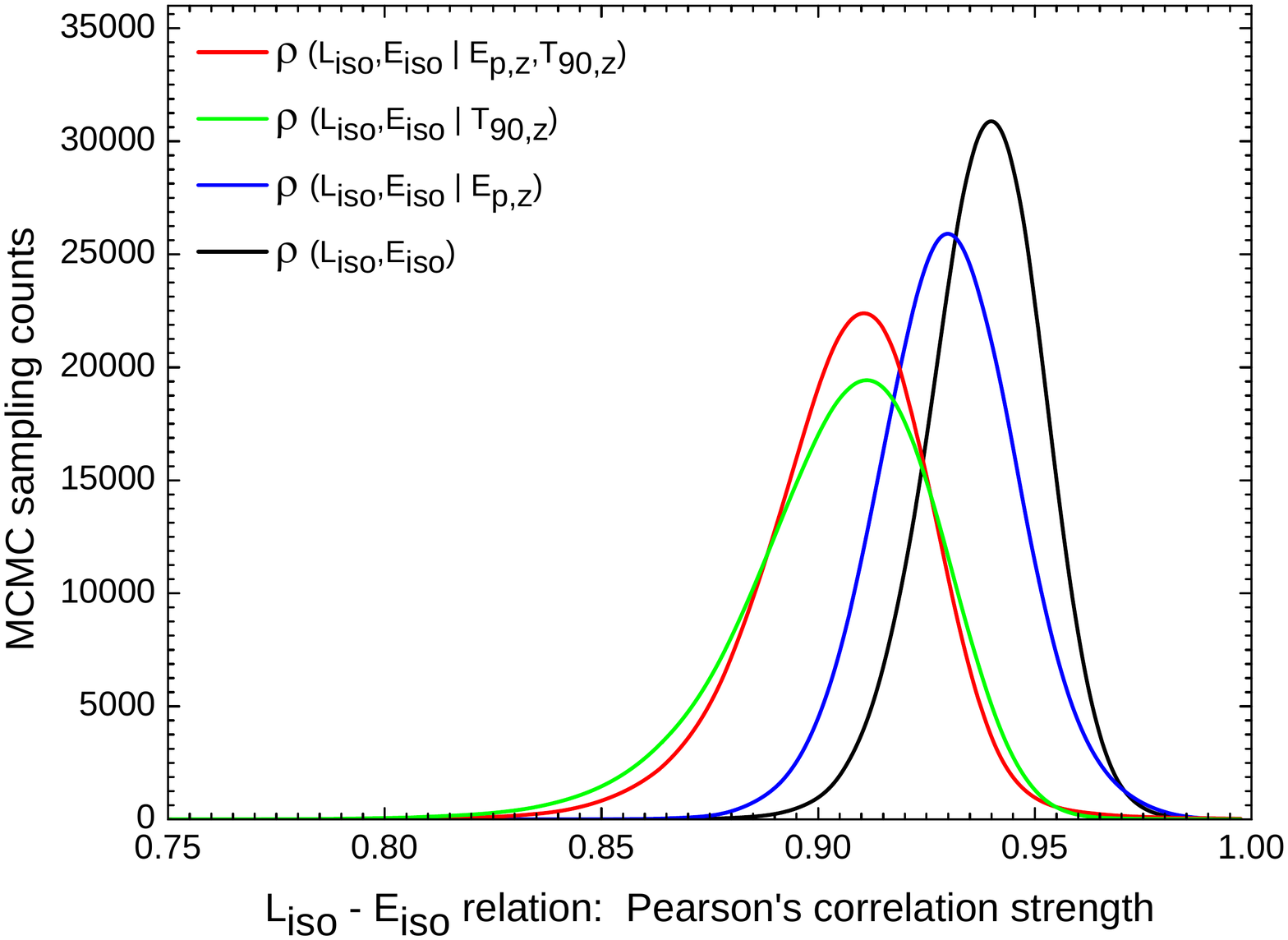}
            \includegraphics[scale=0.31]{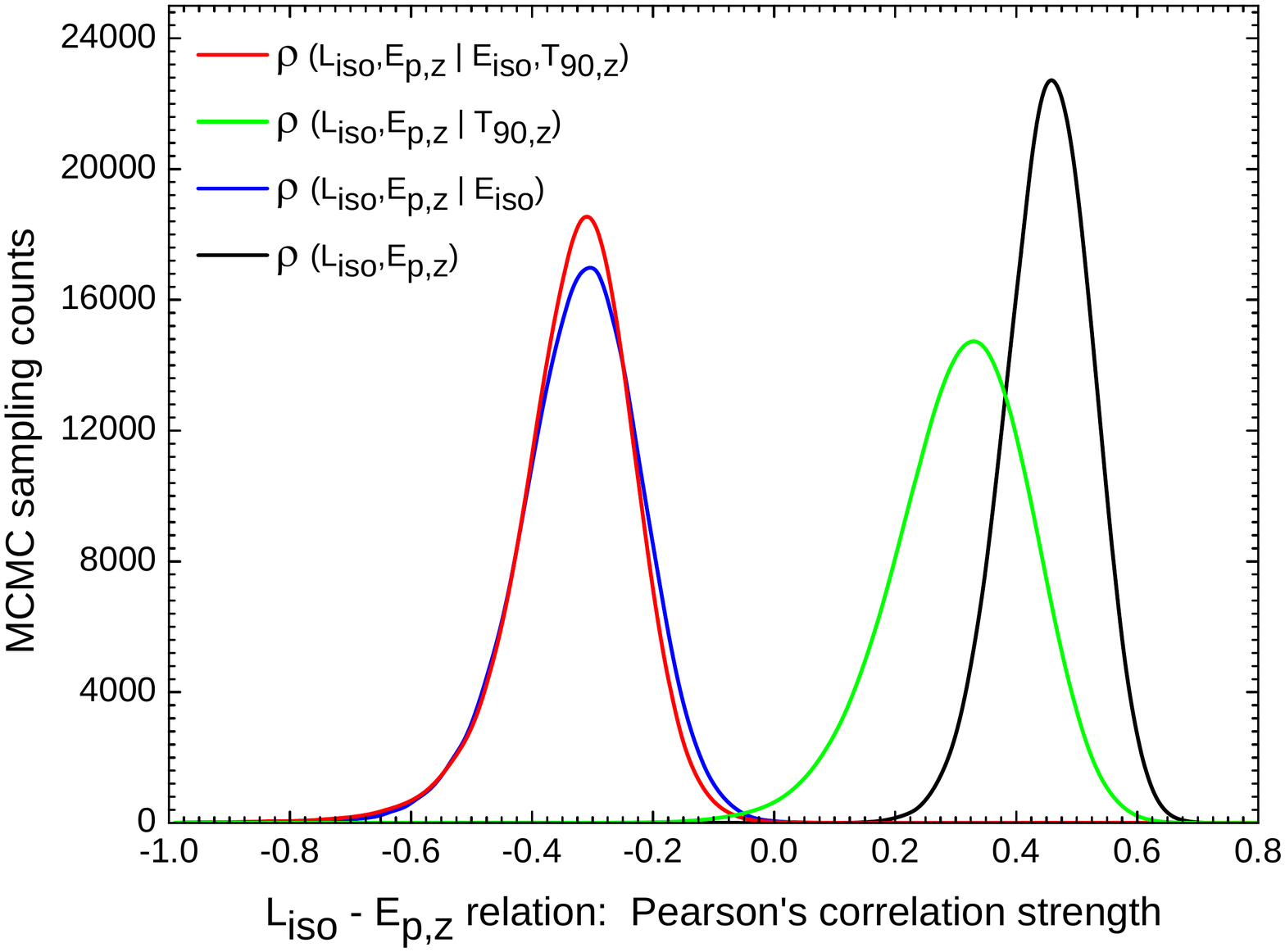}}
            \center{\includegraphics[scale=0.31]{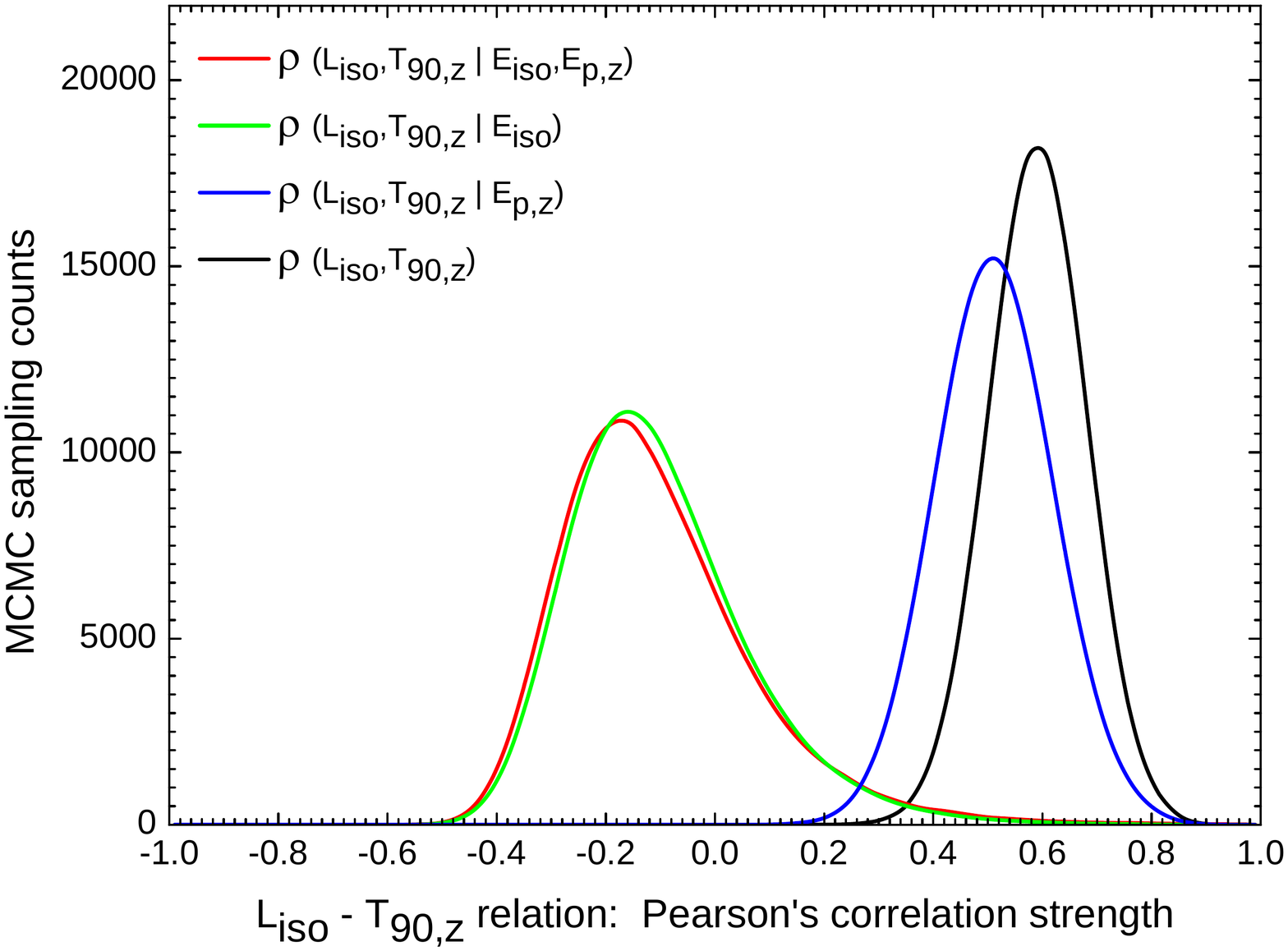}
            \includegraphics[scale=0.31]{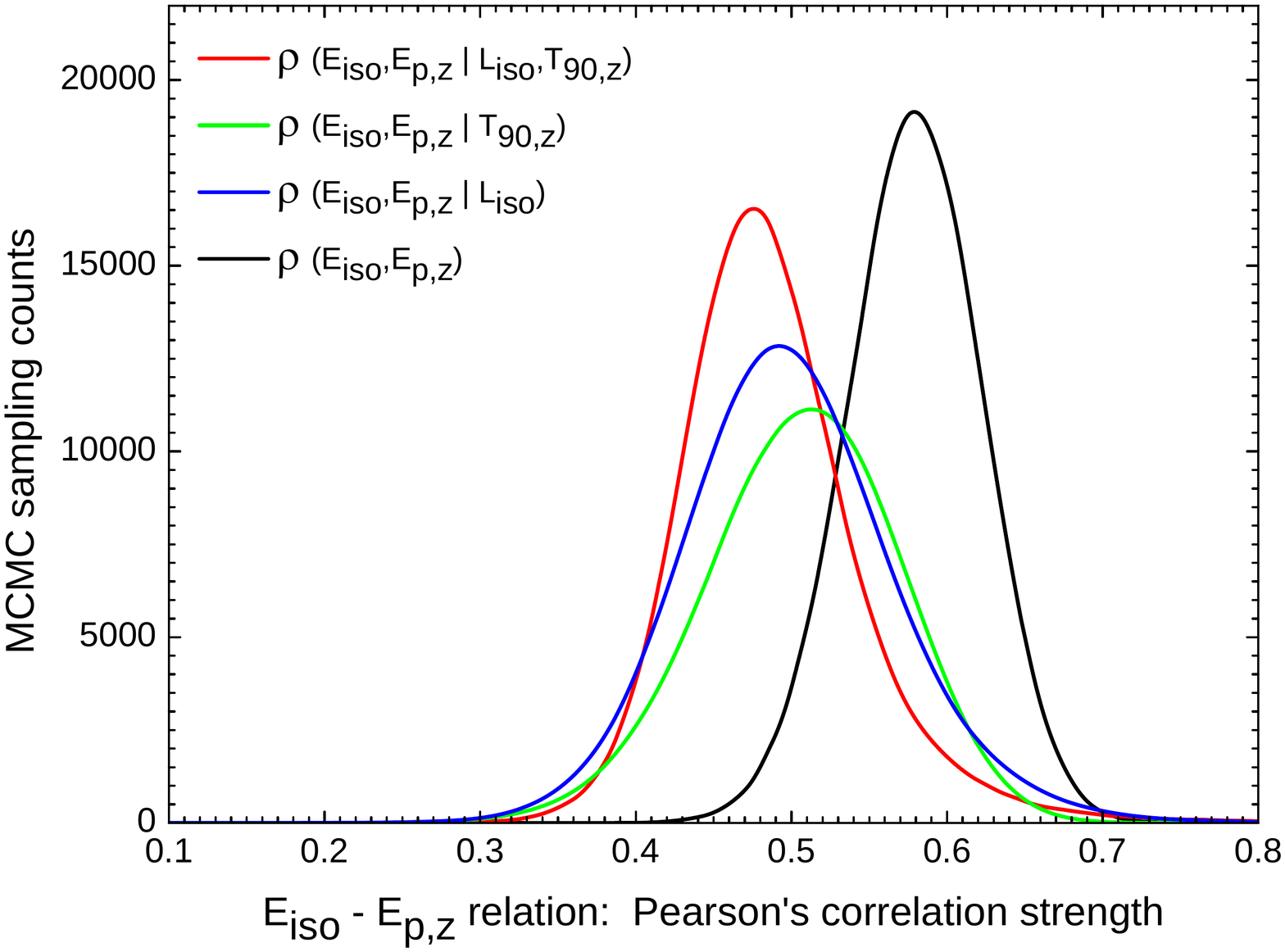}}
            \center{\includegraphics[scale=0.31]{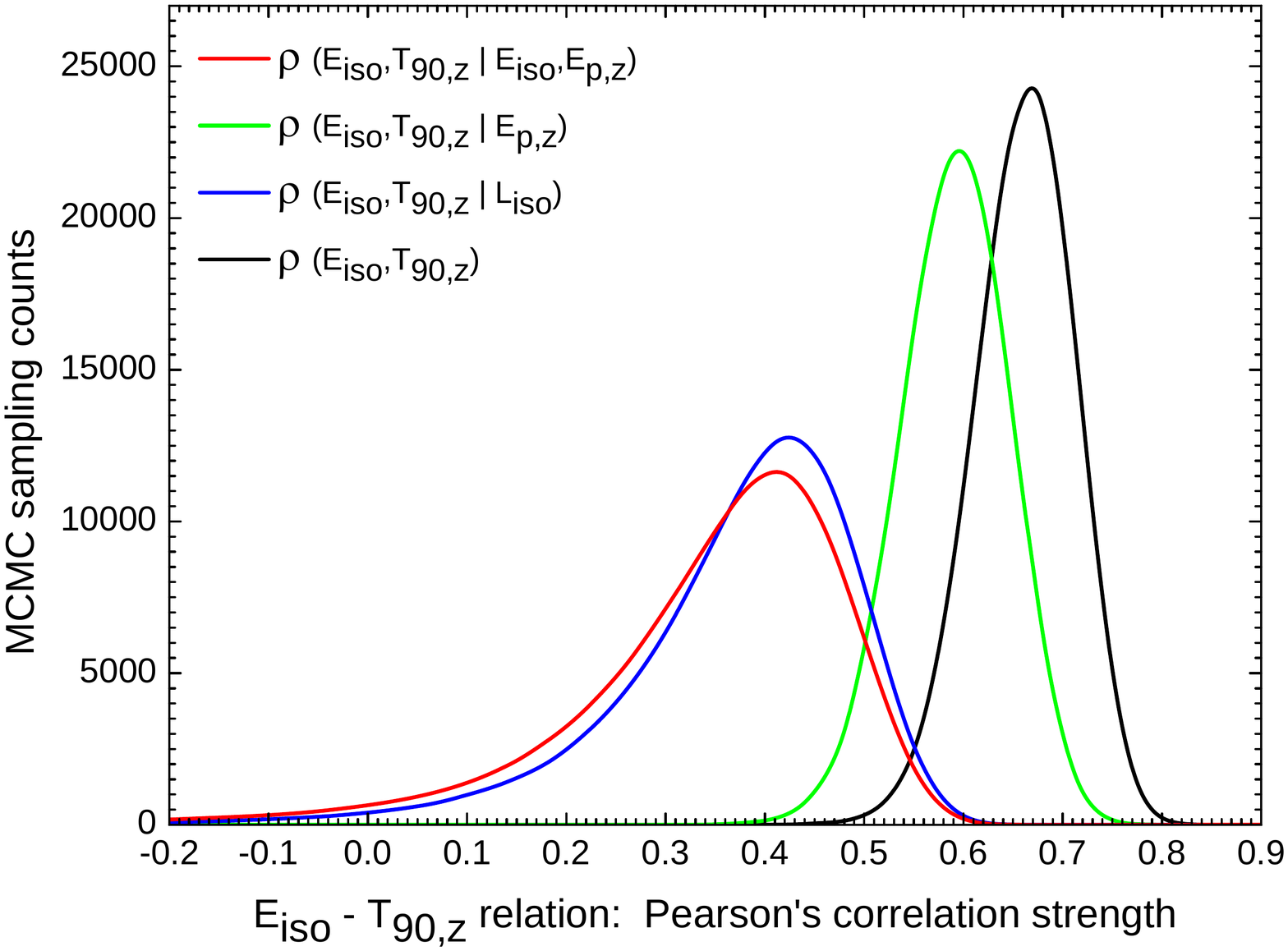}
            \includegraphics[scale=0.31]{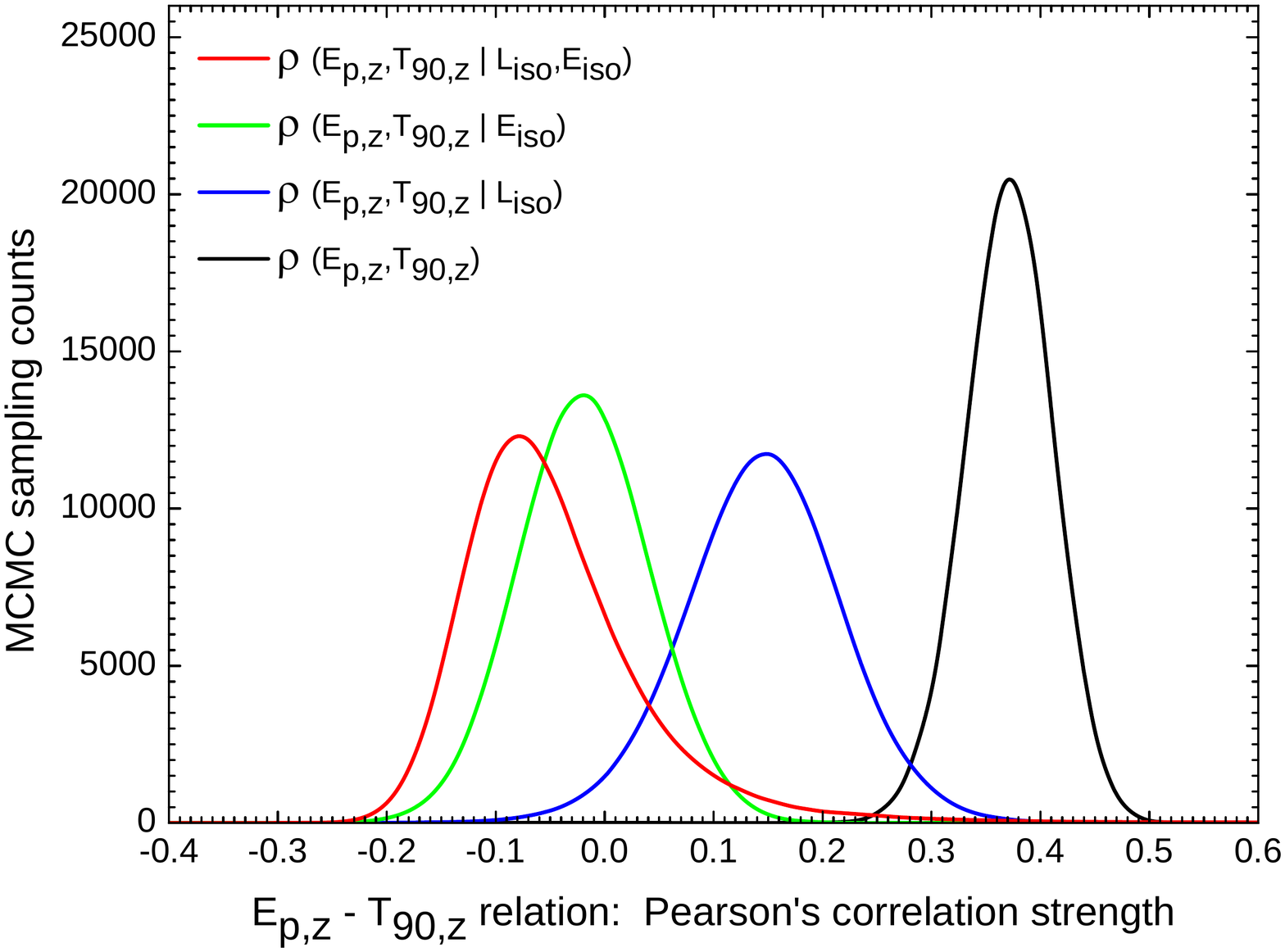}}
            \caption{{\bf Partial correlation analysis of LGRB variables:} Each plot depicts the posterior distributions of the $0^{th}$, $1^{st}$ \& $2^{nd}$ order (partial) correlation coefficients among pairs of LGRB variables, for the median case of LGRB redshift distribution of \citet{li_star_2008}. {\it Top Left:} Keeping $\durz$ \& $\epkz$ fixed, a strong correlation between $\liso$ \& $\eiso$ is still observed, indicating an intrinsic strong association between the total isotropic emission ($\eiso$) and the 1-sec isotropic peak luminosity ($\liso$) of LGRBs . {\it Top Right:} Eliminating the strong dependence of $\liso$ on $\eiso$ reveals that there is indeed a moderate {\it negative} correlation ($\rho\sim-0.32\pm0.1$) between $\liso$ \& $\epkz$ of LGRBs. A similar argument also holds for the correlation of $\liso$ with $\durz$ (as seen in {\it center left} plot). Conversely, {\it center right} plot shows a highly significant positive correlation ($\rho\sim0.49\pm0.06$) of $\eiso$ with $\epkz$ even after dissociation of $\eiso$ \& $\epkz$ from the two other parameters of the model: $\liso$ \& $\durz$, {\it suggesting the existence of a true underlying link between the total isotropic prompt emission ($\eiso$) and hardness ($\epkz$) of LGRBs}. Posterior distributions of the correlation coefficients in the case of the two other LGRB redshift distributions of HB06 and B10 also exhibit similar behavior. \label{fig:parcor}}.
        \end{figure*}

        A partial correlation analysis of the two $\liso-\epkz$ \& $\eiso-\epkz$ relations (Figure \ref{fig:parcor}) reveals that {\it the moderate correlation of the isotropic peak luminosity with the time-integrated peak energy is entirely due to the strong association of $\liso$ with the total isotropic emission from the burst}. As seen in the {\it top left} plot of Figure \ref{fig:parcor} {\it there is indeed a negative correlation between $\liso$ \& $\epkz$ of GRBs for a fixed isotropic emission $\eiso$ and burst duration $\durz$}. Conversely, the model indicates a highly significant correlation of $\eiso$ with the time-integrated $\epkz$, even after elimination of the effects of $\liso$ and $\durz$ on $\eiso-\epkz$ relation.

        As observed in Table (\ref{tab:BFP}), the model predicts positive correlation among all four LGRB variables. The isotropic peak luminosity ($\liso$) and the total isotropic emission ($\eiso$) appear to be strong indicators of each other reciprocally. Surprisingly, it is also observed that the rest-frame duration $\durz$ of LGRBs strongly correlates with both $\eiso$ \& $\liso$. The existence of a possible positive correlation between the isotropic emission and the duration of LGRBs has been implied by the analysis of Swift LGRBs (B10), though only weakly present in there. Such correlations can be enlightening for the early studies of time dilation signatures in BATSE GRBs (Shahmoradi \& Nemiroff, in preparation). A positive duration-brightness correlation is also opposite to -- but not necessarily in contradiction with -- the negative duration-brightness correlation in pulses of individual GRBs \citep[e.g.,][]{fenimore_gamma-ray_1995,nemiroff_pulse_2000,ramirez-ruiz_pulse_2000}. Combination of the two correlations implies that {\it the number of pulses in individual LGRBs should be positively correlated with the peak luminosity (or equivalently, the total isotropic emission) of the bursts}. This is indeed in qualitative agreement with the observed inequality relation between the isotropic peak luminosity and the number of pulses in Swift LGRBs \citep[Figure (6) in][]{schaefer_hubble_2007}.  The strength of the correlations found, encourages the search for the underlying physical mechanism that could give rise to these relations. This is however, beyond the scope of this manuscript \citep[c.f.][for example discussions]{rees_dissipative_2005, ryde_gamma-ray_2006, thompson_thermalization_2007, giannios_peak_2012, dado_kinematic_2012}.

        It is also worth mentioning that the $\dur$ duration of BATSE LGRBs strongly correlates with the bolometric Fluence to bolometric 1-second Peak flux Ratio (FPR), with Pearson's correlation coefficient $\rho_{\text{FPR}-T90}\approx0.67$. A comparison of BATSE data with the predictions of the model for the bivariate distribution of FPR and $\dur$ is given in Figure \ref{fig:FPR-T90} ({\it left panel}). Interestingly, the model predicts the same correlation strength of $\rho_{\text{FPR}-T90}\approx0.67$ for the entire LGRB population, implying that the detection process does not bias the $\text{FPR}-\dur$ relation. Such strong correlation indicates an underlying intrinsic interrelation between the three variables: $\pbol$, $\sbol$ \& $\dur$, also among their corresponding rest-frame counterparts. In fact, B10 use a variant of this trivariate correlation to define an effective peak flux in terms of fluence and duration, discarding the traditional definition of peak flux as the peak photon counts in $1$-second time interval.

\section{Summary \& Concluding Remarks}
    \label{sec:SCR}

    The primary goal of the presented analysis was to model and constrain the luminosity function, temporal and spectral correlations and energetics of Long-duration class of GRBs by exploiting the wealth of information that has been buried and untouched in BATSE GRB catalog to this date. Below is a summary of steps taken to construct the LGRB world model, detailed in Section (\ref{sec:GWM}):

    \begin{itemize}

        \item   A sample of $1366$ LGRBs is carefully selected by the use of Fuzzy C-means clustering algorithm based on the temporal and spectral parameters and inspection of the individual lightcurves of BATSE catalog GRBs (Figure \ref{fig:classification}; Section \ref{sec:samsel}).

        \item   It is proposed that the BATSE LGRB data might be very well consistent with being drawn from a multivariate log-normal population of LGRBs in four rest-frame LGRB variables: the bolometric isotropic 1-second peak luminosity ($\liso$), the bolometric isotropic emission ($\eiso$), the spectral peak energy ($\epkz$) and the duration ($\durz$). Therefore, the observed joint distribution of the four LGRB variables: the bolometric 1-sec peak flux ($\pbol$), the bolometric fluence ($\sbol$), the observed spectral peak energy ($\epk$) and the observed duration ($\dur$), results from the convolution of the rest-frame multivariate log-normal population with the cosmic rate (i.e., the redshift distribution) of LGRBs, truncated by the complex LGRB trigger threshold of BATSE Large Area Detectors (LADs), as illustrated in Section (\ref{sec:MC}), Equations (\ref{eq:cosmicrate}--\ref{eq:obsrate}). A prescription for modeling BATSE LAD detection efficiency is given in Appendix \ref{sec:appB}.

        \item   The LGRB model (Equation \ref{eq:cosmicrate}) is fit to BATSE data by maximizing the likelihood function of the model (Section \ref{sec:MF} \& Appendix \ref{sec:appC}: Equation \ref{eq:likelihood}). In order to derive the best-fit parameters of the model and their corresponding uncertainties, an Adaptive Metropolis-Hastings Markov Chain Monte Carlo (AMH-MCMC) algorithm is set up to efficiently sample from the $16$-dimensional likelihood function. The best-fit parameters are obtained for three LGRB cosmic rates: SFR of \citet{hopkins_normalization_2006}, SFR of \citet{li_star_2008} \& the predicted LGRB redshift distribution of \citet{butler_cosmic_2010} which is consistent with LGRB rate tracing cosmic metallicity with a cutoff ${Z/Z_{\Sun}}\sim0.2-0.5$.

        \item   To ensure the model provides adequate fit to observational data, multivariate goodness-of-fit tests are presented (Section \ref{sec:GOF} \& Figure sets \ref{fig:OFmarginals}, \ref{fig:bivariates1}, \ref{fig:bivariates2}, \ref{fig:bivariates3}).
    \end{itemize}

    \begin{figure*}
             \center{\includegraphics[scale=0.31]{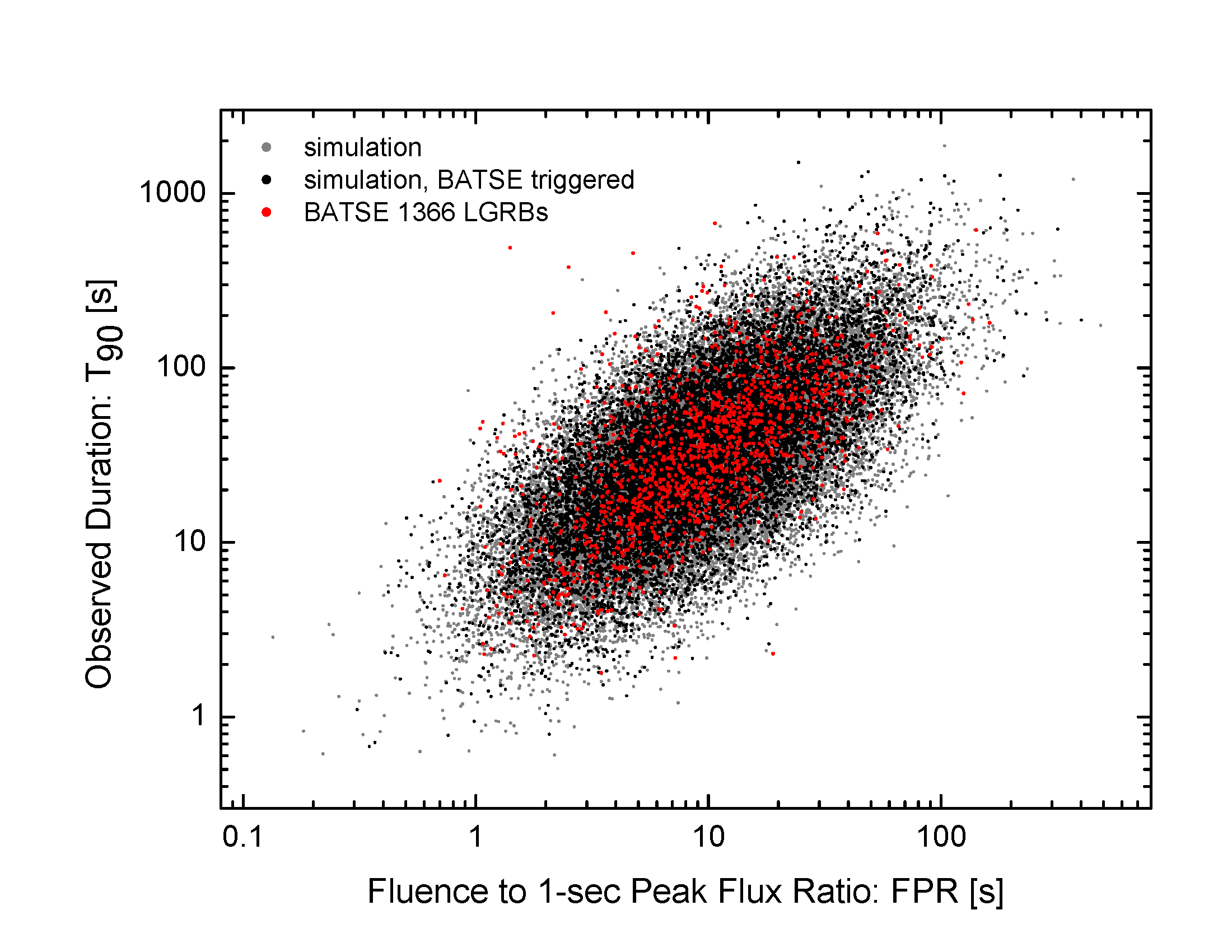}
                     \includegraphics[scale=0.31]{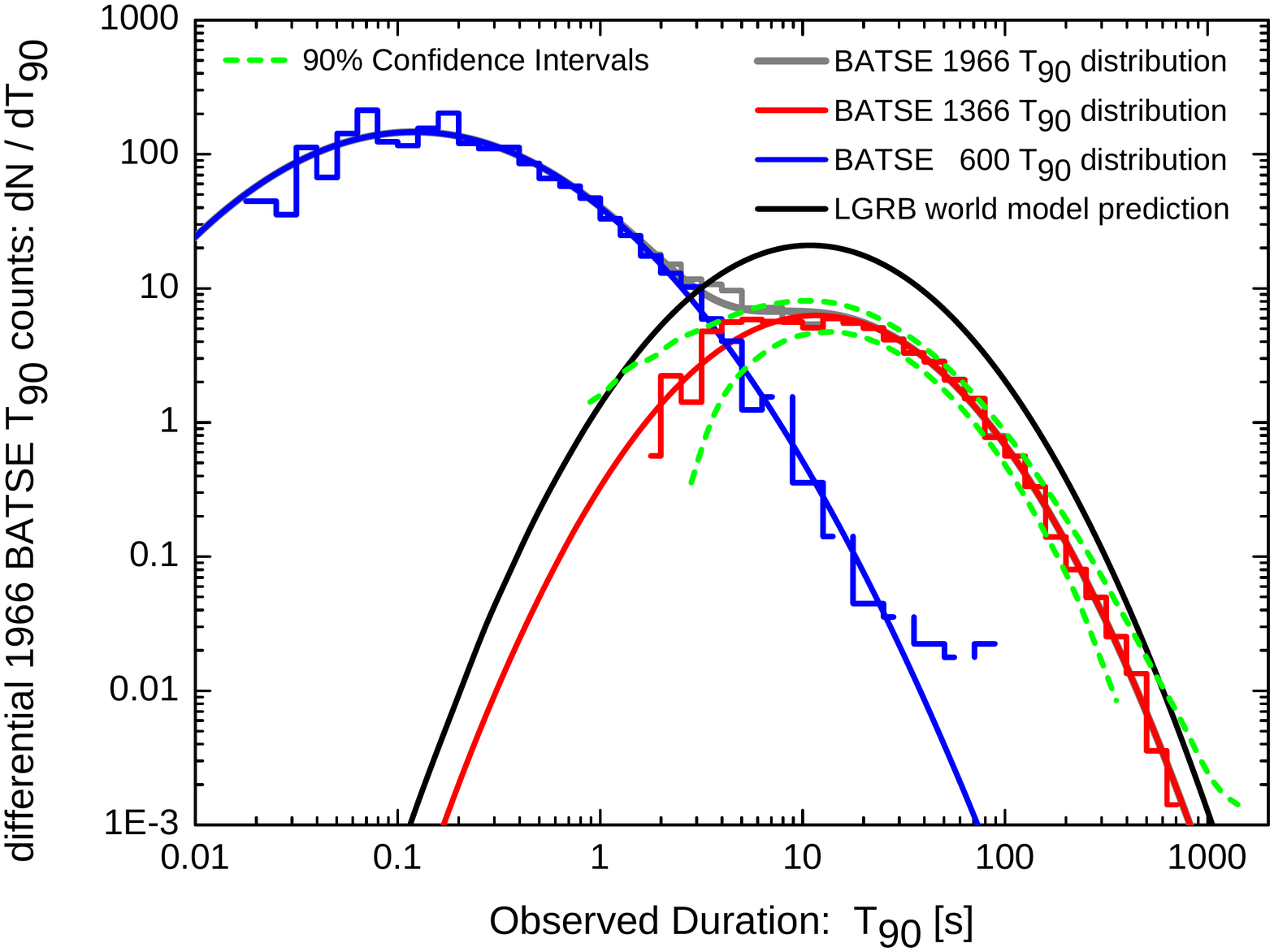}}
            \caption{{\it Left Panel}: The joint bivariate distribution of the observed duration ($\dur$) vs. the Ratio of bolometric Fluence to 1-second bolometric Peak flux (FPR) of $1366$ BATSE LGRBs (\textcolor{red}{red dots}) superposed on the predictions of the LGRB world model (\textcolor{black}{black dots}) for BATSE LGRB detection efficiency. The background \textcolor{gray}{grey dots} represent model predictions for the entire LGRB population (detected and undetected). Both BATSE LGRB data and the LGRB world model exhibit the same (Pearson's) correlation strength of $\rho_{\text{FPR}-\dur}\approx0.67$ indicating that the observed joint distribution is not a byproduct of selection effects in the detection process. {\it Right Panel}: {\bf Differential duration distribution ($dN/d\dur$) of BATSE $1966$ GRBs}: The \textcolor{gray}{grey background histogram \& curve} represents the entire sample of $1966$ BATSE GRBs and the bimodal log-normal fit to this sample, while the \textcolor{blue}{blue} \& \textcolor{red}{red} histograms \& curves represent the the two classes of \textcolor{blue}{short-hard} \& \textcolor{red}{long-soft} GRBs. The \textcolor{black}{black curve} represents the `{\it asymptotic}' prediction of the LGRB world model for the underlying distribution of LGRBs' duration ($\dur$) distribution in the observer frame. The \textcolor{green}{dashed green lines} represent the $90\%$ confidence interval on the `{\it asymptotic}' prediction of LGRB world model for BATSE $1366$ LGRBs (\textcolor{red}{red curve}). The observed flatness in the duration distribution of LGRBs (and also SGRBs) can be attributed to sample incompleteness and the skewed nature of log-normal distribution when plotted as $dN/d\dur$ on logarithmic axes. This interpretation is inconsistent with the argument provided by \citet{bromberg_observational_2012} who point to the observed flatness in $\dur$ distribution of BATSE GRBs as a direct evidence of the Collapsar model of LGRBs. \label{fig:FPR-T90}}
    \end{figure*}

    Summarized below, are the principal conclusions drawn from the analysis based on the proposed LGRB world model:

    \begin{enumerate}
        \item   {\it Energetics}:  It is expected that the peak brightness distribution of LGRBs has effective range of $\log(\pbol [erg/s/cm^2])\in[-7.11\pm2.66]$ corresponding to $\pbol [erg/s/cm^2]\in[1.70\times10^{-10},3.58\times10^{-5}]$. This translates to a dynamic $3\sigma$ range -- in the rest-frame -- of $\log(\liso [erg/s])\in[51.53\pm1.99]$ corresponding to $\liso [erg/s]\in[3.46\times10^{49},3.38\times10^{53}]$. In addition, a turnover is predicted in the differential $\log(N)-\log(P)$ diagram of LGRBs at $\pph\sim0.1~[photons/s/cm^2]$ in BATSE nominal detection energy range ($50-300~keV$). This is consistent with and further extends the apparent flattening in the cumulative $\log(N)-\log(P)$ diagram of Swift LGRBs reported recently by \citet{butler_cosmic_2010}.

            As for the bolometric fluence and the total isotropic emission distributions, a range of $\log(\sbol [erg/cm^2])\in[-6.16\pm3.01]$ corresponding to $\sbol [erg/cm^2]\in[6.82\times10^{-10},7.01\times10^{-4}]$ is indicated. This translates to an average dynamic $3\sigma$ range -- in the rest-frame -- of $\log(\eiso [erg])\in[51.93\pm2.71]$ corresponding to $\eiso [erg]\in[1.66\times10^{49},4.46\times10^{54}]$ (Sections \ref{sec:EEPE} \& \ref{sec:LF}; Table \ref{tab:BFP}). \\

        \item   {\it Durations \& Spectral Peak Energies}: The rest-frame spectral peak energies ($\epkz$) of LGRBs, is likely well described by a log-normal distribution with an average $3\sigma$ range of $\log(\epkz [keV])\in[2.48\pm1.12]$ corresponding to $\epkz [keV]\in[23,4006]$ with peak LGRB rate at $\epkz\sim300~[keV]$. This translates to an effective observer-frame peak energy range of $\log(\epk [keV])\in[1.93\pm1.22]$ corresponding to $\epk [keV]\in[5,1427]$ with peak LGRB rate at $\epk\sim85~[keV]$. It is also observed that the observer-frame $\dur$ durations of LGRBs peaks at $\dur\sim30~[s]$ with a $3\sigma$ range of $\dur [s]\in[1.4,620]$. This translates to an average $3\sigma$ range of rest-frame $\log(\durz [s])\in[0.92\pm1.24]$ corresponding to $\durz [s]\in[0.47,145]$ with a peak rate at $\durz\sim10~[s]$ (Section \ref{sec:durdist}; Table \ref{tab:BFP}).

            Recently, \citet{bromberg_observational_2012} proposed the apparent flatness in the duration distribution of BATSE LGRBs -- when plotted in the form of $dN/d\dur$ instead of $dN/d\log(\dur)$ -- as the first direct evidence of the Collapsar model of LGRBs. The results of presented analysis are inconsistent with a flat $\dur$ distribution of LGRBs at short durations (Figures \ref{fig:OFmarginals}, {\it center right} panel \& \ref{fig:FPR-T90}, {\it right} panel). The observed flat $\dur$ distribution of LGRBs at short durations can be explained away in terms of the skewed nature of log-normal distribution subject to sample incompleteness. It is therefore expected that a significantly larger sample of LGRBs that will be detected by future gamma-ray satellites will smear out the apparent flatness at the short tail of the duration distribution of LGRBs. A similar flat distribution is also observed for SGRBs at very short durations (Figure \ref{fig:FPR-T90}, {\it right panel}) which might be hard to reconcile with the Collapsar interpretation of the observed flatness in LGRBs $\dur$ distribution, proposed by \citet{bromberg_observational_2012}.

        \item   {\it Temporal \& Spectral Correlations}: All four LGRB variables: $\liso$, $\eiso$, $\epkz$ \& $\durz$ appear to be either moderately or strongly positively correlated with each other. In particular, a relatively strong and `{\it broad}' but highly significant correlation strength (Pearson's correlation coefficient $\rho_{\eiso-\epkz}=0.58\pm0.04$) is predicted between $\eiso$ \& $\epkz$ of Long-duration class of GRBs. Surprisingly, $\durz$ appears to evolve with $\liso$ \& $\eiso$ such that brighter bursts generally tend to have longer durations (Section \ref{sec:corcoef}; Table \ref{tab:BFP}). This prediction of the model, together with the previously reported negative correlation of the brightness and the duration of individual pulses in LGRBs \citep[e.g.,][]{fenimore_gamma-ray_1995,nemiroff_pulse_2000,ramirez-ruiz_pulse_2000} might possibly indicate that intrinsically brighter LGRBs contain, on average, higher numbers of pulses. \\
            There is a slight chance that a small fraction ($<50$) of BATSE LGRBs were misclassified as SGRBs by the automated pattern recognition methods exploited in this analysis (c.f. Figure \ref{fig:bivariates1}, {\it center right} panel). If true, it will most likely affect (if significant at all) the constraints derived on the luminosity function of LGRBs and the correlation of $\epkz$ with $\liso$.

        \item   {\it Redshift Distribution}: The lack redshift information for the BATSE GRBs strongly limits the prediction power of the presented analysis for the cosmic rate of LGRBs. Nevertheless, based on the Markov Chain sampling of the likelihood function for the three LGRB redshift distributions considered here (Section \ref{sec:MC} \& Figure \ref{fig:EffLike}), it is observed that BATSE data potentially, but {\it not necessarily}, favors an LGRB rate consistent with cosmic metallicity evolution with a cutoff ${Z/Z_{\Sun}}\sim0.2-0.5$ \citep[c.f.][]{butler_cosmic_2010}, with no luminosity-redshift evolution.

            Assuming LGRBs track Star Formation Rate, only a tiny fraction (i.e., $\sim2-3$) of $1366$ BATSE LGRBs are expected to have originated from high redshifts ($z\gtrsim5$). In the case of an LGRB rate tracing cosmic metallicity evolution \citep[e.g.,][]{butler_cosmic_2010}, the fraction increases by one order of magnitude to $\sim2\%$, corresponding to $\sim27$ bursts out of $1366$ BATSE LGRBs. For comparison, the expected fraction of Swift \& EXIST LGRBs with $z\gtrsim5$ are $\sim6\%$ \& $\sim7\%$ \citep{butler_cosmic_2010}. The discrepancy is well explained by the fact that both Swift and EXIST are more sensitive to long-soft bursts -- characteristic of high redshift LGRBs -- due to their lower gamma-ray trigger energy window, compared to BATSE Large Area Detectors \citep[c.f.][]{gehrels_swift_2004, band_exists_2008, grindlay_grb_2009}.

    \end{enumerate}

     Although fitting is performed for the rest-frame variables, it is notable that the overall shape of the resulting observer-frame distribution of the variables also resembles a multivariate log-normal (c.f. Figures \ref{fig:OFmarginals}, \ref{fig:bivariates1}, \ref{fig:bivariates2}, \ref{fig:bivariates3}, \ref{fig:CFmarginals}, \ref{fig:AYrelations}). In other words, the redshift convolution of the rest-frame population distribution approximately acts as a linear transformation from the rest frame of LGRBs to the observer frame. This is primarily due to the narrow redshift distribution of LGRBs -- as compared to the width of the LGRBs rest-frame temporal and spectral distributions -- with almost $90\%$ of the population originating from intermediate redshifts, $z\in[1,4.3]$. \citet{balazs_difference_2003} provide an elegant discussion on the potential effects of redshift convolution on the observed distribution of LGRBs durations and spectral parameters.

     As implied by the model, there is no evidence for a significant population of bright-hard LGRBs that could have been missed in BATSE catalog of GRBs. Conversely, a large population of low-luminosity with moderate-to-low spectral peak energies seem to have gone undetected by BATSE LAD detectors. It should be emphasized that the apparent lack of very bright-soft LGRBs has a true physical origin according to the analysis presented here and is not an artefact of detection process or spectral fitting models (e.g. the Band model, CPL or SBPL models) used by GRB reseachers. Weather the X-Ray Flashes (XRF), X-ray rich and the sub-luminous GRBs \citep[e.g.,][]{strohmayer_x-ray_1998, kippen_spectral_2003, tikhomirova_new_2006} can be incorporated into a unified class of events described by a single model, remains an open question in this work. At present, these events can be either considered as a separate class of cosmological events or as the soft-dim tail of the LGRB world model that have been mostly out of BATSE detection range and missed (c.f. Figure (3) of \citet{kippen_spectral_2003} for a comparison with the predictions of the LGRB world model here in Figures \ref{fig:OFmarginals}, \ref{fig:bivariates1}, \ref{fig:bivariates2} \& \ref{fig:bivariates3}). A definite answer to this question requires knowledge of the true rate of sub-luminous bursts and XRFs based on the observed rates of these events convolved with complex detection thresholds of different instruments used for observations.

    \citet{butler_cosmic_2010} (B10) have presented an elaborate multivariate analysis of Swift LGRBs. While providing reasonable fit to Swift data, the model of B10 is primarily aimed at the discovery of the potential sub-luminous events that mostly go undetected by gamma-ray detectors. Such model, capable of accounting for a large population of undetected sub-luminous bursts, is proposed at the cost of throwing away parts of information stored in the spectral parameters of LGRBs in the analysis of B10 \citep[c.f.][]{ghirlanda_impact_2012}. Nevertheless, the presented analysis indicates that the apparent correlation of the isotropic peak luminosity ($\liso$) with the {\it time-integrated} spectral peak energy ($\epkz$) of LGRBs is peripheral to the more fundamental relation between the total isotropic emission ($\eiso$) \& $\epkz$ and that the relation can be created by defining an {\it effective luminosity} based on the two GRB variables $\eiso$ and duration (e.g. $\durz$) (Figures \ref{fig:parcor} \& \ref{fig:FPR-T90}: {\it left panel}). In the light of the analysis presented by \citet{ghirlanda_impact_2012}, it can be therefore suggested that a new definition of luminosity based on $\eiso$ \& $\durz$ of GRBs drawn from the LGRB world model of B10 will alleviate the apparent discrepancy between the observed $\liso-\epkz$ relation of LGRBs and the predicted relation from the LGRB model of B10.  It is also expected that a better definition of peak luminosity that is not limited to a specific timescale in the observer-frame of the bursts would result in an improvement in the correlations of the {\it time-integrated} spectral peak energy ($\epkz$) and the isotropic emission ($\eiso$) with the luminosity variable ($\liso$). Given the above arguments, the presence of $\liso$ as an independent variable in the LGRB world model -- which was unfortunate due to the dependence of BATSE trigger algorithm on GRB's peak luminosity -- might be viewed as overfitting and unnecessary.

     The proposed multivariate log-normal model while requiring minimal free parameters compared to any other statistical model considered in GRB literature to date, provides an accurate comprehensive description of the largest catalog of Long-duration GRBs, serving as a powerful probe to explore the population properties of a large fraction of LGRBs that are missed in spectral analyses due to low-quality data or the lack of measured redshift, or simply go undetected due to the instrument's gamma-ray detector threshold. Data from future gamma-ray experiments will enable us to further confirm, improve or invalidate predictions of the presented model.

\appendix

    \section{A. GRB Classification}
    \label{sec:appA}

        It is well known that the traditional classification of GRBs based on a sharp cutoff in the observed duration ($\dur$) distribution of GRBs -- usually set at $\dur=2s~[50-300~keV]$ is insufficient and can be misleading close to the sharply defined border. The apparent long-soft-bright to short-hard-dim trend observed in the prompt emission properties of BATSE GRBs (e.g. Figure \ref{fig:classification}, top panel; also Figure \href{http://adsabs.harvard.edu/abs/2010MNRAS.407.2075S}{(8)} of \citet{shahmoradi_hardness_2010}) necessitates the use of a rigorous classification scheme based on {\it all} available spectral properties of GRBs, in addition to duration.

        Despite a rich literature on the classification methodologies for Gamma-Ray Bursts \citep[e.g.,][and references therein]{hakkila_properties_2000, hakkila_how_2003} the choice of a classification method to separate BATSE catalog of GRBs into two subgroups of Long \& Short -duration with minimal misclassifications remains a difficult task. This is primarily due to the significant overlap (or similarity) in all (or some) spectral and temporal properties of the two classes of GRBs, in addition to heterogeneity of {\it objective functions} that might differ considerably from one classification algorithm to another.

        Here, fuzzy (soft) clustering algorithms are preferred over hard clustering methods, since they provide a probability of the event belonging to each specific subgroup, in contrast to hard classifications that return only binary probabilities of either $0$ or $1$. The choice of fuzzy algorithm greatly facilitates identification of bursts that might have been potentially misclassified (Section \ref{sec:samsel}).

        Investigation of different fuzzy algorithms available in the literature leads us to two prominent candidates: The SAND method of \citet{rousseeuw_fuzzy_1996} \& the fuzzy C-means discussed by \citet{dunn_fuzzy_1973} \& \citet{bezdek_pattern_1981}. While fuzzy C-means is specially useful for cases where subgroups are known to be approximately symmetric, the SAND algorithm is superior to C-means for its lack of sensitivity to different subgroup sizes, orientations and asymmetries. Nevertheless, the presence of a handful of Soft-Gamma Repeaters (SGRs) in the BATSE catalog  -- with spectral properties comparable to that of LGRBs -- results in relatively poor classification by the SAND method as compared to C-means. Besides the choice of algorithm, the GRB variables -- by which the classification is done -- are selected such that the resulting relative sizes of the two SGRB and LGRB populations correspond to those found by \citet{shahmoradi_hardness_2010} through a different approach that the author believes to be less prone to biases (c.f. Figure \href{http://adsabs.harvard.edu/abs/2010MNRAS.407.2075S}{(13)} \& Table \href{http://adsabs.harvard.edu/abs/2010MNRAS.407.2075S}{(4)} in Shahmoradi \& Nemiroff 2010).

    \section{B. BATSE Trigger Efficiency}
	\label{sec:appB}

        Before the LGRB world model of Equation (\ref{eq:cosmicrate}) is fit to BATSE observational data, it is necessary to convolve the model with BATSE trigger threshold (Equation \ref{eq:obsrate}). The study of BATSE detection efficiency is well documented in a series of articles by BATSE team \citep [e.g.,][c.f. \citet{shahmoradi_possible_2011} for further discussion and references]{Pendleton_detector_1995, pendleton_batse_1998, paciesas_fourth_1999, hakkila_batse_2003}.  Here, based on the observation that almost all $1366$ BATSE LGRBs have durations of $\dur>1 [s]$, the primary trigger time-scale for BATSE LGRBs is assumed to be $1024~[ms]$. This eliminates the relatively complex dependence of the detection probability ($\eta$ in Equation \ref{eq:obsrate}) on the duration of the events. The probability of detection for a LGRB is then modeled by the Cumulative Density Function (CDF) of log-normal distribution,

        \begin{equation}
            \label{eq:threshmodel}
            \eta(detection|\mu_{thresh},\sigma_{thresh},\liso,\epkz,z) = \frac{1}{2} + \frac{1}{2}{\rm erf}\bigg( \frac{{\log{\big(P(\liso,\epkz,z)\big)} - \mu_{thresh}}}{\sqrt{2}\sigma_{thresh}} \bigg),
        \end{equation}

        where $P(\liso,\epkz,z)$ is 1-second peak photon flux in the BATSE nominal detection energy range: $50-300~[keV]$, and $\mu_{thresh}$ \& $\sigma_{thresh}$ are the detection threshold parameters to be determined by the model. The link between the 1-sec peak photon flux ($P$) and the LGRB rest-frame variables ($\liso,\epkz$) and redshift ($z$) is provided by fitting a smoothly broken power-law known as the Band model \citep{band_batse_1993} to LGRBs differential photon spectra,

        \begin{equation}
            \label{eq:Band}
            \Phi (E) \propto \begin{cases}
                            	 E^{\alpha}~ \exp\big(-\nicefrac{(1+z)(2+\alpha)E}{\epkz}\big) & E\le[\epkz/(1+z)][(\alpha-\beta)/(2+\alpha)] \\
            	                 E^{\beta} & \mbox{otherwise.}
            	             \end{cases}
        \end{equation}

        such that,

        \begin{equation}
            \label{eq:peakflux}
            P(\liso,\epkz,z)=\frac{\liso}{4\pi {D_L}^2(z)} \frac{\int_{50}^{300} \Phi dE}{\int_{0.1/(1+z)}^{20000/(1+z)} E \Phi dE},
        \end{equation}

        It has been shown by B10 that fixing the the high-energy and low-energy photon indices of the Band model (Equation \ref{eq:Band}) to the corresponding population average $\alpha=-1.1$, $\beta=-2.3$ produces only a negligible error of $<0.05~{\rm dex}$ in the resulting flux estimates. Given the uncertainties in the BATSE LGRB variables , in particular $\epk$ estimates, such an assumption provides reasonable approximation for the calculation of the peak fluxes. A more accurate treatment, however, would be to include possible weak correlations that are observed between the Band model photon indices and the spectral peak energies of the bursts (c.f. \citet{shahmoradi_hardness_2010} \& \citet{shahmoradi_possible_2011} for a discussion of the correlations and potential origins).

\begin{figure*}
            \center{\includegraphics[scale=0.31]{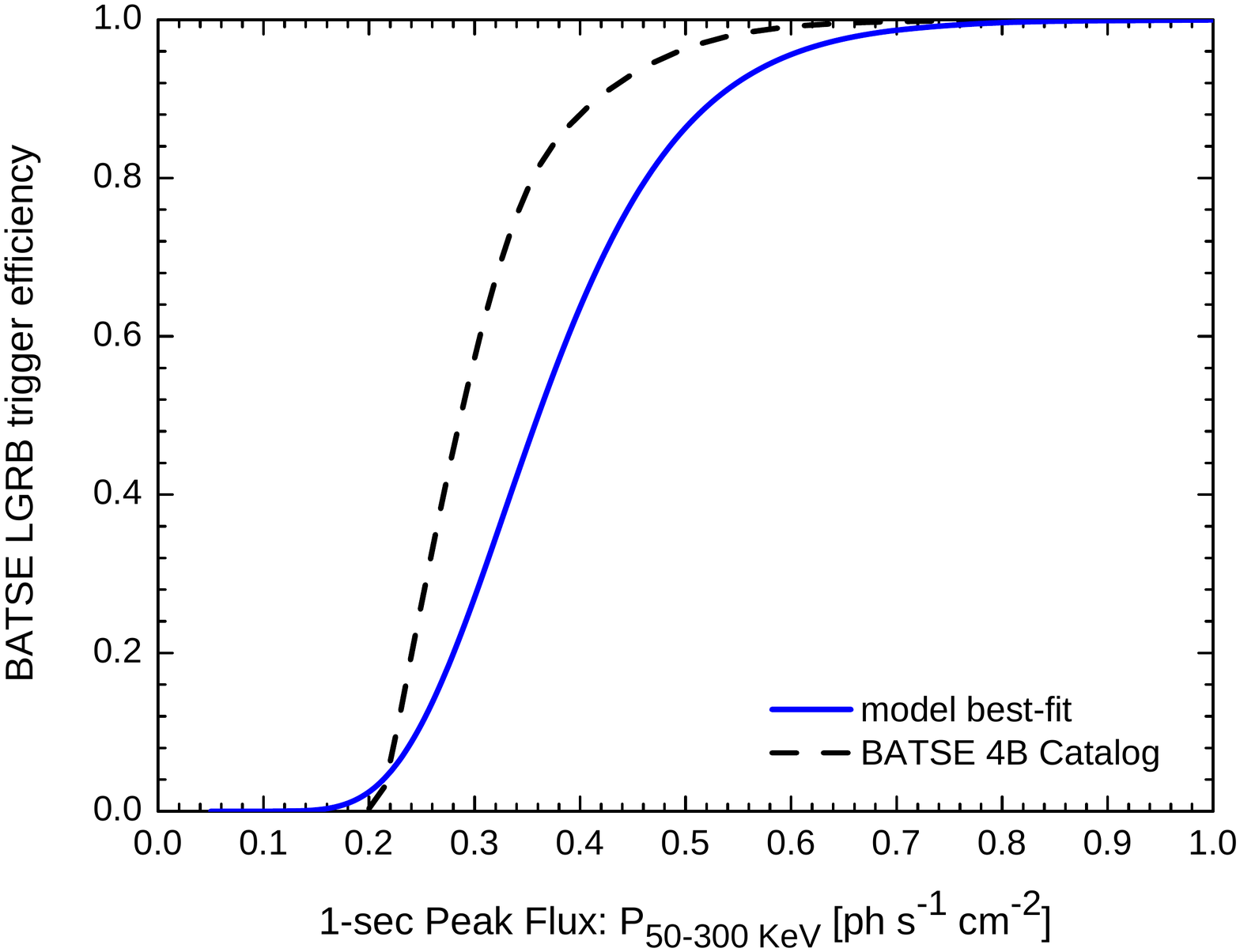}
            \includegraphics[scale=0.31]{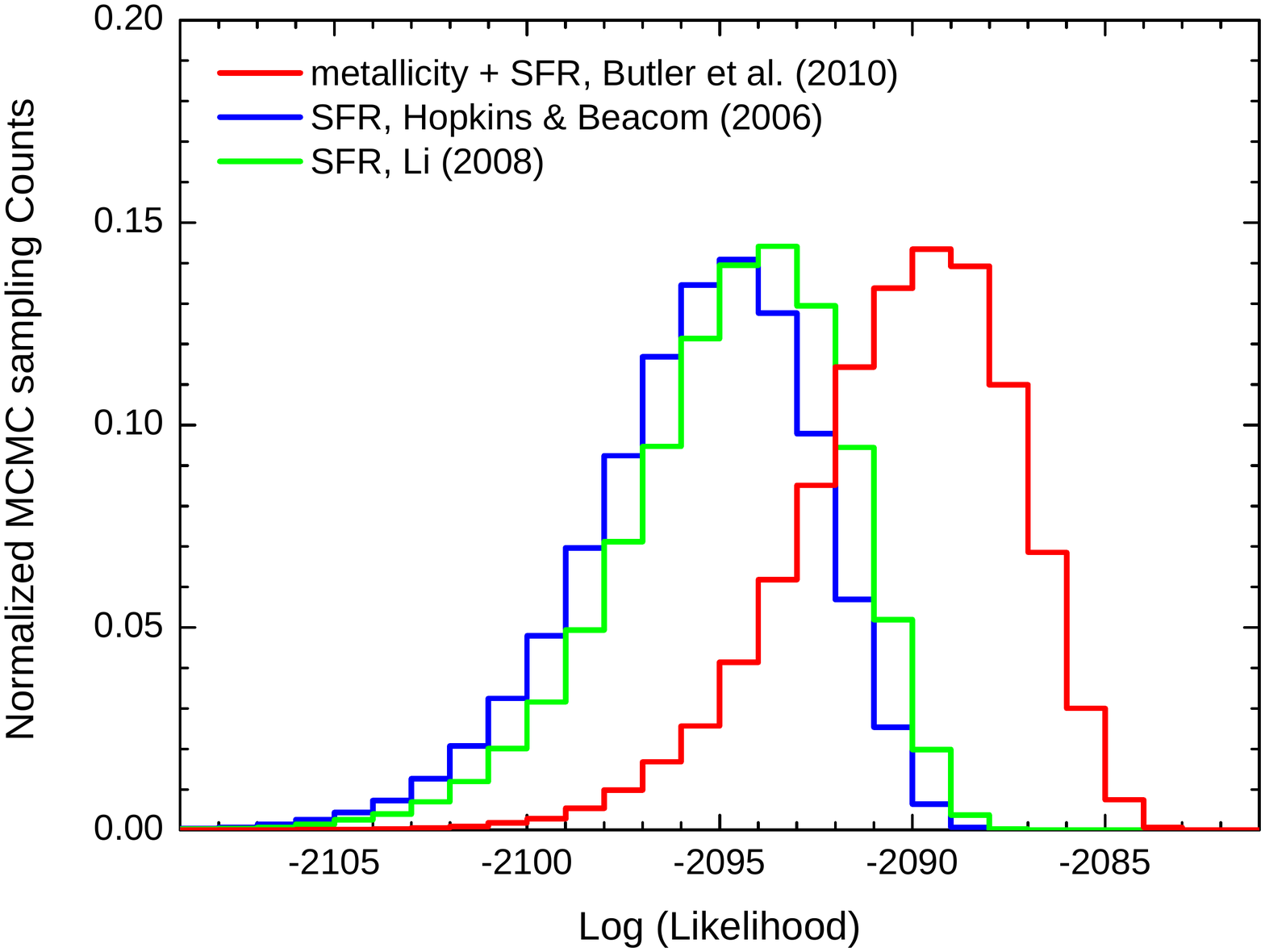}}
            \caption{{\it Left}: A comparison of the model-predicted BATSE trigger efficiency of $1366$ BATSE LGRBs with the nominal trigger efficiency estimates of BATSE 4B Catalog. The discrepancy between the two curves is primarily due to different methodologies and GRB samples used to derive BATSE Catalog efficiency curve\footnote{\url{http://www.batse.msfc.nasa.gov/batse/grb/catalog/4b/tables/4br\_grossc.trig\_sen}} and the trigger efficiency of the LGRB world model. {\it Right}: The normalized sampling distribution of the likelihood function (Equation \ref{eq:likelihood}) from Markov chain for three LGRB redshift distributions (Section \ref{sec:MC}; Equation \ref{eq:zeta}). The fact that the redshift distribution of B10 generally results in larger likelihoods as compared to the two other, can provide evidence -- {\it but does not necessitate in any way} -- that BATSE data favors a LGRB rate tracing cosmic metallicity evolution over SFR. A firm decision can only be made by the complete knowledge of BATSE LGRB redshifts. \label{fig:EffLike}}
        \end{figure*}

        The goodness of the log-normal CDF assumption for BATSE detection efficiency can be checked by a comparison of the resulting model predictions with BATSE $1366$ LGRBs's distribution of peak fluxes (Figure \ref{fig:OFmarginals}). The best-fit model prediction of BATSE trigger efficiency for Long-duration class of gamma-ray bursts is compared to the nominal trigger efficiency of BATSE 4B catalog for the class of soft-long bursts in Figure (\ref{fig:EffLike}, Left Panel). Although the difference between the two curves is significant, it does not necessarily imply a contradiction, given the fact that different methodologies and GRB samples were used to derive the two efficiency curves\footnote{\href{http://www.batse.msfc.nasa.gov/batse/grb/catalog/4b/4br\_efficiency.html}{BATSE 4B Exposure vs Peak Flux Table}}.

    \section{C. Likelihood Function}
	\label{sec:appC}

        To obtain the joint posterior for the unknown parameters of the LGRB world model of Equation (\ref{eq:cosmicrate}) given BATSE data, the likelihood function of the model must be, in principle, constructed by correctly accounting for uncertainties in observational data \citep[e.g.,][]{eddington_formula_1913, jeffreys_correction_1938}. In addition, it is known that astronomical surveys at low Signal-to-Noise Ratios (SNR) close to survey threshold can be potentially biased \citep[e.g.,][]{hogg_maximum_1998}. A Bayesian multilevel methodology \citep[e.g.,][]{hobson_bayesian_2010} can incorporate the above corrections required to construct the likelihood function:
        Under the assumption of normality for the uncertainties of LGRB variables in BATSE catalog, each LGRB event -- denoted by $\vec O_i$, standing for the $i$th LGRB Observation -- has the likelihood $\mathcal{L}_i$ of having the true parameters $\vec O_i\equiv\big[P_{bol,i},S_{bol,i},E_{p,i},T_{90,i}\big]$ that is described by a $4$-dimensional Gaussian Probability Density Function (pdf),

        \begin{equation}
            \mathcal{L}_i(\vec O_i|\vec \mu_{i,0},\Sigma_{i,0})\sim\mathcal{N}(\vec O_i|\vec \mu_{i,0},\Sigma_{i,0}),
        \end{equation}

        where the location ($\mu_{i,0}$) and the scale ($\Sigma_{i,0}$) parameters of the pdf are to be determined by the model and individual events data from BATSE GRB Catalog. Given $\mathcal{L}_i$ and the LGRB world model (Equation \ref{eq:cosmicrate}) convolved with BATSE trigger efficiency (Equation \ref{eq:obsrate}), the full Poisson likelihood function can be written as,

        \begin{eqnarray}
            \label{eq:likelihood}
            {\mathcal L}({\rm Data}|{\rm Model~Parameters})&=&\mathcal{A}^N\exp\bigg(-\mathcal{A}\int_{\vec O space}\mathcal{R}_{obs}(\vec O)~\diff\vec O\bigg) \nonumber \\
            &\times& \prod_{i=1}^{N} \int_{\vec O space} \mathcal{R}_{cosmic}(\vec \mu_{i,0},\Sigma_{i,0},z_i|{\rm Model~Parameters})\mathcal{L}_i(\vec O_i|\vec \mu_{i,0},\Sigma_{i,0})~\diff\vec O,
        \end{eqnarray}

        where $N=1366$, and $\mathcal{A}$ is a factor that properly normalizes the cosmic and the observed rates ($\mathcal{R}_{cosmic}$ \& $\mathcal{R}_{obs}$). The term $\mathcal{R}_{cosmic}$ in Equation \ref{eq:likelihood} acts as a prior for $\mathcal{L}_i$. In the absence of the prior knowledge (i.e., $\mathcal{R}_{cosmic}$, as in the case here), the Empirical Bayes approach can provide an alternative solution, in which, an ad hoc estimate of the model parameters based on the observed data (excluding uncertainties) serves as the prior for same data at the second level of analysis. Calculation of the normalization factor $\mathcal{A}$, involves an integration of the LGRB world model over the 5-dimensional space of LGRB variables and redshift, with a complex integration limit set by BATSE trigger efficiency modeled in Equation \ref{eq:threshmodel}, as a function of $\liso$, $\epkz$ \& $z$. In addition, since almost no redshift information is available for BATSE catalog of GRBs, the probability for each LGRB has to be marginalized over redshift, for which a range of $z\in[0.1,\infty]$ is considered. These integrations make the maximization of the likelihood for all unknown parameters an extremely difficult task, perhaps challenging current computational technologies. Moreover, it has been known that GRB fluences and durations are likely underestimated close to detection threshold due to the so-called fluence-duration bias \citep[e.g.,][]{hakkila_fluence_2000, hakkila_how_2003}. Such bias makes the overestimation-correction of the fluence and duration as prescribed by \citet{hogg_maximum_1998} unjustified before the fluence-duration bias effects are well quantified. The algorithms for calculating peak fluxes, however, appear to result in less biased measurements -- even down to detector threshold -- with negligible uncertainties \citep[e.g.,][]{stern_off-line_2001}. Given the above lines of reasoning and the computational limitations, the uncertainties on three BATSE LGRB variables: $\pbol$, $\sbol$, $\dur$ are excluded from the likelihood function (Equation \ref{eq:likelihood}). The uncertainties on the $\epk$ estimates of BATSE LGRBs are, however, significant compared to the three former variables and must be incorporated in the calculations of the likelihood. Nevertheless, it was realised after likelihood maximization that the exclusion of the $\epk$ uncertainties -- by fixing $\epk$ values to the `bisector' estimates of \citet{shahmoradi_hardness_2010} -- results in only negligible ($\ll 1\sigma$) changes in the model's best-fit parameters. In general, the use of the bisector line of Ordinary Least Squares (OLS) regression lines \citep[e.g.,][]{isobe_linear_1990} for estimation purposes is unfavored due to lack of a maximum likelihood interpretation. The special case of BATSE LGRBs here, however, turns out to be an exception.

        In addition to cosmological time-dilation correction, it is common practice to make an energy correction to the temporal variables of Swift GRBs \citep[e.g.,][]{gehrels_new_2006, butler_cosmic_2010}, such as $\dur$, when translating the variable from the observer-frame to the rest-frame. For BATSE LGRBs, this energy-correction is likely negligible, given the fact that the $\dur$ durations are calculated based on the total photon counts in the BATSE LAD energy range \citep[e.g.,][]{kouveliotou_identification_1993, fishman_first_1994}: $20-2000~keV$ which can be practically considered as bolometric. Nevertheless, it is expected that such an energy correction, if needed, would slightly relax the strong correlation of the rest-frame duration ($\durz$) with the total isotropic emission ($\eiso$) and the peak isotropic luminosity ($\liso$).

        The joint posterior distribution of the model parameters are obtained by the maximizing the likelihood function of Equation (\ref{eq:likelihood}) convolved with non-informative uniform prior on the location parameters and the standard choice of Jeffreys prior on the scale parameters \citep{jeffreys_invariant_1946}. In order to efficiently sample from the 16-dimensional posterior density function, a variant of Markov Chain Monte Carlo (MCMC) methods known as Adaptive Metropolis-Hastings (AMH-MCMC) is employed \citep[e.g.,][]{haario_adaptive_2001}. The choice of an Adaptive (vs. classical) MH algorithm is very important, since the model parameters exhibit strong covariance (Table \ref{tab:correlations}). To reduce the simulation runtime, all algorithms including AMH-MCMC are implemented in Fortran, which is by far the fastest and most efficient programming language for many intensive scientific calculations and number crunching.\footnote{\href{http://portal.acm.org/ft_gateway.cfm?id=1820518\&type=pdf}{The Ideal HPC Programming Language}} In addition, the numerical integration in the definition of the luminosity distance (Equation \ref{eq:lumdis}) -- encountered on the order of $\gtrsim10^9$ times during full MCMC sampling -- is greatly simplified by the analytical approximation method of \citet{wickramasinghe_analytical_2010}. Due to the intrinsic sequential character of MCMC sampling methods, the parallelization of simulation algorithms (on either shared- or distributed- memory architecture machines) is impractical or at best inefficient for a single Markov chain. Nevertheless, to increase the MCMC sample size and more importantly, to ensure convergence to the global (vs. local) extremum, the chain is initiated simultaneously at $10$ random starting points in the parameter space on a $12$-cores desktop CPU. In general, convergence and good mixing occurs within the first few thousands of iterations (burn-in period) given a suitable initial guess for the covariance matrix of the proposal distribution -- here chosen to be multivariate Gaussian. The resulting {\it mean} and $1\sigma$ standard deviations of the model parameters are tabulated in Table (\ref{tab:BFP}).

    \section{D. LGRB Monte Carlo Universe}
	\label{sec:appD}

        The prediction power and consistency of the presented LGRB world model -- based on BATSE data -- can be easily checked against observational data from current and future gamma-ray experiments, in particular Fermi satellite. All it takes is to construct a Monte Carlo universe of LGRBs, based on the best fit parameters of the LGRB world model in Table (\ref{tab:BFP}) and compare the outcome with observational data. Although straightforward, the steps for such simulation and comparison are summarized bellow:

        \begin{enumerate}
            \item
                A random redshift for each simulated LGRB is drawn from the redshift distribution of Equation (\ref{eq:zeta}) with parameters taken from Table (\ref{tab:BFP}). It is recommended to {\it repeatedly} randomly draw the set of model parameters from the full Markov Chain samples\footnote{Available at \url{https://sites.google.com/site/amshportal/research/aca/in-the-news/lgrb-world-model}} instead of fixing the parameters to the mean values reported in Table (\ref{tab:BFP}).

            \item
                The four LGRB variables: $\liso$, $\eiso$, $\epkz$, $\durz$ are randomly drawn from a 4-dimensional log-normal distribution with location ($\vec\mu$) and scale (i.e., the covariance matrix: $\Sigma$) parameters constructed from fitting results in Table (\ref{tab:BFP}).
                This can be easily and quickly done by noting that a multivariate log-normal distribution is equivalent to a multivariate Gaussian distribution in the logarithmic space of the above variables,

                \begin{equation}
                    \log\vec O \equiv \bigg(\log(\liso),\log(\eiso),\log(\epkz),\log(\durz)\bigg),
                \end{equation}

                such that the 4-dimensional log-normal density function , $\mathcal{LN}$, of Equation (\ref{eq:cosmicrate}) can be exactly replaced by a 4-dimensional Gaussian distribution,

                \begin{equation}
                    \label{eq:normaldist}
                    \mathcal{N}\bigg(\log\vec O\big|\vec \mu,\Sigma\bigg) = (2\pi)^{-k/2}|\Sigma|^{\nicefrac{-1}{2}} \times \exp\bigg(-(\log\vec O - \vec\mu)'\Sigma^{-1}(\log\vec O - \vec\mu)/2\bigg),
                \end{equation}

                for which, the cosmic LGRB differential rate ($\mathcal{R}_{cosmic}$) of Equation (\ref{eq:cosmicrate}) will be,

                \begin{equation}
                    \mathcal{R}_{cosmic} = \frac{\diff N}{d\log(\liso)~d\log(\eiso)~d\log(\epkz)~d\log(\durz)~dz},
                \end{equation}

            \item
                The above Monte Carlo universe of LGRBs can be then measured according to the instrument's detection efficiency (Section \ref{sec:MC}: Equation \ref{eq:obsrate}). For the case of BATSE LAD detectors, the trigger efficiency can be modeled according to the prescription in Appendix \ref{sec:appB}.

        \end{enumerate}

\newpage

\acknowledgments

     The author owes special thanks to Robert J. Nemiroff (NASA GSFC Astrophysicist \& professor of Physics at Michigan Tech University) for his generous support and help through numerous discussions over more than a year, from the early stages of developing the basic ideas behind this work to the core of the analysis and final steps of manuscript preparation.\\

    I would like to thank William H. Press (professor of Computer Science and Integrative Biology at The University of Texas at Austin), Lars Koesterke (at Texas Advanced Computing Center) \& Mehdi Mortazavi (at MTU) for helpful comments on some statistical and computational aspects of this work. I am very grateful to Swadesh Mahajan (professor at Institute for Fusion Studies) \& Richard Hazeltine (professor and chair at the Department of Physics, The University of Texas at Austin) for their generous support in the final stages of preparing this manuscript. I would also like to thank Giancarlo Ghirlanda (at INAF-Osservatorio Astronomico di Brera) for his timely feedback and comment on the GRB terminology used in this manuscript. Greatly appreciated were helpful comments and detailed criticisms from the anonymous referee of this manuscript. \\

    This work would have not been accomplished without the vast time and effort spent by many scientists and engineers who designed, built and launched the Compton Gamma-Ray Observatory and were involved in the analysis of GRB data from BATSE Large Area Detectors.

\newpage

\begin{center}
\begin{scriptsize}
\begin{longtable}{cccccccccccccc}
\caption[]{1366 BATSE catalog triggers classified as LGRBs} \label{tab:data} \\
\hline \hline \\[2ex]
Trigger & Trigger & Trigger & Trigger & Trigger & Trigger & Trigger & Trigger & Trigger & Trigger & Trigger & Trigger & Trigger & Trigger \\[0.5ex]
\hline \\[1.8ex]
\endfirsthead
\multicolumn{14}{c}{{\tablename} \thetable{} -- Continued} \\[0.5ex]
  \hline \hline \\[2ex]
Trigger & Trigger & Trigger & Trigger & Trigger & Trigger & Trigger & Trigger & Trigger & Trigger & Trigger & Trigger & Trigger & Trigger\\[0.5ex]
\hline \\[1.8ex]
\endhead
\\[1.8ex] \hline \\[0.5ex]
\multicolumn{14}{r}{{Continued on Next Page \ldots}} \\[0.5ex]
\endfoot
\\[1.8ex] \hline \hline
\endlastfoot
105	&	107	&	109	&	110	&	111	&	114	&	121	&	130	&	133	&	143	&	148	&	160	&	171	&	179	\\ [0.5ex]
204	&	211	&	214	&	219	&	222	&	223	&	226	&	228	&	235	&	237	&	249	&	257	&	288	&	332	\\ [0.5ex]
351	&	394	&	398	&	401	&	404	&	408	&	414	&	451	&	465	&	467	&	469	&	472	&	473	&	493	\\ [0.5ex]
501	&	516	&	526	&	540	&	543	&	548	&	549	&	559	&	563	&	577	&	591	&	594	&	606	&	630	\\ [0.5ex]
647	&	658	&	659	&	660	&	673	&	676	&	678	&	680	&	685	&	686	&	690	&	692	&	704	&	717	\\ [0.5ex]
741	&	752	&	753	&	755	&	761	&	764	&	773	&	795	&	803	&	815	&	816	&	820	&	824	&	825	\\ [0.5ex]
829	&	840	&	841	&	869	&	907	&	914	&	927	&	938	&	946	&	973	&	999	&	1009	&	1036	&	1039	\\ [0.5ex]
1042	&	1046	&	1085	&	1086	&	1087	&	1114	&	1120	&	1122	&	1123	&	1125	&	1126	&	1141	&	1145	&	1148	\\ [0.5ex]
1150	&	1152	&	1153	&	1156	&	1157	&	1159	&	1167	&	1190	&	1192	&	1196	&	1197	&	1200	&	1204	&	1213	\\ [0.5ex]
1218	&	1221	&	1235	&	1244	&	1279	&	1288	&	1291	&	1298	&	1303	&	1306	&	1318	&	1382	&	1384	&	1385	\\ [0.5ex]
1390	&	1396	&	1406	&	1416	&	1419	&	1425	&	1432	&	1439	&	1440	&	1446	&	1447	&	1449	&	1452	&	1456	\\ [0.5ex]
1458	&	1467	&	1468	&	1472	&	1492	&	1515	&	1533	&	1540	&	1541	&	1551	&	1552	&	1558	&	1559	&	1561	\\ [0.5ex]
1567	&	1574	&	1578	&	1579	&	1580	&	1586	&	1590	&	1601	&	1604	&	1606	&	1609	&	1611	&	1614	&	1623	\\ [0.5ex]
1625	&	1626	&	1628	&	1642	&	1646	&	1651	&	1652	&	1653	&	1655	&	1656	&	1657	&	1660	&	1661	&	1663	\\ [0.5ex]
1664	&	1667	&	1676	&	1687	&	1693	&	1700	&	1701	&	1704	&	1711	&	1712	&	1714	&	1717	&	1730	&	1731	\\ [0.5ex]
1733	&	1734	&	1740	&	1742	&	1806	&	1807	&	1815	&	1819	&	1830	&	1883	&	1885	&	1886	&	1922	&	1924	\\ [0.5ex]
1956	&	1967	&	1974	&	1982	&	1989	&	1993	&	1997	&	2018	&	2019	&	2035	&	2047	&	2053	&	2061	&	2067	\\ [0.5ex]
2069	&	2070	&	2074	&	2077	&	2079	&	2080	&	2081	&	2083	&	2087	&	2090	&	2093	&	2101	&	2102	&	2105	\\ [0.5ex]
2106	&	2110	&	2111	&	2112	&	2114	&	2119	&	2122	&	2123	&	2129	&	2133	&	2138	&	2140	&	2143	&	2148	\\ [0.5ex]
2149	&	2151	&	2152	&	2156	&	2181	&	2187	&	2188	&	2189	&	2190	&	2191	&	2193	&	2197	&	2202	&	2203	\\ [0.5ex]
2204	&	2207	&	2211	&	2213	&	2219	&	2228	&	2230	&	2232	&	2233	&	2240	&	2244	&	2252	&	2253	&	2254	\\ [0.5ex]
2267	&	2276	&	2277	&	2287	&	2298	&	2304	&	2306	&	2309	&	2310	&	2311	&	2315	&	2316	&	2321	&	2324	\\ [0.5ex]
2325	&	2328	&	2329	&	2340	&	2344	&	2345	&	2346	&	2347	&	2349	&	2362	&	2367	&	2371	&	2373	&	2375	\\ [0.5ex]
2380	&	2381	&	2383	&	2385	&	2387	&	2391	&	2392	&	2393	&	2394	&	2405	&	2419	&	2423	&	2428	&	2429	\\ [0.5ex]
2430	&	2432	&	2435	&	2436	&	2437	&	2438	&	2440	&	2441	&	2442	&	2443	&	2446	&	2447	&	2450	&	2451	\\ [0.5ex]
2452	&	2453	&	2458	&	2460	&	2472	&	2476	&	2477	&	2482	&	2484	&	2495	&	2496	&	2500	&	2505	&	2508	\\ [0.5ex]
2510	&	2511	&	2515	&	2519	&	2522	&	2528	&	2530	&	2533	&	2537	&	2541	&	2551	&	2560	&	2569	&	2570	\\ [0.5ex]
2581	&	2586	&	2589	&	2593	&	2600	&	2603	&	2606	&	2608	&	2610	&	2611	&	2619	&	2620	&	2628	&	2634	\\ [0.5ex]
2636	&	2640	&	2641	&	2660	&	2662	&	2663	&	2664	&	2665	&	2671	&	2677	&	2681	&	2688	&	2691	&	2695	\\ [0.5ex]
2696	&	2697	&	2700	&	2703	&	2706	&	2709	&	2711	&	2719	&	2725	&	2727	&	2736	&	2749	&	2750	&	2751	\\ [0.5ex]
2753	&	2767	&	2770	&	2774	&	2775	&	2780	&	2790	&	2793	&	2797	&	2798	&	2812	&	2815	&	2825	&	2830	\\ [0.5ex]
2831	&	2843	&	2848	&	2850	&	2852	&	2853	&	2855	&	2856	&	2857	&	2862	&	2863	&	2864	&	2877	&	2880	\\ [0.5ex]
2889	&	2890	&	2891	&	2897	&	2898	&	2900	&	2901	&	2913	&	2916	&	2917	&	2919	&	2922	&	2924	&	2925	\\ [0.5ex]
2927	&	2929	&	2931	&	2932	&	2944	&	2945	&	2947	&	2948	&	2950	&	2951	&	2953	&	2958	&	2961	&	2980	\\ [0.5ex]
2984	&	2985	&	2986	&	2990	&	2992	&	2993	&	2994	&	2996	&	2998	&	3001	&	3003	&	3005	&	3011	&	3012	\\ [0.5ex]
3015	&	3017	&	3026	&	3028	&	3029	&	3032	&	3035	&	3040	&	3042	&	3055	&	3056	&	3057	&	3067	&	3068	\\ [0.5ex]
3070	&	3071	&	3072	&	3074	&	3075	&	3076	&	3080	&	3084	&	3085	&	3088	&	3091	&	3093	&	3096	&	3100	\\ [0.5ex]
3101	&	3102	&	3103	&	3105	&	3109	&	3110	&	3115	&	3119	&	3120	&	3127	&	3128	&	3129	&	3130	&	3131	\\ [0.5ex]
3132	&	3134	&	3135	&	3136	&	3138	&	3139	&	3141	&	3142	&	3143	&	3153	&	3156	&	3159	&	3166	&	3167	\\ [0.5ex]
3168	&	3171	&	3174	&	3177	&	3178	&	3193	&	3212	&	3217	&	3220	&	3227	&	3229	&	3237	&	3238	&	3241	\\ [0.5ex]
3242	&	3245	&	3246	&	3247	&	3255	&	3256	&	3257	&	3259	&	3267	&	3269	&	3276	&	3279	&	3283	&	3284	\\ [0.5ex]
3287	&	3290	&	3292	&	3301	&	3306	&	3307	&	3319	&	3320	&	3321	&	3322	&	3324	&	3330	&	3336	&	3339	\\ [0.5ex]
3345	&	3347	&	3350	&	3351	&	3352	&	3356	&	3358	&	3364	&	3369	&	3370	&	3378	&	3403	&	3405	&	3407	\\ [0.5ex]
3408	&	3415	&	3416	&	3436	&	3439	&	3448	&	3458	&	3465	&	3471	&	3472	&	3480	&	3481	&	3485	&	3486	\\ [0.5ex]
3488	&	3489	&	3491	&	3493	&	3503	&	3505	&	3509	&	3511	&	3512	&	3514	&	3515	&	3516	&	3523	&	3527	\\ [0.5ex]
3528	&	3552	&	3567	&	3569	&	3588	&	3593	&	3598	&	3608	&	3618	&	3634	&	3637	&	3648	&	3649	&	3654	\\ [0.5ex]
3655	&	3658	&	3662	&	3663	&	3664	&	3671	&	3717	&	3733	&	3740	&	3745	&	3765	&	3766	&	3768	&	3771	\\ [0.5ex]
3773	&	3776	&	3779	&	3788	&	3792	&	3800	&	3801	&	3805	&	3807	&	3811	&	3814	&	3815	&	3819	&	3840	\\ [0.5ex]
3843	&	3853	&	3860	&	3869	&	3870	&	3871	&	3875	&	3879	&	3886	&	3890	&	3891	&	3892	&	3893	&	3899	\\ [0.5ex]
3900	&	3901	&	3903	&	3905	&	3906	&	3908	&	3909	&	3912	&	3913	&	3914	&	3916	&	3917	&	3918	&	3924	\\ [0.5ex]
3926	&	3929	&	3930	&	3935	&	3941	&	3954	&	4039	&	4048	&	4095	&	4146	&	4157	&	4216	&	4251	&	4312	\\ [0.5ex]
4350	&	4368	&	4388	&	4556	&	4569	&	4653	&	4701	&	4710	&	4745	&	4814	&	4939	&	4959	&	5080	&	5255	\\ [0.5ex]
5304	&	5305	&	5379	&	5387	&	5389	&	5407	&	5409	&	5411	&	5412	&	5415	&	5416	&	5417	&	5419	&	5420	\\ [0.5ex]
5421	&	5423	&	5428	&	5429	&	5433	&	5434	&	5447	&	5450	&	5451	&	5454	&	5463	&	5464	&	5465	&	5466	\\ [0.5ex]
5470	&	5472	&	5473	&	5474	&	5475	&	5476	&	5477	&	5478	&	5479	&	5480	&	5482	&	5483	&	5484	&	5486	\\ [0.5ex]
5487	&	5489	&	5490	&	5492	&	5493	&	5494	&	5495	&	5497	&	5503	&	5504	&	5507	&	5508	&	5510	&	5512	\\ [0.5ex]
5513	&	5515	&	5516	&	5517	&	5518	&	5523	&	5524	&	5526	&	5530	&	5531	&	5538	&	5539	&	5540	&	5541	\\ [0.5ex]
5542	&	5545	&	5548	&	5551	&	5554	&	5555	&	5559	&	5563	&	5565	&	5566	&	5567	&	5569	&	5571	&	5572	\\ [0.5ex]
5573	&	5574	&	5575	&	5581	&	5585	&	5589	&	5590	&	5591	&	5593	&	5594	&	5597	&	5601	&	5603	&	5604	\\ [0.5ex]
5605	&	5606	&	5608	&	5610	&	5612	&	5614	&	5615	&	5617	&	5618	&	5621	&	5622	&	5624	&	5626	&	5627	\\ [0.5ex]
5628	&	5632	&	5635	&	5637	&	5640	&	5644	&	5645	&	5646	&	5648	&	5654	&	5655	&	5667	&	5697	&	5704	\\ [0.5ex]
5706	&	5713	&	5715	&	5716	&	5718	&	5719	&	5721	&	5723	&	5725	&	5726	&	5729	&	5731	&	5736	&	5773	\\ [0.5ex]
5867	&	5890	&	5955	&	5983	&	5989	&	5995	&	6004	&	6082	&	6083	&	6090	&	6098	&	6100	&	6101	&	6102	\\ [0.5ex]
6103	&	6104	&	6111	&	6113	&	6115	&	6118	&	6119	&	6124	&	6127	&	6128	&	6131	&	6137	&	6139	&	6141	\\ [0.5ex]
6147	&	6151	&	6152	&	6154	&	6158	&	6159	&	6165	&	6167	&	6168	&	6176	&	6186	&	6188	&	6189	&	6190	\\ [0.5ex]
6194	&	6198	&	6206	&	6222	&	6223	&	6225	&	6226	&	6227	&	6228	&	6233	&	6234	&	6241	&	6242	&	6243	\\ [0.5ex]
6244	&	6249	&	6266	&	6267	&	6269	&	6270	&	6271	&	6272	&	6273	&	6274	&	6279	&	6280	&	6283	&	6285	\\ [0.5ex]
6288	&	6295	&	6298	&	6300	&	6303	&	6304	&	6305	&	6306	&	6308	&	6309	&	6315	&	6317	&	6319	&	6320	\\ [0.5ex]
6321	&	6322	&	6323	&	6328	&	6329	&	6330	&	6334	&	6335	&	6337	&	6339	&	6344	&	6345	&	6346	&	6349	\\ [0.5ex]
6351	&	6353	&	6355	&	6369	&	6370	&	6375	&	6380	&	6388	&	6390	&	6395	&	6396	&	6397	&	6399	&	6400	\\ [0.5ex]
6404	&	6405	&	6408	&	6409	&	6413	&	6414	&	6419	&	6422	&	6425	&	6435	&	6437	&	6440	&	6444	&	6446	\\ [0.5ex]
6448	&	6450	&	6451	&	6453	&	6454	&	6472	&	6487	&	6489	&	6490	&	6498	&	6504	&	6519	&	6520	&	6521	\\ [0.5ex]
6522	&	6523	&	6525	&	6528	&	6529	&	6531	&	6533	&	6534	&	6536	&	6538	&	6539	&	6544	&	6546	&	6550	\\ [0.5ex]
6551	&	6552	&	6554	&	6557	&	6560	&	6564	&	6566	&	6576	&	6577	&	6578	&	6582	&	6583	&	6585	&	6587	\\ [0.5ex]
6589	&	6590	&	6592	&	6593	&	6598	&	6600	&	6601	&	6602	&	6605	&	6610	&	6611	&	6613	&	6615	&	6616	\\ [0.5ex]
6619	&	6620	&	6621	&	6622	&	6625	&	6629	&	6630	&	6631	&	6632	&	6642	&	6648	&	6649	&	6655	&	6657	\\ [0.5ex]
6658	&	6665	&	6666	&	6670	&	6672	&	6673	&	6674	&	6676	&	6678	&	6683	&	6686	&	6694	&	6695	&	6698	\\ [0.5ex]
6702	&	6707	&	6708	&	6720	&	6745	&	6762	&	6763	&	6764	&	6767	&	6774	&	6782	&	6796	&	6802	&	6814	\\ [0.5ex]
6816	&	6830	&	6831	&	6853	&	6877	&	6880	&	6882	&	6884	&	6891	&	6892	&	6903	&	6911	&	6914	&	6917	\\ [0.5ex]
6930	&	6935	&	6938	&	6963	&	6987	&	6989	&	7000	&	7012	&	7028	&	7030	&	7064	&	7087	&	7108	&	7110	\\ [0.5ex]
7113	&	7116	&	7130	&	7147	&	7164	&	7167	&	7170	&	7172	&	7178	&	7183	&	7185	&	7191	&	7206	&	7207	\\ [0.5ex]
7209	&	7213	&	7219	&	7228	&	7230	&	7247	&	7250	&	7255	&	7263	&	7285	&	7293	&	7295	&	7298	&	7301	\\ [0.5ex]
7310	&	7318	&	7319	&	7322	&	7323	&	7328	&	7335	&	7343	&	7357	&	7358	&	7360	&	7369	&	7371	&	7374	\\ [0.5ex]
7376	&	7377	&	7379	&	7381	&	7386	&	7387	&	7390	&	7403	&	7404	&	7429	&	7432	&	7433	&	7446	&	7451	\\ [0.5ex]
7452	&	7457	&	7460	&	7464	&	7469	&	7475	&	7477	&	7481	&	7485	&	7486	&	7487	&	7488	&	7491	&	7493	\\ [0.5ex]
7494	&	7497	&	7500	&	7502	&	7503	&	7504	&	7509	&	7515	&	7517	&	7518	&	7520	&	7523	&	7527	&	7528	\\ [0.5ex]
7529	&	7532	&	7533	&	7535	&	7548	&	7549	&	7550	&	7551	&	7552	&	7560	&	7563	&	7564	&	7566	&	7567	\\ [0.5ex]
7568	&	7573	&	7575	&	7576	&	7579	&	7580	&	7587	&	7588	&	7597	&	7598	&	7603	&	7604	&	7605	&	7606	\\ [0.5ex]
7607	&	7608	&	7609	&	7614	&	7615	&	7617	&	7619	&	7625	&	7630	&	7635	&	7638	&	7642	&	7645	&	7648	\\ [0.5ex]
7654	&	7656	&	7657	&	7660	&	7662	&	7677	&	7678	&	7683	&	7684	&	7688	&	7695	&	7701	&	7703	&	7705	\\ [0.5ex]
7707	&	7711	&	7727	&	7729	&	7741	&	7744	&	7749	&	7750	&	7752	&	7762	&	7766	&	7769	&	7770	&	7780	\\ [0.5ex]
7781	&	7785	&	7786	&	7788	&	7790	&	7794	&	7795	&	7798	&	7802	&	7803	&	7810	&	7818	&	7822	&	7825	\\ [0.5ex]
7831	&	7835	&	7838	&	7840	&	7841	&	7843	&	7845	&	7858	&	7862	&	7868	&	7872	&	7884	&	7885	&	7886	\\ [0.5ex]
7888	&	7900	&	7902	&	7903	&	7906	&	7918	&	7923	&	7924	&	7929	&	7932	&	7934	&	7936	&	7938	&	7942	\\ [0.5ex]
7948	&	7954	&	7963	&	7968	&	7969	&	7973	&	7976	&	7984	&	7987	&	7989	&	7992	&	7994	&	7997	&	7998	\\ [0.5ex]
8001	&	8004	&	8008	&	8009	&	8012	&	8019	&	8022	&	8026	&	8030	&	8036	&	8039	&	8045	&	8049	&	8050	\\ [0.5ex]
8054	&	8059	&	8061	&	8062	&	8063	&	8064	&	8066	&	8073	&	8075	&	8084	&	8086	&	8087	&	8098	&	8099	\\ [0.5ex]
8101	&	8102	&	8105	&	8110	&	8111	&	8112	&	8116	&	8121	&	---	&	---	&	---	&	---	&	---	&	---	\\ [0.5ex]
\end{longtable}
{Notes: Temporal \& Spectral data for these triggers are available in \href{http://www.batse.msfc.nasa.gov/batse/grb/catalog/}{BATSE 4B \& Current Catalogs}. The spectral peak energy ($\epk$) estimates of the above triggers and the rest of 2130 BATSE Catalog GRBs are provided by \citet{shahmoradi_hardness_2010}. The full conditional $\epk$ probability density functions are available for download at \url{https://sites.google.com/site/amshportal/research/aca/in-the-news/lgrb-world-model}.}
\end{scriptsize}
\end{center}

\begin{table}[htbp]
\begin{center}
\caption{Correlation Matrix of the parameters of the LGRB World Model, \\
 for the median case of an LGRB cosmic rate tracing Star Formation Rate of Li (2008) \label{tab:correlations}}
\begin{tabular}{|c|c|c|c|c|c|c|c|c|c|c|c|c|c|c|c|c|}
\hline
\rotatebox{55}{Parameter} & \begin{sideways}$\log(\liso)$\end{sideways} & \begin{sideways}$\log(\eiso)$\end{sideways} & \begin{sideways}$\log(\epkz)$\end{sideways} & \begin{sideways}$\log(\durz)$\end{sideways} & \begin{sideways}$\log(\sigma_{\liso})$\end{sideways} & \begin{sideways}$\log(\sigma_{\eiso})$\end{sideways} & \begin{sideways}$\log(\sigma_{\epkz})$\end{sideways} & \begin{sideways}$\log(\sigma_{\durz})$\end{sideways} & \begin{sideways}$\rho_{\liso-\eiso}$\end{sideways} & \begin{sideways}$\rho_{\liso-\epkz}$\end{sideways} & \begin{sideways}$\rho_{\liso-\durz}$\end{sideways} & \begin{sideways}$\rho_{\eiso-\epkz}$\end{sideways} & \begin{sideways}$\rho_{\eiso-\durz}$\end{sideways} & \begin{sideways}$\rho_{\epkz-\durz}$\end{sideways}  & \begin{sideways}$\mu_{thresh}$\end{sideways} & \begin{sideways}$\log(\sigma_{thresh})$\end{sideways} \\
\hline
$\log(\liso)$          & 1.00 &	0.99 &	0.90 &	0.34 &	-0.91 &	-0.86 &	-0.59 &	-0.14 &	-0.10 &	-0.45 &	0.51 &	-0.52 &	0.45 &	0.05 &	-0.68 &	-0.44 \\
\hline
$\log(\eiso)$          & & 1.00 & 0.92 & 0.42 & -0.91 & -0.90 & -0.62 & -0.15 & -0.20 & -0.54 & 0.45 & -0.56 & 0.40 & 0.00 & -0.67 & -0.43 \\
\hline
$\log(\epkz)$          & & & 1.00 & 0.38 & -0.82 & -0.84 & -0.79 & -0.15 & -0.26 & -0.77 & 0.40 & -0.77 & 0.38 & 0.07 & -0.61 & -0.40 \\
\hline
$\log(\durz)$          & & & & 1.00 & -0.37 & -0.52 & -0.32 & -0.16 & -0.52 & -0.32 & -0.48 & -0.33 & -0.50 & -0.56 & -0.17 & -0.09 \\
\hline
$\log(\sigma_{\liso})$ & & & & & 1.00 & 0.94 & 0.59 & 0.12 & 0.14 & 0.53 & -0.53 & 0.57 & -0.46 & 0.03 & 0.50 & 0.30 \\
\hline
$\log(\sigma_{\eiso})$ & & & & & & 1.00 & 0.67 & 0.17 & 0.41 & 0.63 & -0.33 & 0.66 & -0.32 & 0.14 & 0.46 & 0.28 \\
\hline
$\log(\sigma_{\epkz})$ & & & & & & & 1.00 & 0.12 & 0.36 & 0.84 & -0.22 & 0.84 & -0.25 & -0.06 & 0.37 & 0.24 \\
\hline
$\log(\sigma_{\durz})$ & & & & & & & & 1.00 & -0.03 & 0.10 & -0.09 & 0.11 & 0.00 & 0.03 & 0.11 & 0.07 \\
\hline
$\rho_{\liso-\eiso}$  & & & & & & & & & 1.00 & 0.47 & 0.43 & 0.41 & 0.22 & 0.26 & -0.01 & -0.01 \\
\hline
$\rho_{\liso-\epkz}$  & & & & & & & & & & 1.00 & -0.15 & 0.97 & -0.21 & -0.05 & 0.29 & 0.19 \\
\hline
$\rho_{\liso-\durz}$  & & & & & & & & & & & 1.00 & -0.18 & 0.95 & 0.57 & -0.32 & -0.20 \\
\hline
$\rho_{\eiso-\epkz}$  & & & & & & & & & & & & 1.00 & -0.22 & 0.06 & 0.29 & 0.18 \\
\hline
$\rho_{\eiso-\durz}$  & & & & & & & & & & & & & 1.00 & 0.64 & -0.26 & -0.17 \\
\hline
$\rho_{\epkz-\durz}$  & & & & & & & & & & & & & & 1.00 & -0.09 & -0.07 \\
\hline
$\mu_{thresh}$        & & & & & & & & & & & & & & & 1.00 & 0.79 \\
\hline
$\log(\sigma_{thresh})$     & & & & & & & & & & & & & & & & 1.00 \\
\hline
\end{tabular}
\end{center}
\end{table}

\newpage

\bibliographystyle{apj}
\bibliography{LGRB_world_model_references}

\begin{thebibliography}{122}
\expandafter\ifx\csname natexlab\endcsname\relax\def\natexlab#1{#1}\fi

\bibitem[{Amati {et~al.}(2008)Amati, Guidorzi, Frontera, Della~Valle, Finelli,
  Landi, \& Montanari}]{amati_measuring_2008}
Amati, L., Guidorzi, C., Frontera, F., {et~al.} 2008, Monthly Notices of the
  Royal Astronomical Society, 391, 577

\bibitem[{Amati {et~al.}(2002)Amati, Frontera, Tavani, in't Zand, Antonelli,
  Costa, Feroci, Guidorzi, Heise, Masetti, Montanari, Nicastro, Palazzi, Pian,
  Piro, \& Soffitta}]{amati_intrinsic_2002}
Amati, L., Frontera, F., Tavani, M., {et~al.} 2002, Astronomy and Astrophysics,
  390, 81

\bibitem[{Atteia {et~al.}(1987)Atteia, Barat, Hurley, Niel, Vedrenne, Evans,
  Fenimore, Klebesadel, Laros, Cline, Desai, Teegarden, Estulin, Zenchenko,
  Kusnetsov, \& Kurt}]{atteia_second_1987}
Atteia, J.-L., Barat, C., Hurley, K., {et~al.} 1987, The Astrophysical Journal
  Supplement Series, 64, 305

\bibitem[{Azzalini(1985)}]{azzalini_class_1985}
Azzalini, A. 1985, Scandinavian Journal of Statistics, 12, 171, {ArticleType:}
  research-article / Full publication date: 1985 / Copyright © 1985 Board of
  the Foundation of the Scandinavian Journal of Statistics

\bibitem[{Balazs {et~al.}(2003)Balazs, Bagoly, Horváth, Mészáros, \&
  Mészáros}]{balazs_difference_2003}
Balazs, L.~G., Bagoly, Z., Horváth, I., Mészáros, A., \& Mészáros, P.
  2003, Astronomy and Astrophysics, 401, 129

\bibitem[{Band {et~al.}(1993)Band, Matteson, Ford, Schaefer, Palmer, Teegarden,
  Cline, Briggs, Paciesas, Pendleton, Fishman, Kouveliotou, Meegan, Wilson, \&
  Lestrade}]{band_batse_1993}
Band, D., Matteson, J., Ford, L., {et~al.} 1993, The Astrophysical Journal,
  413, 281

\bibitem[{Band(2001)}]{band_energy_2001}
Band, D.~L. 2001, The Astrophysical Journal, 563, 582

\bibitem[{Band(2003)}]{band_comparison_2003}
---. 2003, The Astrophysical Journal, 588, 945

\bibitem[{Band(2006)}]{band_postlaunch_2006}
---. 2006, The Astrophysical Journal, 644, 378

\bibitem[{Band \& Preece(2005)}]{band_testing_2005}
Band, D.~L., \& Preece, R.~D. 2005, The Astrophysical Journal, 627, 319

\bibitem[{Band {et~al.}(2008)Band, Grindlay, Hong, Fishman, Hartmann, Garson,
  Krawczynski, Barthelmy, Gehrels, \& Skinner}]{band_exists_2008}
Band, D.~L., Grindlay, J.~E., Hong, J., {et~al.} 2008, The Astrophysical
  Journal, 673, 1225

\bibitem[{Bezdek(1981)}]{bezdek_pattern_1981}
Bezdek, J.~C. 1981, Pattern Recognition with Fuzzy Objective Function
  Algorithms (Norwell, {MA}, {USA}: Kluwer Academic Publishers)

\bibitem[{Bloom {et~al.}(2008)Bloom, Butler, \& Perley}]{bloom_gamma-ray_2008}
Bloom, J.~S., Butler, N.~R., \& Perley, D.~A. 2008, in , 11--15

\bibitem[{Boella {et~al.}(1997)Boella, Butler, Perola, Piro, Scarsi, \&
  Bleeker}]{boella_bepposax_1997}
Boella, G., Butler, R.~C., Perola, G.~C., {et~al.} 1997, Astronomy and
  Astrophysics Supplement Series, 122, 299

\bibitem[{Brainerd(1997)}]{Brainerd_cosmological_1997}
Brainerd, J.~J. 1997, The Astrophysical Journal, 487, 96

\bibitem[{Briggs(1993)}]{briggs_dipole_1993}
Briggs, M.~S. 1993, The Astrophysical Journal, 407, 126

\bibitem[{Bromberg {et~al.}(2012)Bromberg, Nakar, Piran, \&
  Sari}]{bromberg_observational_2012}
Bromberg, O., Nakar, E., Piran, T., \& Sari, R. 2012, The Astrophysical
  Journal, 749, 110

\bibitem[{Burrows {et~al.}(2005)Burrows, Hill, Nousek, Kennea, Wells, Osborne,
  Abbey, Beardmore, Mukerjee, Short, Chincarini, Campana, Citterio, Moretti,
  Pagani, Tagliaferri, Giommi, Capalbi, Tamburelli, Angelini, Cusumano,
  Bräuninger, Burkert, \& Hartner}]{burrows_swift_2005}
Burrows, D.~N., Hill, J.~E., Nousek, J.~A., {et~al.} 2005, Space Science
  Reviews, 120, 165

\bibitem[{Butler {et~al.}(2010)Butler, Bloom, \&
  Poznanski}]{butler_cosmic_2010}
Butler, N.~R., Bloom, J.~S., \& Poznanski, D. 2010, The Astrophysical Journal,
  711, 495

\bibitem[{Butler {et~al.}(2009)Butler, Kocevski, \&
  Bloom}]{butler_generalized_2009}
Butler, N.~R., Kocevski, D., \& Bloom, J.~S. 2009, The Astrophysical Journal,
  694, 76

\bibitem[{Butler {et~al.}(2007)Butler, Kocevski, Bloom, \&
  Curtis}]{butler_complete_2007}
Butler, N.~R., Kocevski, D., Bloom, J.~S., \& Curtis, J.~L. 2007, The
  Astrophysical Journal, 671, 656

\bibitem[{Campisi {et~al.}(2010)Campisi, Li, \&
  Jakobsson}]{campisi_redshift_2010}
Campisi, M.~A., Li, L.-X., \& Jakobsson, P. 2010, Monthly Notices of the Royal
  Astronomical Society, 407, 1972

\bibitem[{Cochran(1954)}]{cochran_methods_1954}
Cochran, W.~G. 1954, Biometrics, 10, 417, {ArticleType:} research-article /
  Full publication date: Dec., 1954 / Copyright © 1954 International Biometric
  Society

\bibitem[{Dado \& Dar(2012)}]{dado_kinematic_2012}
Dado, S., \& Dar, A. 2012, The Astrophysical Journal, 749, 100

\bibitem[{Dermer(1992)}]{dermer_statistics_1992}
Dermer, C.~D. 1992, Physical Review Letters, 68, 1799

\bibitem[{Dezalay {et~al.}(1997)Dezalay, Atteia, Barat, Boer, Darracq, Goupil,
  Niel, Talon, Vedrenne, Hurley, Terekhov, Sunyaev, \&
  Kuznetsov}]{Dezalay_hardness-intensity_1997}
Dezalay, J.-P., Atteia, J.-L., Barat, C., {et~al.} 1997, The Astrophysical
  Journal, 490, L17

\bibitem[{Dunn(1973)}]{dunn_fuzzy_1973}
Dunn, J.~C. 1973, Journal of Cybernetics, 3, 32

\bibitem[{Eddington(1913)}]{eddington_formula_1913}
Eddington, A.~S. 1913, Monthly Notices of the Royal Astronomical Society, 73,
  359

\bibitem[{Fenimore {et~al.}(1995)Fenimore, in~'t Zand, Norris, Bonnell, \&
  Nemiroff}]{fenimore_gamma-ray_1995}
Fenimore, E.~E., in~'t Zand, J. J.~M., Norris, J.~P., Bonnell, J.~T., \&
  Nemiroff, R.~J. 1995, The Astrophysical Journal Letters, 448, L101

\bibitem[{Fenimore {et~al.}(1988)Fenimore, Conner, Epstein, Klebesadel, Laros,
  Yoshida, Fujii, Hayashida, Itoh, Murakami, Nishimura, Yamagami, Kondo, \&
  Kawai}]{fenimore_interpretations_1988}
Fenimore, E.~E., Conner, J.~P., Epstein, R.~I., {et~al.} 1988, The
  Astrophysical Journal Letters, 335, L71

\bibitem[{Fenimore {et~al.}(1993)Fenimore, Epstein, Ho, Klebesadel, Lacey,
  Laros, Meier, Strohmayer, Pendleton, Fishman, Kouveliotou, \&
  Meegan}]{fenimore_intrinsic_1993}
Fenimore, E.~E., Epstein, R.~I., Ho, C., {et~al.} 1993, , Published online: 04
  November 1993; {\textbar} doi:10.1038/366040a0, 366, 40

\bibitem[{Fisher(1924)}]{fisher_conditions_1924}
Fisher, R.~A. 1924, Journal of the Royal Statistical Society, 87, 442,
  {ArticleType:} misc / Full publication date: May, 1924 / Copyright © 1924
  Royal Statistical Society

\bibitem[{Fishman {et~al.}(1994)Fishman, Meegan, Wilson, Brock, Horack,
  Kouveliotou, Howard, Paciesas, Briggs, Pendleton, Koshut, Mallozzi,
  Stollberg, \& Lestrade}]{fishman_first_1994}
Fishman, G.~J., Meegan, C.~A., Wilson, R.~B., {et~al.} 1994, The Astrophysical
  Journal Supplement Series, 92, 229

\bibitem[{Gehrels {et~al.}(2009)Gehrels, Ramirez-Ruiz, \&
  Fox}]{gehrels_gamma-ray_2009}
Gehrels, N., Ramirez-Ruiz, E., \& Fox, D.~B. 2009, Annual Review of Astronomy
  and Astrophysics, 47, 567

\bibitem[{Gehrels {et~al.}(2004)Gehrels, Chincarini, Giommi, Mason, Nousek,
  Wells, White, Barthelmy, Burrows, Cominsky, Hurley, Marshall, Meszaros,
  Roming, Angelini, Barbier, Belloni, Campana, Caraveo, Chester, Citterio,
  Cline, Cropper, Cummings, Dean, Feigelson, Fenimore, Frail, Fruchter,
  Garmire, Gendreau, Ghisellini, Greiner, Hill, Hunsberger, Krimm, Kulkarni,
  Kumar, Lebrun, {Lloyd‐Ronning}, Markwardt, Mattson, Mushotzky, Norris,
  Osborne, Paczynski, Palmer, Park, Parsons, Paul, Rees, Reynolds, Rhoads,
  Sasseen, Schaefer, Short, Smale, Smith, Stella, Tagliaferri, Takahashi,
  Tashiro, Townsley, Tueller, Turner, Vietri, Voges, Ward, Willingale, Zerbi,
  \& Zhang}]{gehrels_swift_2004}
Gehrels, N., Chincarini, G., Giommi, P., {et~al.} 2004, The Astrophysical
  Journal, 611, 1005

\bibitem[{Gehrels {et~al.}(2006)Gehrels, Norris, Barthelmy, Granot, Kaneko,
  Kouveliotou, Markwardt, Mészáros, Nakar, Nousek, {O'Brien}, Page, Palmer,
  Parsons, Roming, Sakamoto, Sarazin, Schady, Stamatikos, \&
  Woosley}]{gehrels_new_2006}
Gehrels, N., Norris, J.~P., Barthelmy, S.~D., {et~al.} 2006, Nature, 444, 1044

\bibitem[{Ghirlanda {et~al.}(2005)Ghirlanda, Ghisellini, Firmani, Celotti, \&
  Bosnjak}]{ghirlanda_peak_2005}
Ghirlanda, G., Ghisellini, G., Firmani, C., Celotti, A., \& Bosnjak, Z. 2005,
  Monthly Notices of the Royal Astronomical Society, 360, L45

\bibitem[{Ghirlanda {et~al.}(2004)Ghirlanda, Ghisellini, \&
  Lazzati}]{ghirlanda_collimation-corrected_2004}
Ghirlanda, G., Ghisellini, G., \& Lazzati, D. 2004, The Astrophysical Journal,
  616, 331

\bibitem[{Ghirlanda {et~al.}(2008)Ghirlanda, Nava, Ghisellini, Firmani, \&
  Cabrera}]{ghirlanda_e_2008}
Ghirlanda, G., Nava, L., Ghisellini, G., Firmani, C., \& Cabrera, J.~I. 2008,
  Monthly Notices of the Royal Astronomical Society, 387, 319

\bibitem[{Ghirlanda {et~al.}(2012)Ghirlanda, Ghisellini, Nava, Salvaterra,
  Tagliaferri, Campana, Covino, {D'Avanzo}, Fugazza, Melandri, \&
  Vergani}]{ghirlanda_impact_2012}
Ghirlanda, G., Ghisellini, G., Nava, L., {et~al.} 2012, Monthly Notices of the
  Royal Astronomical Society, 422, 2553

\bibitem[{Giannios(2012)}]{giannios_peak_2012}
Giannios, D. 2012, Monthly Notices of the Royal Astronomical Society, 422, 3092

\bibitem[{Grindlay \& Team(2009)}]{grindlay_grb_2009}
Grindlay, J., \& Team, E. 2009, in , 18--24

\bibitem[{Guetta {et~al.}(2005)Guetta, Piran, \&
  Waxman}]{guetta_luminosity_2005}
Guetta, D., Piran, T., \& Waxman, E. 2005, The Astrophysical Journal, 619, 412

\bibitem[{Haario {et~al.}(2001)Haario, Saksman, \&
  Tamminen}]{haario_adaptive_2001}
Haario, H., Saksman, E., \& Tamminen, J. 2001, Bernoulli, 7, 223,
  {ArticleType:} research-article / Full publication date: Apr., 2001 /
  Copyright © 2001 International Statistical Institute {(ISI)} and Bernoulli
  Society for Mathematical Statistics and Probability

\bibitem[{Hakkila(2003)}]{hakkila_batse_2003}
Hakkila, J. 2003, in  ({AIP}), 176--178

\bibitem[{Hakkila {et~al.}(2003)Hakkila, Giblin, Roiger, Haglin, Paciesas, \&
  Meegan}]{hakkila_how_2003}
Hakkila, J., Giblin, T.~W., Roiger, R.~J., {et~al.} 2003, The Astrophysical
  Journal, 582, 320

\bibitem[{Hakkila {et~al.}(2000{\natexlab{a}})Hakkila, Haglin, Roiger,
  Mallozzi, Pendleton, \& Meegan}]{hakkila_properties_2000}
Hakkila, J., Haglin, D.~J., Roiger, R.~J., {et~al.} 2000{\natexlab{a}}, in ,
  33--37

\bibitem[{Hakkila {et~al.}(2000{\natexlab{b}})Hakkila, Meegan, Pendleton,
  Mallozzi, Haglin, \& Roiger}]{hakkila_fluence_2000}
Hakkila, J., Meegan, C.~A., Pendleton, G.~N., {et~al.} 2000{\natexlab{b}}, in ,
  48--52

\bibitem[{Hobson {et~al.}(2010)Hobson, Jaffe, Liddle, Mukeherjee, \&
  Parkinson}]{hobson_bayesian_2010}
Hobson, M.~P., Jaffe, A.~H., Liddle, A.~R., Mukeherjee, P., \& Parkinson, D.
  2010, Bayesian Methods in Cosmology

\bibitem[{Hogg \& Turner(1998)}]{hogg_maximum_1998}
Hogg, D.~W., \& Turner, E.~L. 1998, Publications of the Astronomical Society of
  the Pacific, 110, 727, {ArticleType:} research-article / Full publication
  date: June 1998 / Copyright © 1998 The University of Chicago Press

\bibitem[{Hopkins \& Beacom(2006)}]{hopkins_normalization_2006}
Hopkins, A.~M., \& Beacom, J.~F. 2006, The Astrophysical Journal, 651, 142

\bibitem[{Isobe {et~al.}(1990)Isobe, Feigelson, Akritas, \&
  Babu}]{isobe_linear_1990}
Isobe, T., Feigelson, E.~D., Akritas, M.~G., \& Babu, G.~J. 1990, The
  Astrophysical Journal, 364, 104

\bibitem[{Jeffreys(1938)}]{jeffreys_correction_1938}
Jeffreys, H. 1938, Monthly Notices of the Royal Astronomical Society, 98, 190

\bibitem[{Jeffreys(1946)}]{jeffreys_invariant_1946}
---. 1946, Proceedings of the Royal Society of London. Series A. Mathematical
  and Physical Sciences, 186, 453

\bibitem[{Justel {et~al.}(1997)Justel, Peña, \&
  Zamar}]{justel_multivariate_1997}
Justel, A., Peña, D., \& Zamar, R. 1997, Statistics \& Probability Letters,
  35, 251–259

\bibitem[{Kippen {et~al.}(2003)Kippen, Woods, Heise, in't Zand, Briggs, \&
  Preece}]{kippen_spectral_2003}
Kippen, R.~M., Woods, P.~M., Heise, J., {et~al.} 2003, in , 244--247

\bibitem[{Klebesadel {et~al.}(1973)Klebesadel, Strong, \&
  Olson}]{klebesadel_Observations_1973}
Klebesadel, R.~W., Strong, I.~B., \& Olson, R.~A. 1973, The Astrophysical
  Journal Letters, 182, L85

\bibitem[{Kolmogoroff(1941)}]{kolmogoroff_confidence_1941}
Kolmogoroff, A. 1941, The Annals of Mathematical Statistics, 12, 461,
  {ArticleType:} research-article / Full publication date: Dec., 1941 /
  Copyright © 1941 Institute of Mathematical Statistics

\bibitem[{Kommers {et~al.}(2000)Kommers, Lewin, Kouveliotou, van Paradijs,
  Pendleton, Meegan, \& Fishman}]{kommers_intensity_2000}
Kommers, J.~M., Lewin, W. H.~G., Kouveliotou, C., {et~al.} 2000, The
  Astrophysical Journal, 533, 696

\bibitem[{Kouveliotou {et~al.}(1993)Kouveliotou, Meegan, Fishman, Bhat, Briggs,
  Koshut, Paciesas, \& Pendleton}]{kouveliotou_identification_1993}
Kouveliotou, C., Meegan, C.~A., Fishman, G.~J., {et~al.} 1993, The
  Astrophysical Journal Letters, 413, L101

\bibitem[{Kumar \& Piran(2000)}]{kumar_energetics_2000}
Kumar, P., \& Piran, T. 2000, The Astrophysical Journal, 535, 152

\bibitem[{Kutner {et~al.}(2004)Kutner, Neter, Nachtsheim, \&
  Li}]{kutner_applied_2004}
Kutner, M.~H., Neter, J., Nachtsheim, C.~J., \& Li, W. 2004, Applied Linear
  Statistical Models {w/Student} {CD-ROM}, 5th edn. ({McGraw-Hill} Education)

\bibitem[{Levesque(2012)}]{levesque_host_2012}
Levesque, E.~M. 2012, Host Galaxies of Gamma-Ray Bursts

\bibitem[{Levesque {et~al.}(2010{\natexlab{a}})Levesque, Berger, Kewley, \&
  Bagley}]{levesque_host_2010}
Levesque, E.~M., Berger, E., Kewley, L.~J., \& Bagley, M.~M.
  2010{\natexlab{a}}, The Astronomical Journal, 139, 694

\bibitem[{Levesque {et~al.}(2010{\natexlab{b}})Levesque, Kewley, Graham, \&
  Fruchter}]{levesque_high-metallicity_2010}
Levesque, E.~M., Kewley, L.~J., Graham, J.~F., \& Fruchter, A.~S.
  2010{\natexlab{b}}, The Astrophysical Journal Letters, 712, L26

\bibitem[{Levesque {et~al.}(2010{\natexlab{c}})Levesque, Soderberg, Foley,
  Berger, Kewley, Chakraborti, Ray, Torres, Challis, Kirshner, Barthelmy,
  Bietenholz, Chandra, Chaplin, Chevalier, Chugai, Connaughton, Copete, Fox,
  Fransson, Grindlay, Hamuy, Milne, Pignata, Stritzinger, \&
  Wieringa}]{levesque_high-metallicity_2010-1}
Levesque, E.~M., Soderberg, A.~M., Foley, R.~J., {et~al.} 2010{\natexlab{c}},
  The Astrophysical Journal Letters, 709, L26

\bibitem[{Li(2007)}]{li_redshift_2007}
Li, L.-X. 2007, Monthly Notices of the Royal Astronomical Society, 374, L20

\bibitem[{Li(2008)}]{li_star_2008}
---. 2008, Monthly Notices of the Royal Astronomical Society, 388, 1487

\bibitem[{Lloyd {et~al.}(2000)Lloyd, Petrosian, \&
  Mallozzi}]{lloyd_cosmological_2000}
Lloyd, N.~M., Petrosian, V., \& Mallozzi, R.~S. 2000, The Astrophysical
  Journal, 534, 227

\bibitem[{Mallozzi {et~al.}(1995)Mallozzi, Paciesas, Pendleton, Briggs, Preece,
  Meegan, \& Fishman}]{mallozzi_nu_1995}
Mallozzi, R.~S., Paciesas, W.~S., Pendleton, G.~N., {et~al.} 1995, The
  Astrophysical Journal, 454, 597

\bibitem[{Mazets \& Golenetskii(1981)}]{mazets_recent_1981}
Mazets, E.~P., \& Golenetskii, S.~V. 1981, Astrophysics and Space Science, 75,
  47

\bibitem[{Meegan {et~al.}(1992)Meegan, Fishman, Wilson, Horack, Brock,
  Paciesas, Pendleton, \& Kouveliotou}]{meegan_spatial_1992}
Meegan, C.~A., Fishman, G.~J., Wilson, R.~B., {et~al.} 1992, Nature, 355, 143

\bibitem[{Metzger {et~al.}(1997)Metzger, Djorgovski, Kulkarni, Steidel,
  Adelberger, Frail, Costa, \& Frontera}]{metzger_spectral_1997}
Metzger, M.~R., Djorgovski, S.~G., Kulkarni, S.~R., {et~al.} 1997, Nature, 387,
  878

\bibitem[{Nakar(2007)}]{nakar_short-hard_2007}
Nakar, E. 2007, Physics Reports, 442, 166

\bibitem[{Nakar \& Piran(2005)}]{nakar_outliers_2005}
Nakar, E., \& Piran, T. 2005, Monthly Notices of the Royal Astronomical
  Society, 360, L73

\bibitem[{Nava {et~al.}(2008)Nava, Ghirlanda, Ghisellini, \&
  Firmani}]{nava_peak_2008}
Nava, L., Ghirlanda, G., Ghisellini, G., \& Firmani, C. 2008, Monthly Notices
  of the Royal Astronomical Society, 391, 639

\bibitem[{Nemiroff(2000)}]{nemiroff_pulse_2000}
Nemiroff, R.~J. 2000, The Astrophysical Journal, 544, 805

\bibitem[{Nemiroff {et~al.}(1994)Nemiroff, Norris, Bonnell, Wickramasinghe,
  Kouveliotou, Paciesas, Fishman, \& Meegan}]{nemiroff_gross_1994}
Nemiroff, R.~J., Norris, J.~P., Bonnell, J.~T., {et~al.} 1994, The
  Astrophysical Journal Letters, 435, L133

\bibitem[{Nishimura(1988)}]{nishimura_what_1988}
Nishimura, J. 1988, 413--425

\bibitem[{Norris {et~al.}(2005)Norris, Bonnell, Kazanas, Scargle, Hakkila, \&
  Giblin}]{norris_long-lag_2005}
Norris, J.~P., Bonnell, J.~T., Kazanas, D., {et~al.} 2005, The Astrophysical
  Journal, 627, 324

\bibitem[{Paciesas {et~al.}(1999)Paciesas, Meegan, Pendleton, Briggs,
  Kouveliotou, Koshut, Lestrade, {McCollough}, Brainerd, Hakkila, Henze,
  Preece, Connaughton, Kippen, Mallozzi, Fishman, Richardson, \&
  Sahi}]{paciesas_fourth_1999}
Paciesas, W.~S., Meegan, C.~A., Pendleton, G.~N., {et~al.} 1999, The
  Astrophysical Journal Supplement Series, 122, 465

\bibitem[{Paczynski(1986)}]{paczynski_gamma-ray_1986}
Paczynski, B. 1986, The Astrophysical Journal Letters, 308, L43

\bibitem[{Peacock(1983)}]{peacock_two-dimensional_1983}
Peacock, J.~A. 1983, Monthly Notices of the Royal Astronomical Society, 202,
  615

\bibitem[{Pendleton {et~al.}(1998)Pendleton, Hakkila, \&
  Meegan}]{pendleton_batse_1998}
Pendleton, G.~N., Hakkila, J., \& Meegan, C.~A. 1998 ({ASCE}), 899--903

\bibitem[{Pendleton {et~al.}(1995)Pendleton, Paciesas, Mallozzi, Koshut,
  Fishman, Meegan, Wilson, Horack, \& Lestrade}]{Pendleton_detector_1995}
Pendleton, G.~N., Paciesas, W.~S., Mallozzi, R.~S., {et~al.} 1995, Nuclear
  Instruments and Methods in Physics Research A, 364, 567

\bibitem[{Petrosian(1993)}]{petrosian_interpretation_1993}
Petrosian, V. 1993, The Astrophysical Journal Letters, 402, L33

\bibitem[{Petrosian \& Lee(1996)}]{Petrosian_fluence_1996}
Petrosian, V., \& Lee, T.~T. 1996, The Astrophysical Journal Letters, 467, L29

\bibitem[{Petrosian {et~al.}(1999)Petrosian, Lloyd, \&
  Lee}]{Petrosian_cosmological_1999}
Petrosian, V., Lloyd, N., \& Lee, A. 1999, in , 235

\bibitem[{Porciani \& Madau(2001)}]{porciani_association_2001}
Porciani, C., \& Madau, P. 2001, The Astrophysical Journal, 548, 522

\bibitem[{Press(1972)}]{press_multivariate_1972}
Press, S. 1972, Journal of Multivariate Analysis, 2, 444

\bibitem[{Press {et~al.}(1992)Press, Teukolsky, Vetterling, \&
  Flannery}]{press_numerical_1992}
Press, W.~H., Teukolsky, S.~A., Vetterling, W.~T., \& Flannery, B.~P. 1992,
  Numerical recipes in {FORTRAN.} The art of scientific computing

\bibitem[{Racusin {et~al.}(2011)Racusin, Oates, Schady, Burrows, de~Pasquale,
  Donato, Gehrels, Koch, {McEnery}, Piran, Roming, Sakamoto, Swenson, Troja,
  Vasileiou, Virgili, Wanderman, \& Zhang}]{racusin_fermi_2011}
Racusin, J.~L., Oates, S.~R., Schady, P., {et~al.} 2011, The Astrophysical
  Journal, 738, 138

\bibitem[{Ramirez-Ruiz \& Fenimore(2000)}]{ramirez-ruiz_pulse_2000}
Ramirez-Ruiz, E., \& Fenimore, E.~E. 2000, The Astrophysical Journal, 539, 712

\bibitem[{Rees \& Mészáros(2005)}]{rees_dissipative_2005}
Rees, M.~J., \& Mészáros, P. 2005, The Astrophysical Journal, 628, 847

\bibitem[{Reichart \& Lamb(2001)}]{reichart_construction_2001}
Reichart, D.~E., \& Lamb, D.~Q. 2001, {AIP} Conference Proceedings, 586, 599

\bibitem[{Rousseeuw {et~al.}(1996)Rousseeuw, Kaufman, \&
  Trauwaert}]{rousseeuw_fuzzy_1996}
Rousseeuw, P., Kaufman, L., \& Trauwaert, E. 1996, Computational Statistics \&
  Data Analysis, 23, 135

\bibitem[{Ryde {et~al.}(2006)Ryde, Björnsson, Kaneko, Mészáros, Preece, \&
  Battelino}]{ryde_gamma-ray_2006}
Ryde, F., Björnsson, C.-I., Kaneko, Y., {et~al.} 2006, The Astrophysical
  Journal, 652, 1400

\bibitem[{Salvaterra \& Chincarini(2007)}]{salvaterra_gamma-ray_2007}
Salvaterra, R., \& Chincarini, G. 2007, The Astrophysical Journal Letters, 656,
  L49

\bibitem[{Salvaterra {et~al.}(2009)Salvaterra, Guidorzi, Campana, Chincarini,
  \& Tagliaferri}]{salvaterra_evidence_2009}
Salvaterra, R., Guidorzi, C., Campana, S., Chincarini, G., \& Tagliaferri, G.
  2009, Monthly Notices of the Royal Astronomical Society, 396, 299

\bibitem[{Salvaterra {et~al.}(2012)Salvaterra, Campana, Vergani, Covino,
  {D'Avanzo}, Fugazza, Ghirlanda, Ghisellini, Melandri, Nava, Sbarufatti,
  Flores, Piranomonte, \& Tagliaferri}]{salvaterra_complete_2012}
Salvaterra, R., Campana, S., Vergani, S.~D., {et~al.} 2012, The Astrophysical
  Journal, 749, 68

\bibitem[{Schaefer(2007)}]{schaefer_hubble_2007}
Schaefer, B.~E. 2007, The Astrophysical Journal, 660, 16

\bibitem[{Schmidt(1999)}]{schmidt_luminosities_1999}
Schmidt, M. 1999, The Astrophysical Journal Letters, 523, L117

\bibitem[{Schmidt(2001)}]{schmidt_luminosity_2001}
---. 2001, The Astrophysical Journal, 552, 36

\bibitem[{Schmidt(2009)}]{schmidt_gamma-ray_2009}
---. 2009, The Astrophysical Journal, 700, 633

\bibitem[{Sethi \& Bhargavi(2001)}]{sethi_luminosity_2001}
Sethi, S., \& Bhargavi, S.~G. 2001, Astronomy and Astrophysics, 376, 10

\bibitem[{Shahmoradi \& Nemiroff(2009)}]{shahmoradi_how_2009}
Shahmoradi, A., \& Nemiroff, R. 2009, {AIP} Conference Proceedings, 1133, 425

\bibitem[{Shahmoradi \& Nemiroff(2010)}]{shahmoradi_hardness_2010}
Shahmoradi, A., \& Nemiroff, R.~J. 2010, Monthly Notices of the Royal
  Astronomical Society, 407, 2075

\bibitem[{Shahmoradi \& Nemiroff(2011)}]{shahmoradi_possible_2011}
---. 2011, Monthly Notices of the Royal Astronomical Society, 411, 1843

\bibitem[{Smirnov(1948)}]{smirnov_table_1948}
Smirnov, N. 1948, The Annals of Mathematical Statistics, 19, 279,
  {ArticleType:} research-article / Full publication date: Jun., 1948 /
  Copyright © 1948 Institute of Mathematical Statistics

\bibitem[{Stanek {et~al.}(2006)Stanek, Gnedin, Beacom, Gould, Johnson,
  Kollmeier, Modjaz, Pinsonneault, Pogge, \& Weinberg}]{stanek_protecting_2006}
Stanek, K.~Z., Gnedin, O.~Y., Beacom, J.~F., {et~al.} 2006, Acta Astronomica,
  56, 333

\bibitem[{Stern {et~al.}(2002)Stern, Atteia, \& Hurley}]{stern_evidence_2002}
Stern, B.~E., Atteia, J.-L., \& Hurley, K. 2002, The Astrophysical Journal,
  578, 304

\bibitem[{Stern {et~al.}(2001)Stern, Tikhomirova, Kompaneets, Svensson, \&
  Poutanen}]{stern_off-line_2001}
Stern, B.~E., Tikhomirova, Y., Kompaneets, D., Svensson, R., \& Poutanen, J.
  2001, The Astrophysical Journal, 563, 80

\bibitem[{Strohmayer {et~al.}(1998)Strohmayer, Fenimore, Murakami, \&
  Yoshida}]{strohmayer_x-ray_1998}
Strohmayer, T.~E., Fenimore, E.~E., Murakami, T., \& Yoshida, A. 1998, The
  Astrophysical Journal, 500, 873

\bibitem[{Thompson {et~al.}(2007)Thompson, Mészáros, \&
  Rees}]{thompson_thermalization_2007}
Thompson, C., Mészáros, P., \& Rees, M.~J. 2007, The Astrophysical Journal,
  666, 1012

\bibitem[{Tikhomirova {et~al.}(2006)Tikhomirova, Stern, Kozyreva, \&
  Poutanen}]{tikhomirova_new_2006}
Tikhomirova, Y., Stern, B.~E., Kozyreva, A., \& Poutanen, J. 2006, Monthly
  Notices of the Royal Astronomical Society, 367, 1473

\bibitem[{Wanderman \& Piran(2010)}]{wanderman_luminosity_2010}
Wanderman, D., \& Piran, T. 2010, Monthly Notices of the Royal Astronomical
  Society, 406, 1944

\bibitem[{Wang \& Dai(2011)}]{wang_evolving_2011}
Wang, F.~Y., \& Dai, Z.~G. 2011, The Astrophysical Journal Letters, 727, L34

\bibitem[{Wickramasinghe \& Ukwatta(2010)}]{wickramasinghe_analytical_2010}
Wickramasinghe, T., \& Ukwatta, T.~N. 2010, Monthly Notices of the Royal
  Astronomical Society, 406, 548

\bibitem[{Woosley(1993)}]{woosley_gamma-ray_1993}
Woosley, S.~E. 1993, The Astrophysical Journal, 405, 273

\bibitem[{Woosley \& Heger(2006)}]{woosley_progenitor_2006}
Woosley, S.~E., \& Heger, A. 2006, The Astrophysical Journal, 637, 914

\bibitem[{Yonetoku {et~al.}(2004)Yonetoku, Murakami, Nakamura, Yamazaki, Inoue,
  \& Ioka}]{yonetoku_gamma-ray_2004}
Yonetoku, D., Murakami, T., Nakamura, T., {et~al.} 2004, The Astrophysical
  Journal, 609, 935

\bibitem[{Zhang {et~al.}(2007)Zhang, Zhang, Liang, Gehrels, Burrows, \&
  Mészáros}]{zhang_making_2007}
Zhang, B., Zhang, B.-B., Liang, E.-W., {et~al.} 2007, The Astrophysical Journal
  Letters, 655, L25

\end{thebibliography}

\end{document}